\documentclass{elsart}

\usepackage[square,comma]{natbib}
\usepackage{graphicx}
\usepackage{pxfonts}

\usepackage{amssymb}

\usepackage{epstopdf}

\journal{}

\begin{document}

\thispagestyle{empty}
\begin{Large}
\textbf{DEUTSCHES ELEKTRONEN-SYNCHROTRON}

\textbf{\large{Ein Forschungszentrum der
Helmholtz-Gemeinschaft}\\}
\end{Large}

DESY 14-129

July 2014

\begin{eqnarray}
\nonumber &&\cr \nonumber && \cr \nonumber &&\cr
\end{eqnarray}
\begin{eqnarray}
\nonumber
\end{eqnarray}
\begin{center}
\begin{Large}
\textbf{Brightness of Synchrotron radiation from Undulators and Bending Magnets}
\end{Large}
\begin{eqnarray}
\nonumber &&\cr \nonumber && \cr
\end{eqnarray}

\begin{large}
Gianluca Geloni,
\end{large}
\textsl{\\European XFEL GmbH, Hamburg}
\begin{large}

Vitali Kocharyan and Evgeni Saldin
\end{large}
\textsl{\\Deutsches Elektronen-Synchrotron DESY, Hamburg}
\begin{eqnarray}
\nonumber
\end{eqnarray}
\begin{eqnarray}
\nonumber
\end{eqnarray}
ISSN 0418-9833
\begin{eqnarray}
\nonumber
\end{eqnarray}
\begin{large}
\textbf{NOTKESTRASSE 85 - 22607 HAMBURG}
\end{large}
\end{center}
\clearpage
\newpage

\begin{frontmatter}

\title{Brightness of Synchrotron radiation from Undulators and Bending Magnets}


\author[XFEL]{Gianluca Geloni}
\author[DESY]{Vitali Kocharyan}
\author[DESY]{Evgeni Saldin}

\address[XFEL]{European XFEL GmbH, Hamburg, Germany}
\address[DESY]{Deutsches Elektronen-Synchrotron (DESY), Hamburg,
Germany}

\begin{abstract}
We consider the maximum of the Wigner distribution (WD) of synchrotron radiation (SR) fields as a possible definition of SR source brightness. Such figure of merit was originally introduced in the SR community by Kim. The brightness defined in this way is always positive and, in the geometrical optics limit, can be interpreted as maximum density of photon flux in phase space. For undulator and bending magnet radiation from a single electron, the WD function can be explicitly calculated. In the case of an electron beam with a finite emittance the brightness is given by the maximum of the convolution of a single electron WD function and the probability distribution of the electrons in phase space. In the particular case when both electron beam size and electron beam divergence dominate over the diffraction size and the diffraction angle, one can use a geometrical optics approach. However, there are intermediate regimes when only the electron beam size or the electron beam divergence dominate. In this asymptotic cases the geometrical optics approach is still applicable, and the brightness definition used here yields back once more the maximum photon flux density in phase space. In these intermediate regimes we find a significant numerical disagreement between exact calculations and the approximation for undulator brightness currently used in literature. We extend the WD formalism to a satisfactory theory for the brightness of a bending magnet. We find that in the intermediate regimes the usually accepted approximation for bending magnet brightness turns out to be inconsistent even parametrically.

\end{abstract}




\end{frontmatter}


\clearpage

\section{\label{sec:intro} Introduction}

Ultra-relativistic electrons accelerated through magnetic fields generate electromagnetic radiation, called Synchrotron Radiation (SR). Emission from SR sources may range from far infrared (FIR) to X-ray frequencies. The use of proper physical quantities and figures of merit enable their adequate characterization and deep understanding. The properties of a SR source are usually described using three quantities: the spectral  flux, the maximum of the angular spectral flux and the brightness \cite{GREE}-\cite{TALM}. Among these quantities the brightness\footnote{What we really refer to, here and in the following, with the term "brightness" is actually "the spectral brightness" or "the brilliance". For simplicity though, we will call it "brightness" throughout the text.} has a very important role.

The physical meaning of brightness can be best understood by considering the imaging of the source on any experimental sample. Typically, only a small fraction of the photons in the beam can be effectively focused. The brightness is an appropriate figure of merit for estimating the photon flux density on the sample, and was originally defined with the help of traditional radiometry. Traditional radiometry is based on a geometrical optics approach. The basic quantity in this discipline is the radiance, which measures a spectral photon flux per unit area per unit projection solid angle. Since the radiance can be evaluated at any point along a photon beamline, it is associated  with specific locations within an optical system, including an image plane where, usually, an experimental sample is placed. All other radiometric units can be derived from the radiance integrating over area or solid angle. Integrating over the solid angle yields a spatial flux. Integrating over the area yields the angular flux. Integrating over both area and solid angle yields total flux \cite{PRIV,PALM}.  In this picture, the radiance is the photon flux density in phase space. In non-dissipative cases where Liouville theorem holds, the radiance is an invariant quantity down the photon beamline.

The main issue with this concept stems from the fact that traditional radiometry only provides a natural description of the properties of light from incoherent sources.  Second generation SR sources are characterized by poor transverse coherence, and application of the concept of radiance to SR allows for phase space analysis of the X-ray radiation. For these sources the brightness is nothing more than the radiance. However, with the advent of third generation SR sources, electron beams began to have ultra-small emittances and quantities used in traditional radiometry needed to be generalized to SR sources of arbitrary state of transverse coherence. The question then arises whether it is possible to find a definition of brightness that, irrespective of the state of coherence of the SR source, has all properties that one normally associates with it in the geometrical optics limit.

The basic question concerning the characterization of third generation SR sources is the definition of brightness in terms of electromagnetic fields, and their statistical properties (i.e on the basis of classical relativistic electrodynamics and statistical optics). We begin our discussion with the generalization of the brightness definition first proposed by Kim \cite{KIM1,KIM3}, which is essentially the Wigner distribution (WD) \cite{WIGN} of the SR electric field. In \cite{KIM1,KIM3}, the characteristics of the odd harmonics field from undulator sources in resonance approximation was analyzed in detail using the WD formalism. The WD was explicitly derived both in the diffraction limited regime and in the electron beam size- and divergence-dominated regime.  After Kim's pioneering papers \cite{KIM1,KIM3} no further theoretical progress was made. In particular, the considerations by Kim \cite{KIM1,KIM3} on bending magnet and wiggler brightness in terms of the WD formalism were not further developed into a satisfactory theory.

In literature, the brightness of a SR source is sometimes defined as the Wigner distribution itself, i.e. as a phase space quasi-probability function \cite{KIM1,KIM3,CIOC,ELLE}. As such,  for SR sources of arbitrary state of transverse coherence, it is not guaranteed to be positive. Moreover it is convenient to introduce a figure of merit which always gives back a single, positive number and can serve, at the same time, as a measure for the WD. In this article we shall define the brightness for any synchrotron source as the maximum value of the WD. Note that the brightness defined in this way is always a positive quantity. We can regard it as a self-evident generalization of Kim's choice of the on-axis peak value of the WD as a figure of merit for the undulator case but, surprisingly, we failed to find it in literature\footnote{Several alternative definitions of brightness can be found in literature, see e.g. \cite{BAZA, ELLE}. Some of them are formulated in terms of integrals of the squared WD. Others keep to the definition of brightness as the on-axis  value of the WD. We will not consider them here, as these are in contrast with the concept of brightness in the geometrical optics limit, which stems from classical radiometry and coincides with the maximum flux density in phase space. Formulating the theory of brightness in the language of Wigner distributions has only one guideline, a particular correspondence principle, which is based on the assumption that the formalism involved in the calculation of the brightness must include radiometry as a limiting case.}.


In the following of this section we present an introductory discussion of brightness of synchrotron radiation sources in general. Then, in section \ref{oldres} we give a summary of results in literature, and an overview of novel findings. In section \ref{sec:due} we discuss an analysis of undulator brightness, and of the approximations proposed by Kim \cite{KIM1,KIM3}. In the limiting cases where the geometrical optics treatment can still be applied, but only the electron beam size or the divergence dominate on the diffraction size and angle we find a significant numerical disagreement between exact and approximated calculations. In section \ref{sec:tre} we come to the new results and extend the WD formalism to a satisfactory theory of bending magnet brightness.

\section{\label{sec:wiggen} Wigner distribution and SR sources}

As already discussed, we will focus on the description of a SR source in terms of WD. We will be interested in the case of an ultra relativistic electron beam going through a certain magnetic system. We will discuss of an undulator to illustrate our reasoning, but the considerations in this section, being fully general, apply to any other magnetic system (wiggler, bending magnet) as well. SR theory is naturally developed in the space-frequency domain, as one is usually interested into radiation properties at a given position in space at a certain frequency. In this article we define the relation between temporal and frequency domain via the following definition of Fourier transform pairs:

\begin{eqnarray}
&&\bar{f}(\omega) = \int_{-\infty}^{\infty} dt~ f(t) \exp(i \omega t ) \leftrightarrow
f(t) = \frac{1}{2\pi}\int_{-\infty}^{\infty} d\omega \bar{f}(\omega) \exp(-i \omega t) ~.
\label{ftdef2}
\end{eqnarray}
We call $z$ the observation distance along the optical axis of the system and $\vec{r}$ fixes the transverse position of the observer. The contribution of the $k$-th  electron to the field depends on the transverse offset $\vec{l}_k$ and deflection angles $\vec{\eta}_k$ that the electron has at some reference point on the optical axis $z$ e.g. the center of the undulator, that we will take as $z=0$. Moreover, the arrival time $t_k$ at position $z=0$ has the effect of multiplying the electric field in space-frequency domain by a phase factor $\exp(i \omega t_k)$, $\omega$ being the frequency.  Any fixed polarization component of the total field in the space frequency domain, which is a scalar quantity, can therefore be written as

\begin{eqnarray}
\bar{E}_\mathrm{tot}(z,\vec{r},\omega) = \sum_{k=1}^{N_e} \bar{E}(\vec{\eta}_k, \vec{l}_k, \vec{r}, z, \omega) \exp(i\omega t_k)~,
\label{total}
\end{eqnarray}
where $\vec{\eta}_k$, $\vec{l}_k$ and $t_k$ are random variables, and $N_e$ is the number of electrons in the bunch. Note that the single-electron field $\bar{E}$ in Eq. (\ref{total}) is a complex quantity, and can be written as $\bar{E} = A_k \exp(i \phi_k)$, with $A_k>0$ and $\phi_k$ real numbers. It follows that the SR field at a fixed frequency and position is a sum of many phasors, one for each electron, of the form $A_k \exp(i \phi_k + i \omega t_k)$.

Elementary phasors composing the sum obey three important statistical properties, that are satisfied in all SR problems of interest. First, random variables $t_k$ are statistically independent of each other, and of variables $\vec{\eta}_k$ and $\vec{l}_k$. This assumption  follows from the properties of shot noise in a storage ring, which is a fundamental effect related with quantum fluctuations. Second, the amplitudes $A_k$  have obviously finite mean $\langle A_k \rangle$ and finite second moment $\langle A_k^2 \rangle$. Third, we assume that the electron bunch duration $\sigma_T$  is large enough so that $\omega \sigma_T \gg 1$. Under these non-restrictive assumptions  the phases $\omega t_k$  can be regarded as uniformly distributed on the interval $(0, 2\pi)$. Moreover, with the help of the central limit theorem, it can be demonstrated that real and imaginary parts of $E_\mathrm{tot}$ are distributed in accordance to a Gaussian law. It follows that SR is Gaussian random process. Moreover, since one deals with pulses of finite duration, the process is intrinsically non-stationary\footnote{Non-stationarity may or may not be detected, depending on the the monochromator bandwidth in the actual setup.}. An important consequence of the fact that the SR random process can be considered Gaussian is that higher-order correlation functions can be expressed in terms of the second order correlation function with the help of the moment theorem \cite{GOOD}.

As a result, the knowledge of the second-order correlation function in the space-frequency domain is all we need to completely characterize the signal from a statistical viewpoint. The following definition holds:

\begin{eqnarray}
\Gamma_\omega(z, \vec{r}_1, \vec{r}_2, \omega_1, \omega_2) = \langle \bar{E}_\mathrm{tot}(z, \vec{r}_1, \omega_1) \bar{E}^*_\mathrm{tot}(z, \vec{r}_2,\omega_2) \rangle ~,
\label{gamma}
\end{eqnarray}
where brackets $\langle...\rangle$ indicate ensemble average over electron bunches. For any given function $w\left(\vec{\eta}_k, \vec{l}_k, t_k\right)$, the ensemble average is defined as

\begin{eqnarray}
\left<w\left(\vec{\eta}_k, \vec{l}_k, t_k\right)\right> = \int_{-\infty}^{\infty}
d\vec{\eta}_k \int_{-\infty}^{\infty} d\vec{l}_k \int_{-\infty}^{\infty} d t_k
w\left(\vec{\eta}_k, \vec{l}_k, t_k\right) P\left(\vec{\eta}_k,
\vec{l}_k, t_k\right) \label{ensembledef}~,
\end{eqnarray}
where integrals in $d\vec{l}_k$ and $d\vec{\eta}_k$ span over all offsets and deflections, and $P=P(\vec{\eta}_k, \vec{l}_k, t_k)$ indicates the probability density distribution in the joint random variables $\vec{\eta}_k$, $\vec{l}_k$, and $t_k$. The already discussed independence of $t_k$ from $\vec{l}_k$ and $\vec{\eta}_k$ allows to write $P$ as

\begin{eqnarray}
P\left(\vec{\eta}_k, \vec{l}_k, t_k\right) =
f_\bot\left(\vec{l}_k,\vec{\eta}_k\right) f(t_k)~.
\label{independence}
\end{eqnarray}
Here $f$ is the longitudinal bunch profile of the electron beam, while $f_\bot$ is the transverse phase space distribution.

Substituting Eq. (\ref{total}) in Eq. (\ref{gamma}) one has

\begin{eqnarray}
\Gamma_{\omega} = \left<\sum_{m,n=1}^{N_e} \bar{E}
\left(\vec{\eta}_m ,\vec{l}_m , z ,\vec{r}_{{} 1}, \omega_1\right)
\bar{E}^* \left(\vec{\eta}_n,\vec{l}_n, z,\vec{r}_{{} 2},
\omega_2\right) \exp{[i(\omega_1 t_m- \omega_2 t_n)]} \right> .\cr
&& \label{gamma2}
\end{eqnarray}
Expansion of Eq. (\ref{gamma2}) gives

\begin{eqnarray}
&&\Gamma_{\omega} = \sum_{m=1}^{N_e} \left\langle \bar{E}
\left(\vec{\eta}_m,\vec{l}_m,z,\vec{r}_{{} 1}, \omega_1\right)
\bar{E}^*\left(\vec{\eta}_m,\vec{l}_m,z,\vec{r}_{{} 2},
\omega_2\right) \exp{[i( \omega_1- \omega_2) t_m]} \right\rangle
\cr && + \sum_{m\ne n} \left\langle
\bar{E}\left(\vec{\eta}_m,\vec{l}_m,z,\vec{r}_{{} 1},
\omega_1\right) \exp{(i \omega_1
t_m)}\right\rangle\left\langle\bar{E}^*
\left(\vec{\eta}_n,\vec{l}_n,z,\vec{r}_{{} 2}, \omega_2\right)
\exp{(- i \omega_2 t_n)} \right\rangle.\cr && \label{gamma3}
\end{eqnarray}
With the help of Eq. (\ref{ensembledef}) and Eq. (\ref{independence}) one sees that the ensemble average $\langle \exp{(i \omega t_k)} \rangle$ is essentially the Fourier transform\footnote{Eq. (\ref{FTlong}) coincides with our definition of temporal Fourier transform, Eq. (\ref{ftdef2}), but it will not change if one uses another convention for the Fourier Transform definition.} of the longitudinal bunch profile function $f$, that is

\begin{equation}
\left\langle \exp{(i \omega t_k)} \right\rangle =
\int_{-\infty}^{\infty} d t_k f(t_k) \exp(i\omega t_k) \equiv
\bar{f}(\omega)~. \label{FTlong}
\end{equation}
Using Eq. (\ref{FTlong}), Eq. (\ref{gamma3}) can be written as

\begin{eqnarray}
\Gamma_{\omega} &&= \sum_{m=1}^{N_e} \bar{f}( \omega_1- \omega_2)
\left \langle \bar{E} \left(\vec{\eta}_m,\vec{l}_m, z, \vec{r}_{{}
1}, \omega_1\right) \bar{E}^*
\left(\vec{\eta}_m,\vec{l}_m,z,\vec{r}_{{} 2}, \omega_2\right)
\right\rangle \cr && + \sum_{m\ne n} \bar{f}( \omega_1)\bar{f}(-
\omega_2)  \left\langle
\bar{E}\left(\vec{\eta}_m,\vec{l}_m,z,\vec{r}_{{} 1},
\omega_1\right) \right\rangle
\left\langle\bar{E}^*\left(\vec{\eta}_n, \vec{l}_n,z,\vec{r}_{{}
2}, \omega_2\right) \right\rangle ~, \label{gamma4}
\end{eqnarray}
where it is interesting to note that $\bar{f}(- \omega_2) = \bar{f}^*(\omega_2)$ because ${f}$ is a real function. When the radiation wavelengths of interest are much shorter than the bunch length we can safely neglect the second term on the right hand side of Eq. (\ref{gamma4}) since the form factor $\bar{f}(\omega)$ goes rapidly to zero for frequencies larger than the characteristic frequency associated with the bunch length: think for instance, at a millimeter long bunch compared with radiation in the Angstrom wavelength range\footnote{When the radiation wavelength of interested is comparable or longer than the bunch length, the second term in Eq. (\ref{gamma4}) is dominant with respect to the first, because it scales with the number of particles \textit{squared}: in this case, analysis of the second term leads to a treatment of Coherent Synchrotron Radiation (CSR) phenomena. In this paper we will not be concerned with CSR and we will neglect the second term in Eq. (\ref{gamma4}), assuming that the radiation wavelength of interest is shorter than the bunch length. Also note that $\bar{f}( \omega_1- \omega_2)$ depends on the \textit{difference} between $ \omega_1$ and $ \omega_2$, and the first term cannot be neglected.}. Therefore we write

\begin{eqnarray}
\Gamma_{\omega} &&= \sum_{m=1}^{N_e} \bar{f}( \omega_1- \omega_2)
\left \langle \bar{E} \left(\vec{\eta}_m,\vec{l}_m, z, \vec{r}_{{}
1}, \omega_1\right) \bar{E}^*
\left(\vec{\eta}_m,\vec{l}_m,z,\vec{r}_{{} 2}, \omega_2\right)
\right\rangle \cr && = N_e \bar{f}( \omega_1- \omega_2) \left
\langle \bar{E}\left(\vec{\eta},\vec{l}, z, \vec{r}_{{} 1},
\omega_1\right) \bar{E}^* \left(\vec{\eta},\vec{l},z,\vec{r}_{{}
2}, \omega_2\right) \right\rangle  ~. \label{gamma5}
\end{eqnarray}
As one can see from Eq. (\ref{gamma5}) each electron is correlated just with itself: cross-correlation terms between different electrons was included in the second term on the right hand side of Eq. (\ref{gamma4}), which has been dropped.

On the one hand, for an electron bunch with rms duration $\sigma_T$ the characteristic scale of $\bar{f}$ is given by $1/\sigma_T$. On the other hand,  in all our cases of interest  $\bar{E}$ has a much slower dependence on $\omega$ than $1/\sigma_T$. As a result, $\bar{E}$ does not vary appreciably on the characteristic scale of $\bar{f}$\footnote{SR radiation expands into the bandwidth $\Delta \omega \sim \omega$ and $\Delta \omega \sim \omega/N_w$ for bending magnet sources and undulator sources, respectively. Here $N_w$ is the number of undulator periods. In all cases of practical interest $\omega \sigma_T/N_w \gg 1$.}. We can therefore simplify Eq. (\ref{gamma5}) to

\begin{eqnarray}
\Gamma_\omega (z, \vec{r}_1, \vec{r}_2, \omega_1, \omega_2) = N_e \bar{f}(\omega_1 - \omega_2) G(z, \vec{r}_1, \vec{r}_2, \omega_1)
\label{breakGamma}
\end{eqnarray}

where

\begin{equation}
G(z,\vec{r}_{{} 1},\vec{r}_{{} 2}, \omega) \equiv \left\langle
\bar{E} \left(\vec{\eta},\vec{l},z,\vec{r}_{{} 1}, \omega\right)
\bar{E}^*\left(\vec{\eta},\vec{l},z,\vec{r}_{{} 2}, \omega\right)
\right\rangle~\label{coore}
\end{equation}
is known as cross-spectral density. Before proceeding we introduce, for future reference, the notion of spectral degree of coherence, $g$, that can be presented as a function of $\vec{r}_1$ and $\vec{r}_2$ as

\begin{eqnarray}
g(\vec{r}_1, \vec{r}_2) = \frac{G(\vec{r}_1, \vec{r}_2)}{\left[G(\vec{r}_1, \vec{r}_1)G(\vec{r}_2, \vec{r}_2)\right]^{1/2}} = \frac{G(\vec{r}_1, \vec{r}_2)}{\left[\langle |\bar{E}(\vec{r}_1)|^2\rangle \langle |\bar{E}(\vec{r}_2)|^2\rangle \right]^{1/2}},
\label{gsmall}
\end{eqnarray}
where, for notational simplicity, we did not indicate the dependence of the single particle fields on $\vec{\eta},\vec{l}, z$ and $\omega$.  The function $g$ is normalized to unity by definition, i.e. $g(\vec{r},\vec{r}) = 1$.

Eq. (\ref{breakGamma}) fully characterizes the system under study from a statistical viewpoint. Correlation in frequency and space are expressed by two separate factors. In particular, spatial correlation is expressed by the cross-spectral density function under non-restrictive assumptions on the SR bandwidth and on the bunch duration.

Among all possible SR sources, there exists an important class called quasi-homogeneous. Quasi-homogeneity is nothing but the spatial analogue of quasi-stationarity. In general, quasi-homogeneous sources are particular class of Schell's model sources \cite{MAND}. Schell's model sources are defined by the condition that there is some position down the beamline, which we will call $z=0$ without loss of generality, such that their cross-spectral density (note that Eq. (\ref{coore}) is defined at any position $z$ down the beamline) is of the form

\begin{eqnarray}
G(r_1, r_2) = \langle |\bar{E}(\vec{r}_1)|^2\rangle^{1/2} \langle |\bar{E}(\vec{r}_2)|^2\rangle^{1/2}  g(\vec{r}_1 - \vec{r}_2)  ~.
\label{SchM}
\end{eqnarray}
Equivalently, one may simply define Schell's model sources using the condition that the spectral degree of coherence depends on the positions across the source only through the difference $\Delta \vec{r} = \vec{r}_1 - \vec{r}_2$. Quasi-homogeneous sources are Schell's sources obeying the extra assumption that $\langle |\bar{E}(\vec{r})|\rangle^2 $  varies so slowly with position that it is approximately constant over distances across the source, which are of the order of the correlation length, which is the effective width of $g(\Delta \vec{r})$. Because of this, for quasi-homogeneous sources we are allowed to make the approximation

\begin{eqnarray}
G(\vec{r}, \Delta \vec{r}) = I(\vec{r}) g(\Delta \vec{r})~ ,
\label{qh}
\end{eqnarray}
where now $I(\vec{r}) \equiv \langle |\bar{E}(\vec{r})|\rangle^2$  is proportional to the intensity distribution at the source. By definition, quasi-homogeneous synchrotron sources obey a very particular kind of random process that is spatially ergodic. Qualitatively, a given random process is spatially ergodic when all ensemble averages can be substituted by 2D space averages. Remembering this property one can derive the expression for the spectral degree of coherence

\begin{eqnarray}
g(\Delta \vec{r}) \sim  \int  \bar{E}(\vec{r}) \bar{E}^*(\vec{r}+ \Delta \vec{r}) d\vec{r} ~,
\label{gquh2}
\end{eqnarray}
which is the 2D autocorrelation function of a particular realization of the total electric field $\bar{E}(\vec{r})$ calculated at the source position.  From this we come to the conclusion, to be used in the following of this section, that the 2D Fourier transform of $g(\Delta \vec{r})$ is always a positive quantity\footnote{The proof is based on the autocorrelation theorem, which states that the Fourier transform of the 2D autocorrelation function of $\bar{E}_0(x, y)$ as function of variables $\theta_x$, $\theta_y$ is  given by $|w(\theta_x, \theta_y)|^2$. Here $w(\theta_x, \theta_y)$ is the Fourier transform of $\bar{E}(x, y)$ as a function of the same variables $\theta_x$ and $\theta_y$.}.

For our purposes it is preferable to express the cross-spectral density in symmetrized form. We therefore introduce the new variables $\vec{r}$, given by

\begin{eqnarray}
\vec{\bar{r}} \equiv (\vec{r}_1+ \vec{r}_2)/2 ~ ,
\label{rave}
\end{eqnarray}

together with the previously defined difference $\Delta \vec{r}$

\begin{eqnarray}
\Delta \vec{r} \equiv \vec{r}_1 - \vec{r}_2~.
\label{dr}
\end{eqnarray}
We then have the inverse relations

\begin{eqnarray}
\vec{r}_1 = \vec{r} + \Delta \vec{r}/2
\label{r1}
\end{eqnarray}
and

\begin{eqnarray}
\vec{r}_2 = \vec{r} - \Delta \vec{r}/2~.
\label{r2}
\end{eqnarray}
If we now change variables in Eq. (\ref{coore}) according to Eq. (\ref{rave}) and Eq, (\ref{dr}), we find that the general expression for the cross-spectral density can be written as

\begin{eqnarray}
G(\vec{r}, \Delta \vec{r}) = \langle \bar{E}(\vec{r}+\Delta \vec{r}/2) \bar{E}^*(\vec{r} - \Delta \vec{r}/2) \rangle ~.
\label{GGG}
\end{eqnarray}
Let us consider a certain phase space distribution for an electron beam with a given transverse phase space distribution $f_\bot(\vec{l},\vec{\eta})$, which is a function of offset $\vec{l}$ and deflection $\vec{\eta}$. At the source position one can write

\begin{eqnarray}
G\left(\vec{r}, \Delta \vec{r}\right) = \int d\vec{l}~d\vec{\eta}~f_\bot\left(\vec{l},\vec{\eta}\right)  \bar{E}\left(\vec{l},\vec{\eta},\vec{r}+ \frac{\Delta \vec{r}}{2}\right) \bar{E}^*\left(\vec{l},\vec{\eta},\vec{r} - \frac{\Delta \vec{r}}{2}\right) ~.
\label{GGG2}
\end{eqnarray}
Aside for a normalization constant $\mathcal{A}$, the inverse Fourier transform of the cross-spectral density with respect to $\Delta x$ and $\Delta y$ can be written as:

\begin{eqnarray}
W(\vec{r}, \vec{\theta}) = \mathcal{A} \int d\Delta\vec{r} ~G(\vec{r}, \Delta \vec{r})  \exp\left(-i \omega \vec{\theta} \cdot \Delta \vec{r}/c \right)~.
\label{Wig1}
\end{eqnarray}
This is the expression for the Wigner distribution in terms of the cross-spectral density. We regard it as a distribution function defined over the four dimensions $(\vec{r}, \vec{\theta})$ and parameterized by $z$. Let us now specialize the expression in Eq. (\ref{Wig1}) to the case when the source is quasi-homogeneous. Since in that case the cross-spectral density function $G$ can be factorized into the product of  intensity $I(\vec{r})$ and spectral degree of coherence $g(\Delta \vec{r})$, the Wigner distribution function also factorizes as

\begin{eqnarray}
W(\vec{r}, \vec{\theta}) = I(\vec{r}) \mathcal{I}(\vec{\theta})~,
\label{Wig2}
\end{eqnarray}
having recognized that $\mathcal{I}(\vec{\theta})$ is proportional to the Fourier transform of the spectral degree of coherence $g(\Delta \vec{r})$, and can be identified with the angular distribution of radiation intensity.  Since the distribution $W$ is the product of two positive quantities, never assumes negative values, and it can always be interpreted as a phase space distribution. This analysis shows that quasi-homogeneous sources can always be characterized in terms of geometrical optics. It also shows that in this particular case the coordinates in phase space, $\vec{r}$ and $\vec{\theta}$, are separable. In the case of non quasi-homogeneous sources one can still define $W$ using Eq. (\ref{Wig1}). It can be shown that $W$ always assumes real values. However, the Wigner distribution is not always a positive function. As a consequence, it cannot always be interpreted as a phase space distribution. Yet, the integral over $\vec{r}$ and $\vec{\theta}$ can be shown to be positive, and therefore the maximum of the Wigner distribution is also bound to be positive, so that we can take this value as a natural definition for the brightness of SR sources.

The basic properties of the Wigner distribution include statements about its 2D projections. If we make use of Eq. (\ref{Wig1}) we obtain the following expression for the $(x,y)$ projection:

\begin{eqnarray}
&& \int d\vec{\theta} ~W(\vec{r}, \vec{\theta}) = \cr && \mathcal{A} \int d\vec{\theta}  \int d\Delta\vec{r} ~\exp\left(-i \omega \vec{\theta} \cdot \Delta \vec{r}/c \right) \langle \bar{E}(\vec{r}+\Delta \vec{r}/2) \bar{E}^*(\vec{r} - \Delta \vec{r}/2) \rangle~.
\label{projxy}
\end{eqnarray}
Changing the order of integration and using the fact that the integral of the exponential function essentially results in a Dirac $\delta$-function according to

\begin{eqnarray}
&& \int d\vec{\theta}  \exp\left(-i \omega \vec{\theta} \cdot \Delta \vec{r}/c \right) = (2 \pi)^2 \delta\left(- \frac{\omega \Delta \vec{r}}{c}\right)
\label{ddelta}
\end{eqnarray}
we see that

\begin{eqnarray}
\int d\vec{\theta} ~W(\vec{r}, \vec{\theta}) = (2\pi)^2 \frac{c^2}{\omega^2}\mathcal{A} \left \langle \left| \bar{E}(\vec{r})\right|^2 \right \rangle~.
\label{partialint}
\end{eqnarray}
In the quasi-homogeneous limit, as we just discussed, $W$ can be interpreted as photon distribution in phase space. Then, for consistency with this limit, one should require that integrating the Wigner distribution function over the solid angle $d\Omega = d \vec{\theta}$ yields the  photon spectral and spatial flux density: 

\begin{eqnarray}
\int d\vec{\theta}~ W(\vec{r}, \vec{\theta}) = \frac{d \dot{N}_{ph}}{dS (d\omega/\omega)} = \frac{I}{e \hbar}  \frac{c}{4 \pi^2} \left\langle\left|\bar{E}(\vec{r})\right|^2 \right \rangle ~,
\label{intW}
\end{eqnarray}
where $I$ is  electron beam current, $e$ is charge of the electron taken without sign, $c$ is the speed of light in vacuum and $\hbar$ is the reduced Planck constant.  Here we have used Parseval theorem, and included an additional factor two on the right-hand side of Eq. (\ref{intW}),  indicating that we use positive frequencies only. Comparison of the requirement in Eq. (\ref{intW}) with the mathematical property in Eq. (\ref{partialint}) fixes univocally the normalization constant $\mathcal{A}$ as

\begin{eqnarray}
\mathcal{A} = \frac{c}{(2 \pi)^4} \frac{I}{e \hbar} \left(\frac{\omega}{c}\right)^2 ~.
\label{norm}
\end{eqnarray}
Note that $\mathcal{A}$ depends on the units chosen (in this case Gaussian units) and on our definition of Fourier transformation (Eq. (\ref{ftdef2})). The Wigner distribution $W$ is also univocally defined as

\begin{eqnarray}
W\left(\vec{r}, \vec{\theta}\right) = \frac{c}{(2 \pi)^4} \frac{I}{e \hbar} \left(\frac{\omega}{c}\right)^2 \int d\Delta\vec{r} ~G(\vec{r}, \Delta \vec{r})  \exp\left(-i \omega \vec{\theta} \cdot \Delta \vec{r}/c \right)~.
\label{Wig13}
\end{eqnarray}
%
In the quasi-homogeneous asymptotic case, the brightness is a conserved quantity for perfect optical systems, and can be interpreted as maximum density of photon flux in phase space\footnote{In the more general case of non quasi-homogeneous sources, the brightness is conserved only in those cases where the evolution equation for $W$ follows Liouville equation. This is noticeably the case for free-space.}.

When one needs to specify the Wigner distribution or the cross-spectral density at any position down the beamline, one needs to calculate the field at any position down the beamline. In order to do so, we first calculate the field from a single relativistic electron moving along an arbitrary trajectory in the far zone, and then we solve the propagation problem in paraxial approximation. This last step allows calculation of the field at any position by backward-propagation in free-space with the help of the paraxial Green's function, that is the Fresnel propagator.

Suppose we are interested in the radiation generated by an electron and observed far away from it. In this case it is possible to find a relatively simple expression for the electric field \cite{JACK}. We indicate the electron velocity in units of $c$ with $\vec{\beta}(t)$, the Lorentz factor (that will be considered fixed throughout this paper) with $\gamma$, the electron trajectory in three dimensions with $\vec{R}(t)$ and the observation position with $\vec{R}_0 \equiv (z_0, \vec{r}_0)$. Finally, we introduce the unit vector

\begin{equation}
\vec{n} =
\frac{\vec{R}_0-\vec{R}(t)}{|\vec{R}_0-\vec{R}(t)|}
\label{enne}
\end{equation}
pointing from the retarded position of the electron to the observer. In the far zone, by definition, the unit vector $\vec{n}$ is nearly constant in time. If the position of the observer is far away enough from the charge, one can make the expansion

\begin{eqnarray}
\left| \vec{R}_0-\vec{R}(t) \right|= R_0 - \vec{n} \cdot \vec{R}(t)~.
\label{next}
\end{eqnarray}
We then obtain the following approximate expression for the the radiation field  in the frequency domain\footnote{For a better understanding of the physics involved one can refer to the textbooks \cite{ELLE, JACK}. A different constant of proportionality in Eq. (\ref{revwied}) compared to textbooks is to be ascribed to the use of different units and definition of the Fourier transform.}:

\begin{eqnarray}
\vec{\bar{E}}(\vec{R}_0,\omega) &=& -{i\omega e\over{c
R_0}}\exp\left[\frac{i \omega}{c}\vec{n}\cdot\vec{R}_0\right]
\int_{-\infty}^{\infty}
dt~{\vec{n}\times\left[\vec{n}\times{\vec{\beta}(t)}\right]}\exp
\left[i\omega\left(t-\frac{\vec{n}\cdot
\vec{R}(t)}{c}\right)\right] ~,\cr && \label{revwied}
\end{eqnarray}
where $(-e)$ is the negative electron charge. Using the complex notation, in this and in the following sections we assume, in agreement with Eq. (\ref{ftdef2}), that the temporal dependence of fields with a certain frequency is of the form:

\begin{eqnarray}
\vec{E} \sim \vec{\bar{E}}(z,\vec{r},\omega) \exp(-i \omega t)~.
\label{eoft}
\end{eqnarray}
With this choice for the temporal dependence we can describe a plane wave traveling along the positive  $z$-axis with

\begin{eqnarray}
\vec{E} = \vec{E}_0 \exp\left(\frac{i\omega}{c}z -i \omega t\right)~.
\label{eoftrav}
\end{eqnarray}
In the following we will always assume that the ultra-relativistic approximation is satisfied, which is the case for SR setups. As a consequence, the paraxial approximation applies too. The paraxial approximation implies a slowly varying envelope of the field with respect to the wavelength. It is therefore convenient to introduce the slowly varying envelope of the transverse field components as

\begin{equation}
\vec{\widetilde{E}}(z,\vec{r},\omega) = \vec{\bar{E}}(z,\vec{r},\omega) \exp{\left(-i\omega z/c\right)}~. \label{vtilde}
\end{equation}
Introducing angles $\theta_x = x_0/z_0$ and $\theta_y = y_0/z_0$, the transverse components of the envelope of the field in Eq. (\ref{revwied}) in the far zone and in paraxial approximation can be written as

\begin{eqnarray}
\vec{\widetilde{{E}}}(z_0, \vec{r}_0,\omega) &=& -{i
\omega e\over{c^2}z_0} \int_{-\infty}^{\infty} dz' {\exp{\left[i
\Phi_T\right]}}  \left[\left({v_x(z')\over{c}}
-\theta_x\right){\vec{e_x}}
+\left({v_y(z')\over{c}}-\theta_y\right){\vec{e_y}}\right]
~,\cr && \label{generalfin}
\end{eqnarray}
where the total phase $\Phi_T$ is

\begin{eqnarray}
&&\Phi_T = \omega \left[{s(z')\over{v}}-{z'\over{c}}\right] \cr &&
+ \frac{\omega}{2c}\left[z_0 (\theta_x^2+\theta_y^2) - 2 \theta_x x(z') - 2 \theta_y y(z') + z(\theta_x^2+\theta_y^2)\right]~
. \label{totph}
\end{eqnarray}
Here $v_x(z')$ and $v_y(z')$ are the horizontal and the vertical components of the transverse velocity of the electron,  $x(z')$ and $y(z')$ specify the transverse position of the electron as a function of the longitudinal position, $\vec{e}_x$ and $\vec{e}_y$ are unit vectors along the transverse coordinate axis. Finally, $s(z')$ is longitudinal coordinate along the trajectory. The electron is moving with velocity $\vec{v}$, whose magnitude is constant and equal to $v = ds/dt$.

Eq. (\ref{generalfin}) can be used to characterize the far field from an electron moving on any trajectory.  When the single-electron fields inside the ensemble average brackets $\langle...\rangle$ are specified at a certain position $z_1$, the fields at any other position $z_2$ can be found by propagating forward or backward in free-space according to the paraxial law

\begin{eqnarray}
&&\widetilde{E} \left(\vec{\eta},\vec{l},z_2,\vec{r}_{2}, \omega\right) = \frac{i \omega}{2
\pi c( {z}_2- {z}_1)} \int d \vec{ {r}_1}~\widetilde{E} \left(\vec{\eta},\vec{l},z_1,\vec{r}_{1}, \omega\right) \exp{\left[\frac{i \omega
\left|{\vec{ {r}}}_{2}-\vec{ {r}_{1}}\right|^2}{2 c (
{z}_2- {z}_1)}\right]}~. \cr && \label{fieldpropback}
\end{eqnarray}
In particular, one may decide to backpropagate the field even at positions well inside the magnetic structure under study. In this case, the field distribution is obviously virtual in nature, because it is not actually there, but it fully characterizes the radiation field from a single electron with given offset and deflection. Within the paraxial approximation, single-electron fields are fully characterized when they are known on a transverse plane at one arbitrary position $z$. Because of this, all positions $z$ are actually equivalent. As we will see there can be, however, a privileged position $z=z_s$ where the electric field assumes a particularly simple form: at this position, in many cases of practical interest including undulator and bending magnet radiation, the field wavefront from a single electron is simply plane\footnote{In the undulator this position is just in the middle of the setup. In the bending magnet case it is at the point of the trajectory tangent to the $z$ axis.}.  Without loss of generality one can set $z_s =0$ for simplicity and call this the source position. Then, the relation between the field from a single electron at the source $\widetilde{E} \left(\vec{\eta},\vec{l},0,\vec{r}, \omega\right)$  and the field in the far zone, $\widetilde{E} \left(\vec{\eta},\vec{l},z_0,\vec{\theta}, \omega\right) $, follows once more from Eq. (\ref{fieldpropback}):

\begin{eqnarray}
&&\widetilde{E} \left(\vec{\eta},\vec{l},0,\vec{r}, \omega\right) = \frac{i z_0 \omega}{2\pi c} \int d \vec{\theta}~ \widetilde{E} \left(\vec{\eta},\vec{l},z_0,\vec{\theta}, \omega\right) \exp\left(-\frac{i\theta^2 z_0 \omega}{2c}\right)\exp \left(\frac{i \omega \vec{r}\cdot\vec{\theta}}{c}\right)~\cr &&
\label{farzone}
\end{eqnarray}
\begin{eqnarray}
&&\widetilde{E} \left(\vec{\eta},\vec{l},z_0,\vec{\theta}, \omega\right) = \frac{i \omega}{2\pi c z_0} \exp\left(\frac{i\theta^2 z_0 \omega}{2c}\right) \int d \vec{r} ~\widetilde{E} \left(\vec{\eta},\vec{l},0,\vec{r}, \omega\right)  \exp \left(-\frac{i \omega\vec{r}\cdot\vec{\theta}}{c}\right)\cr &&
\label{farzonef}
\end{eqnarray}
We assume that a plane wave traveling along the positive $z$-axis can be expressed as in Eq. (\ref{eoftrav}). Then, the negative sign in the exponential factor $\exp(-i \omega z/c)$ in Eq. (\ref{vtilde}) determines the sign of the exponential in Eq. (\ref{fieldpropback}) and consequently the sign of the exponential that appears in the integrand in Eq. (\ref{farzonef}), which is the solution of the propagation problem in the far zone. Returning to the definition of Wigner distribution, we see that in order to be able to give a physical interpretation to the WD variables $\theta_x$ and $\theta_y$ as angles of plane-wave propagation modes we must choose the same (negative) sign in the exponential that appear in Eq. (\ref{Wig1}) and in the integrand in Eq. (\ref{farzonef}).

If we insert Eq. (\ref{GGG2}) into Eq. (\ref{Wig13}) we will obtain the most general expression for the Wigner distribution, accounting for a given phase space distribution of the electron beam. There are practical situations when offset and deflection of an electron lead to the same offset and deflection of the radiation beam from that electron. This is the case for undulator and bending magnet setups without focusing elements. In such situations, the expression for the Wigner distribution can be greatly simplified. The presence of an electron offset $\vec{l}$ shifts the single-electron field source, while a deflection $\vec{\eta}$ tilts the source. Therefore

\begin{eqnarray}
\widetilde{E}\left(\vec{l},\vec{\eta},0,\vec{r},\omega\right) = \widetilde{E}_0\left(\vec{r}-\vec{l}\right)\exp\left[i \omega \vec{\eta} \cdot \left(\vec{r}-\vec{l}\right)/c \right]
\label{tiltshift}
\end{eqnarray}
where we set $\widetilde{E}_0\left(\vec{r}\right)\equiv \widetilde{E}\left(0,0,0,\vec{r},\omega\right)$. Note that, due to duality between source and far zone plane, an electron offset tilts the single-electron far-zone field, while a deflection shifts the beam angle offset according to:

\begin{eqnarray}
\widetilde{E}\left(\vec{l},\vec{\eta},z_0,\vec{\theta},\omega\right) =\widetilde{E}_{F}\left(\vec{\theta}-\vec{\eta}\right)\exp\left[-i \omega \vec{l} \cdot (\vec{\theta}-\vec{\eta})/c \right]
\label{tiltshiftfar}
\end{eqnarray}

where we set $\widetilde{E}_{F}\left(\vec{\theta}\right) \equiv \widetilde{E}\left(0,0,z_0,\vec{\theta},\omega\right)$.

From Eq. (\ref{tiltshift}) one obtains

\begin{eqnarray}
G\left(\vec{r},\Delta \vec{r}\right) = \int d \vec{l} ~ G_0\left(\vec{r}-\vec{l},\Delta \vec{r}\right)  \int d\vec{\eta} f_\bot\left(\vec{l},\vec{\eta}\right) \exp\left(i \omega \vec{\eta}\cdot \Delta \vec{r}/c\right)
\label{GG0}
\end{eqnarray}
where $G_0(\vec{r}, \Delta \vec{r}) \equiv E_0(\vec{r}+\Delta \vec{r}/2)E_0^*(\vec{r}-\Delta \vec{r}/2)$.

A very useful addition theorem \cite{KIM1} can then be obtained from direct calculations:

\begin{eqnarray}
W\left(\vec{r},\vec{\theta}\right) =  \int d \vec{l} d\vec{\eta} ~W_0\left(\vec{r}-\vec{l}, \vec{\theta} - \vec{\eta} \right)   f_\bot\left(\vec{l},\vec{\eta}\right)
\label{Wcorr}
\end{eqnarray}
with $W_0$ defined as the Wigner distribution associated to $G_0$. This can be summarized by saying that that electron offset and deflection correspond to an offset in position and angle of the corresponding Wigner distribution $W_0$, and that the overall Wigner distribution $W$ can be found by addition over single-electron contributions. We will apply this knowledge to the special cases of undulator and bending magnet sources respectively in section \ref{sec:due} and section \ref{sec:tre}. Before that, in the next section we will present a summary of previous results and an overview of novel findings.


\section{\label{oldres} Overview and early results}

%

\subsection{Undulators}

It is helpful to start our investigations by examining the brightness for quasi-homogenous sources. The simplest case study is the undulator source. Here we take advantage of the particular but important situation of perfect resonance, when the undulator field can be presented in terms of analytical functions. Moreover, we will consider the practical case of a Gaussian electron beam, which yields further analytical simplifications.  In the quasi-homogeneous case of beam size- and divergence-dominated regime, the undulator brightness at the fundamental harmonic is easily shown to be

\begin{eqnarray}
B = \frac{F}{4 \pi^2 \epsilon_x \epsilon_y } ~.
\label{BU0}
\end{eqnarray}
where $F$ is the total spectral photon flux, and $\epsilon_{x,y}$ are the rms geometrical electron beam emittance in the horizontal and vertical directions. Usually the electron beam waist is located in the middle of the undulator, where $\epsilon_{x,y} = \sigma_{x,y} \sigma_{x',y'}$, with $\sigma_{x,y}$ the electron beam rms sizes and $\sigma_{x',y'}$ the electron beam rms divergences at that position. Therefore one obtains the well-known expression

\begin{eqnarray}
B = \frac{F}{4 \pi^2 \sigma_x \sigma_y \sigma_{x'} \sigma_{y'}} ~.
\label{BU1}
\end{eqnarray}
This result has, of course, a very simple physical interpretation in terms of maximum density of photon spectral flux in phase space.  In fact, in the beam size- and divergence-dominated regime the photon beam can be modeled as a collection of rays with the same phase space of the electron beam, and also the photon beam has its waist in the middle of the undulator. Another way of stating the same concept is by saying that the radiation source is located in the middle of the undulator. One word of caution should be spent commenting this last result. From beam dynamics considerations it is known that, if focusing elements are absent, the beam size varies along the undulator like

\begin{eqnarray}
\sigma^2_{x,y}(z) = \sigma^2_{x,y} + z^2\sigma^2_{x',y'}~,
\label{sigprop}
\end{eqnarray}
where $-L/2 < z < L/2$ is the distance from the waist, with $L$ is the undulator length. The average beam size along the undulator length is then

\begin{eqnarray}
\langle \sigma^2_{x,y} \rangle = \sigma^2_{x,y} + \frac{L^2}{12}\sigma^2_{x',y'}~.
\label{text}
\end{eqnarray}
%


Textbook \cite{WIED}  regards Eq. (\ref{text}) as a clear evidence of the fact that our undulator source possesses a finite longitudinal dimension: the size of this extended source changes along the undulator, and this effect tends to increase the effective source size and reduce the undulator brightness, as one can see from the average beam size Eq. (\ref{text}), which is larger than that at the waist.  However, this is a misconception because, according to the electrodynamics of ultrarelativistic charged particles, the source size is not widened at all, which is demonstrated by the fact that  Eq. (\ref{BU1}) for the brightness only includes the electron beam size at the waist.

As discussed before, our source, placed in the middle of the undulator, is obviously virtual in nature. However, it is no mathematical abstraction. Synchrotron radiation is often used to measure the size of the electron beam. For example, in the electron beam size-dominated regime, undulator radiation can be used to form an image of the cross-section of the electron beam. We can take a single focusing mirror of focal length $f = z_1/2$ at distance $z_1$ from the source to form a $1:1$ image at the same distance $z_1$ downstream of the lens. We can set the object plane at the undulator center. If, for example, the resolution is limited by diffraction effects only, the rms  size of the image will be equal to the rms of the electron beam size at the waist, and it is not affected by variations of the parabolically shaped beta function along the undulator. This image is also a visualization of the virtual source.

We should remark that statistical optics is the only mean to deal, in general, with the stochastic nature of SR. Only in those particular cases when SR can be treated in terms of geometrical optics beamline scientists can take advantage of ray-tracing techniques. One of these cases is described above for the electron beam size- and divergence-dominated regime.

In all generality, in order to decide whether geometrical optics or wave optics is applicable, one should separately compare the electron beam sizes and divergences at the electron beam waist position\footnote{They could be actually compared at any position down the beamline. Here, however, we present an analysis for the source position only, where the radiation wavefront from a single electron is plane.} with the radiation diffraction sizes and diffraction angles, which are quantities pertaining the single-electron radiation. When at least the electron beam size or divergence dominates, one can use a geometrical optics approach. As we have seen, quasi-homogeneity is a necessary and sufficient condition for geometrical optics to be used in the representation of any SR source. A source is quasi-homogeneous if and only if it is possible to factorize the cross-spectral density $G$ in the product of two factors separately depending on $\vec{r}$ and $\Delta \vec{r}$. For example, in the particular case when the electron beam size dominates over diffraction, the cross-spectral density $G$ (and hence also the Wigner distribution) admits factorization, and the source can still be described with the help of geometrical optics even if the divergence is dominated by diffraction effects: in that case, since the source is quasi-homogeneous, the Fourier transform of the spectral degree of coherence $g(\Delta \vec{r})$ yields a diffraction limited intensity distribution in the far zone\footnote{This statement pertains the symmetry between spatial and angular domain, and can be seen as the inverse of the van Cittert-Zernike theorem.}.

According to our definition, when we can treat SR in terms of geometrical optics, the brightness is always the maximum of the phase space density of the photon beam. In the beam size-dominated or beam divergence-dominated limits, the undulator brightness can be determined analytically yielding the following cases (see section \ref{sec:due} for details) :

\begin{itemize}

\item{Beam divergence-dominated regime, for $\sigma^2_{x',y'} \gg  \lambdabar/L$, $\sigma^2_{x,y} \ll \lambdabar L$. In this case}

\begin{eqnarray}
B = \frac{F}{2 \sigma_{x'} \sigma_{y'}\lambda L}
\label{BQH1}
\end{eqnarray}

\item{Beam size-dominated regime, for $\sigma_{x,y}^2  \gg \lambdabar L$,  $\sigma_{x',y'}^2 \ll \lambdabar/L$. In this case}

\begin{eqnarray}
B = \frac{F L}{2 \pi^2 \sigma_x \sigma_y \lambda}
\label{BQH2}
\end{eqnarray}

\end{itemize}

Having dealt with quasi-homogeneous sources we now turn our attention to diffraction limited undulator sources. In the case of a zero-emittance electron beam, the undulator brightness can also be determined analytically yielding

\begin{eqnarray}
B = \frac{4}{\lambda^2} F ~.
\label{Bfil}
\end{eqnarray}
In literature it is often noted that the $\lambda^2/4$ factor is the volume of a diffraction limited beam in the photon phase space\footnote{Although, strictly speaking, one cannot talk of phase space for a diffraction limited beam.}. The numerical factor four is dictated by the normalization condition Eq. (\ref{intW}) and by the axial symmetry properties of undulator radiation at the fundamental harmonic. This point is discussed more in detail in \cite{ELLE}.

The following expression (originally proposed by Kim in \cite{KIM1}) is the usual estimate of the undulator brightness:

\begin{eqnarray}
B = \frac{F}{4 \pi^2} \frac{1}{[(\sigma_x^2 +\sigma_r^2)(\sigma_y^2 + \sigma_r^2)
(\sigma_{x'}^2 +\sigma_{r'}^2)(\sigma_{y'}^2 + \sigma_{r'}^2)]^{1/2}} ~.
\label{Bundug}
\end{eqnarray}
Eq. (\ref{Bundug}) can be obtained by approximating the radiation from a single electron by a Gaussian laser mode with rms divergence and source size $\sigma_{r'}$ and $\sigma_r$. Then, one can use the addition theorem to obtain the brightness for a beam of electrons. The integral in Eq. (\ref{Wcorr}), in this case, is a convolution of two Gaussian functions. For large electron beam emittances, this expression is in agreement with the geometrical optics limit pertaining to the beam size- and divergence-dominated regime, Eq. (\ref{BU1}). Once the electron sizes and divergences are fixed, in Eq. (\ref{Bundug}) there are still two independent parameters, $\sigma_r$ and $\sigma_{r'}$. In \cite{KIM1}, these two parameters are chosen in such a way that the diffraction limit in Eq. (\ref{Bfil}) is satisfied. This leaves only one degree of freedom to be fixed. In other words, according to Eq. (\ref{Bundug}), when the product $\sigma_r \sigma_{r'}$ is fixed, there is only one adjustable parameter to be fitted to the exact result for the brightness\footnote{In the case of third generation light sources, one always has $\sigma_x^2 \gg \sigma_r^2$ and $\sigma_{x'}^2 \gg \sigma_{r'}^2$. In this case, the geometrical optics limit is reached for $\sigma_{y}^2 \gg \sigma_{r}^2$ or $\sigma_{y'}^2 \gg \sigma_{r'}^2$. In section \ref{TGS} we will show that for both these asymptotes Eq. (\ref{Bundug}) is in disagreement with the exact result.} in the beam divergence-dominated regime, Eq. (\ref{BQH1}), and in the beam size-dominated regime, Eq. (\ref{BQH2}). In \cite{KIM1} the following definition fixes both the degrees of freedom available in Eq. (\ref{Bundug}):

\begin{eqnarray}
&& \sigma_r = \frac{\sqrt{\lambda L}}{4\pi}~,
\cr && \sigma_{r'} = \sqrt{\frac{\lambda}{L}}~.
\label{sigrrp0}
\end{eqnarray}
Such choice has also been adopted in some articles \cite{HUL1,HOW1} and books \cite{CIOC, CLAR}. The approximate Eq. (\ref{Bundug}), together with the definitions in (\ref{sigrrp0}), gives a qualitative agreement with exact results in all geometrical optics limiting cases, but detailed quantitative agreement is, in some case, rather poor.  In particular, it should be remarked that the approximate expression in Eq. (\ref{Bundug}) overestimates the exact value of the brightness by eight times in the beam divergence-dominated regime, Eq. (\ref{BQH1}), and underestimates it by two times in the beam size-dominated regime, Eq. (\ref{BQH2}). However, it should be noted that the choice of $\sigma_r$ and $\sigma_{r'}$ in (\ref{sigrrp0}) is not the optimal choice\footnote{In literature one can also find reasonable arguments in favor of the choice (\ref{sigrrp0}), see e.g. \cite{HULA}. The point here is that the exact expressions Eq. (\ref{BQH1}), Eq. (\ref{BQH2}) are novel results that we could not find in literature. Because of this, only now we can optimize the last degree of freedom in Eq. (\ref{Bundug}) by comparing the approximated expression with the exact results.} for the approximation in Eq. (\ref{Bundug}). In fact, the previous remark indicates that larger values of $\sigma_r$ and smaller values of $\sigma_{r'}$ should be chosen. More quantitatively, choosing

\begin{eqnarray}
&& \sigma_r = \frac{\sqrt{2 \lambda L}}{4\pi}~,
\cr && \sigma_{r'} = \sqrt{\frac{\lambda}{2 L}}~
\label{sigrrp}
\end{eqnarray}
we can obtain perfect numerical agreement with the exact result for the beam size-dominated regime Eq. (\ref{BQH2}) and an overestimation of a factor four in the beam divergence-dominated regime, Eq. (\ref{BQH1}). The choice of parameters in (\ref{sigrrp}) was first introduced in \cite{KIME} and is used, nowadays, in most calculations exploiting the approximate Eq. (\ref{Bundug}) \cite{WIED, ELLE, HUAN}.

\subsection{Bending magnets}

Up to this point our analysis followed, in its general lines, the one given by Kim in \cite{KIM1}. As the next step we turn to consider the brightness from bending magnets, which was not described, up to now, in a satisfactory way within the formalism of the Wigner distribution. This problem is still under discussion from the early days of SR theory \cite{GREE, KRIN, KIM1, HUL1, HUL2}, and it is reflected in the fact that many textbooks devoted to SR theory, like e.g. \cite{WIED}, do not discuss the brightness from bending magnets. Other books like \cite{DUKE} discuss only the geometrical limit, and \cite{TALM} presents a qualitative treatment. Only \cite{ELLE,CLAR} try to give a complete analysis of the problem. Part of the difficulty of applying the concept of brightness to the bending magnet case can be traced back to attempts using a mixture of intuitive geometrical optics and wave optics considerations, instead of exact results as was done in the undulator case. A typical example of such intuitive description can be found in \cite{CLAR}: "To calculate the bending magnet brightness we need to consider the effective phase space which the photon flux is being emitted taking into account of both the finite electron and photon beam sizes and divergences. First, there is no need to consider any horizontal angle effects as the light is emitted smoothly over the full horizontal $2\pi$ radians". A quantitative definition of bending magnet brightness according to \cite{CLAR} is then given by

\begin{eqnarray}
B = \frac{dF}{d\theta_x} \frac{1}{(2\pi)^{3/2} \Sigma_x \Sigma_y \Sigma_{y'}}~   ,
\label{Btext}
\end{eqnarray}
where $dF/d\theta_x$ is the photon flux per unit horizontal angle (on the bending plane). The effective horizontal and vertical source size and effective vertical divergence are then calculated to be

\begin{eqnarray}
&&\Sigma_x = \sqrt{\sigma^2_x + \sigma^2_r}  ~,
\cr && \Sigma_y = \sqrt{\sigma^2_y + \sigma^2_r}  ~,
\cr && \Sigma_{y'} = \sqrt{\sigma^2_{y'} + \sigma^2_{r'}}~,
\label{sigmas}
\end{eqnarray}
where the vertical opening angle $\sigma_{r'}$ may be determined from the equality $(2\pi)^{1/2} \sigma_{r'} (dF/d\Omega)_{|_{\theta_y=0}} = dF/d\theta_x$, $(dF/d\Omega)_{|_{\theta_y=0}}$ being the on-axis photon flux density per unit solid angle \cite{CLAR}. Since, usually, only the horizontal polarization component of SR radiation from a bending magnet is important, here we discuss SR beam brightness only for the $E_x$ electric field component.

%
Using this method, considering the horizontally-polarized component of the field one finds that $\sigma_{r'} = 0.67/\gamma$ at the critical wavelength $\lambda = \lambda_c = 2\pi R/\gamma^3$, where $R$ is the radius of curvature of the electron orbit in the bending magnet\footnote{Our definition of critical wavelength does not match the conventional definition introduced in textbooks, see e.g.  \cite{DUKE}. The numerical factor (our critical wavelength is $3/2$ times longer) has been chosen so that  this is a convenient combination  to be used in section \ref{sec:tre} for dimensional analysis of SR from a bending magnet in the space frequency domain.}. Similarly to the undulator case, the angular divergence and source size for the radiation emitted by a single electron is chosen to satisfy $2\pi \sigma_r \sigma_{r'} = \lambda/2$. It is instructive  to examine this expression in diffraction limited case. From Eq. (\ref{Btext}) we then find

\begin{eqnarray}
B = \frac{2}{\lambda} \frac{dF}{d \theta_x} \frac{1}{\sqrt{2\pi}\sigma_r}  =  \frac{4}{\lambda^2} \frac{dF}{d
\theta_x} \sqrt{2\pi} \sigma_{r'}~ .
\label{BMX}
\end{eqnarray}
If the horizontal opening angle of the beamline, that is the half-angle subtended by the aperture in horizontal direction $\theta_a$, is  larger than $\sigma_{r'}$, then the flux increases proportionally to $\theta_a$. However, in the limit for $\theta_a \gg \sigma_{r'}$ the brightness becomes independent of the beamline opening angle. It is often noted that the factor $\lambda^2/4$ is the minimal phase space volume of a diffraction limited beam. Therefore one can summarize by saying that SR emitted from a bending magnet within a solid angle of order $\sigma^2_{r'}$ occupies the minimal phase space volume $\lambda^2/4$.  The physical interpretation of the effect mentioned above can be found, among other references, in the review paper \cite{WILL}, where the brightness of the radiation from a bending magnet is described in the language of geometrical optics. This language is intrinsically inadequate to describe the focusing of diffraction-limited radiation from a bending magnet. In fact, this physical phenomenon fully pertains wave optics. It is however possible to use geometrical optics reasoning and obtain an intuitive understanding of the situation. Later on in this work we will show that such intuitive understanding is in qualitative agreement with an analysis fully based on wave optics. A complication in determining the brightness is the correlation between the longitudinal position of the source along the bending magnet and the (horizontal) angle of observation in the far zone. An important parameter that is required to calculate the brightness is the source area. In general, citing \cite{WILL} almost literally: "there are three contributions to the horizontal source size and these are (1) the intrinsic size of the electron beam itself (in our notation $\sigma_x$); (2) the projected (observed) size due to the large horizontal sweep angle $\theta_a$, leading to an extended source (equal to the sagitta $R \theta_a^2/8$ ); and (3) the diffraction limited source size. We can estimate the diffraction limited source size as $\lambda/\theta_a$". The effective source size is then found by adding the three sizes  in quadrature, thus

\begin{eqnarray}
\Sigma_x = \left[\sigma_x^2 + (R \theta_a^2/8)^2 + \lambda^2/\theta_a^2\right]^{1/2} ~.
\label{totsize}
\end{eqnarray}
Finally, it is worth emphasizing that if the horizontal opening angles $\theta_a$ is larger than the diffraction angle $\sigma_{r'}$, the projected source size increases more rapidly $( \sim \theta_a^2)$ than the flux $( \sim \theta_a)$ \cite{WILL}.

In literature one can often find that the brightness is strictly linked to transverse coherence properties of the radiation and that the coherent flux  $F_\mathrm{coh}$ can be defined as

\begin{eqnarray}
F_\mathrm{coh} = \frac{\lambda^2}{4} B ~.
\label{Fcoh}
\end{eqnarray}
Note that the common understanding that the coherence flux available after spatial filtering can always be written as Eq. (\ref{Fcoh}) is erroneous. For example, as we discussed above, the brightness of the radiation from a bending magnet in the diffraction limited case is constant when $\theta_a$ is larger than the diffraction angle $\sigma_{r'}$. However, in this situation  the coherent flux increases proportionally to the horizontal  opening angle $\theta_a$. In fact, the entire photon flux collected from a diffraction limited source is fully transversely coherent. A Young's double pinhole interferometer can be used for demonstrating this fact. In the case of a diffraction limited source, the interference pattern recorded by the interferometer is always characterized by a $100 \%$ fringe contrast.

The physical meaning of brightness can be best understood by considering the imaging of the source on an experimental sample. Radiation is usually concentrated by using a wide-aperture focusing optical system. The brightness is a figure of merit that quantifies how well a SR beam can be focused. For this purpose one considers an ideal optical system. In fact, in general, the maximum photon flux density onto the image plane is altered by optical elements along the setup. We may interpret the brightness $B$ as the theoretical maximum concentration of the SR photon flux on the image-receiving surface where, usually, the sample is placed.

The source size can be affected by the presence of a finite electron beam. It is often noted that the coherent part of the total flux can be ultimately focused down to a spot-size of dimension $\lambda^2/4$. Such consideration thus leads, at least qualitatively, to the relation between brightness and coherent fraction of total flux given by Eq. (\ref{Fcoh}). It is clear that even a fully coherent beam can be of `bad quality' in the sense that a only small fraction of photons in the beam can be focused to a spot-size of dimension $\lambda^2/4$. This is the case when the radiation source is characterized by a complicated wavefront, which has an effect similar to optics abberation, when there is a departure of the far field wavefront from the ideal spherical form. Then, one cannot reach an effective focusing on the sample even in the diffraction limited case. It is well-known that, in general, there are two different characteristics of the radiation beam. The first one is the degree of transverse coherence  $\xi = F_\mathrm{coh}/F$, which reflects the statistical properties of radiation fields \cite{ATTW}. The second one is the well-known $M^2$ factor, which is widely used in the laser community to quantify how well a deterministic laser beam can be focused \cite{SIEG}.  The diffraction-limited photon beam from an undulator has an $M^2$ factor close to unity and  Eq. (\ref{Fcoh}) is correct. At variance, the diffraction-limited photon beam from a bending magnet  has an $M^2$ factor close to unity only within a solid angle of about $\sigma_{r'}^2$. The brightness is a useful figure of merit that incorporates, simultaneously, both the statistical properties and the wavefront qualities of the radiation pulse.

When the electron beam has zero emittance we are dealing with perfectly coherent wavefronts. Intuitively, in this situation one would apply methods from wave optics in order to solve the image formation problem. At variance, in literature it is often discussed an estimation of the source size for the case of diffraction limited radiation from a bending magnet based on geometrical optics \cite{GREE, KRIN, HUL1, HUL2, WILL}. A situation where one deals with a similar problem is in the calculation of a laser beam focus through a lens, when severe aberrations are present. Although the laser beam is coherent, when diffraction effects are negligible compared to aberration effects, the beam focusing can be calculated with the help of geometrical optics considerations. In the geometrical optics limit, the wave equation can be replaced with the Eikonal equation, which should be solved for surfaces of equal phase. Once the function of equal phase is known, one can apply usual ray-tracing techniques remembering that rays are, at any point, normal to the surface with equal phase. The similarity between the two situations is highlighted by the essential feature of a diffraction limited SR beam from a bending magnet: at beamline opening angles $\theta_a \gg \sigma_{r'}$ wavefront `aberrations'  are present in the sense discussed above\footnote{In the particular case of bending magnet radiation from a single electron, for a beamline opening angle $\theta_a \gg \sigma_{r'}$, departure of the far field wavefront from the ideal spherical form can be considered as a coma-like aberration.}, and are severe, meaning that $M^2 \gg 1$. Therefore, a geometrical optics approximation, leading in particular to ray-tracing techniques, can be applied to the analysis of the image formation problem.

Before proceeding we should make a few remarks concerning the terminology used in relation to brightness treatments. The theory of brightness for a bending magnet is much more difficult than that for an undulator. One can now see a net distinction between the geometrical optics limit in the framework of statistical optics when one discusses about an incoherent (i.e. quasi-homogeneous) SR source and the geometrical optics limit in the framework of coherent Fourier optics, when one discusses about highly `aberrated' beams radiated from a single electron in the bending magnet setup. An example where terminology is not accurately used can be found in \cite{KIM1}: "Let us now turn to a more rigorous derivation of the source brightness of the synchrotron radiation due to a single electron. ... According to Eq. (31), photons are emitted incoherently in the tangential direction at each point of the trajectory".  It should be clear that one can talk about `incoherently emitted' photons only in the framework of statistical optics, when one deals with SR as random process. In the case of a single electron case we are always dealing with coherently emitted photons at each point of the trajectory. Actually, the discussion in \cite{KIM1} must be understood as an application of the geometrical optics approximation to the bending magnet radiation from a single electron in the case of large open angle, i.e when wavefront `aberrations' effects are dominant compared to diffraction effects.

Up to this point our analysis of bending magnet brightness has followed that given in reviews \cite{KRIN, WILL} and books \cite{ELLE,CLAR} on SR theory.  It is now instructive to consider our approach, and examine its results in the diffraction limited case. With our definition of brightness we do not need to worry about how to account for the effect of SR beam `aberrations'. In fact, the essential feature of our method is that we derive the brightness of a diffraction limited SR beam from the expression for the maximum Wigner distribution. Clearly, the brightness defined in this way automatically includes a factor that characterizes the possibility of focusing the SR beam.

Let us indicate with $B_0$ the brightness given by the approximated expression Eq. (\ref{BMX}). Mathematically, our calculation is based on Eq. (\ref{Wig13}). One obtains that the bending magnet brightness in the diffraction limited case is given by $B_0$ times a function which contains the only variable $\lambda_c/\lambda$. In particular, at $\lambda = \lambda_c$ the exact result for $B = \max(W)$ is

\begin{eqnarray}
B= 1.90 B_0~.
\label{BB00}
\end{eqnarray}
The approximated result that can be obtained from Eq. (\ref{BMX}) is naturally different from the exact one, although the difference is not large. However, the fact remains that, at variance with the approximated case for undulator brightness, the usual estimate for the brightness of radiation from a bending magnet does not coincide with the exact result in the limit for a zero beam emittance. The best way to avoid this kind of difficulties is to use the Wigner function formalism as in the undulator case. Only in this way it is possible to give an expression for the brightness of radiation from a bending magnet that is logically consistent and a directly applicable formulation that can be used by SR beamline scientists.

The most serious objection to approximation in Eq. (\ref{Btext}) is that this expression does not include the  electron beam divergence in the horizontal direction. In fact, this is in contrast with results obtained within the Wigner function formalism.  In this respect, let us examine  Eq. (\ref{Btext})  in the following limiting case of beam divergence-dominated regime:

\begin{eqnarray}
&& \sigma_x^2    \ll \sigma_r^2 ~, \cr
&& \sigma_y^2    \ll \sigma_r^2 ~, \cr
&& \sigma_{x'}^2 \gg \sigma_{r'}^2 ~, \cr
&& \sigma_{y'}^2 \gg \sigma_{r'}^2 ~.
\label{siglim}
\end{eqnarray}
Eq. (\ref{Btext})  simplifies to

\begin{eqnarray}
B = \frac{dF}{d \theta_x}  \frac{1}{(2\pi)^{3/2} \sigma_r^2 \sigma_{y'}}~ ,
\label{simptetx}
\end{eqnarray}
which  can also be written as

\begin{eqnarray}
B = B_0 \frac{\sigma_{r'}}{\sigma_{y'}}~  .
\label{BB0}
\end{eqnarray}
Here $B_0$ is given, as before, by Eq. (\ref{BMX}).  The contrast with the result obtained by exploiting the Wigner function formalism can be seen by a straightforward application of Eq. (\ref{Wig13}), which yields, in the same notations (at $\lambda = \lambda_c$):

\begin{eqnarray}
B = 2.45 B_0 \frac{\sigma_{r'}}{\sigma_{x'}} \frac{\sigma_{r'}}{\sigma_{y'}} ~.
\label{Btocomp}
\end{eqnarray}
The horizontal electron beam divergence is a problem parameter. We deduced this result by applying a rigorous mathematical method, without any intuitive arguments.  An intuitive way of understanding this property is to recall that due to wavefront `aberrations', one has an $M^2$ factor close to unity only when considering radiation from a single electron emitted around the axis within a solid angle of about $\sigma_{r'}^2$.  Suppose that the electron beam has a divergence characterized by $\sigma_{x'}^2 \gg \sigma_{r'}^2$ and $\sigma_{y'}^2 \gg \sigma_{r'}^2$. In this case, from geometrical considerations it is evident that only a photon flux of order of

\begin{eqnarray}
\left(\frac{dF}{d \Omega}\right)_{|_{\theta_y=0}} \sigma_{r'}^2 \frac{\sigma_{r'}}{\sigma_{x'}} \frac{\sigma_{r'}}{\sigma_{y'}}~
\label{maxfoc}
\end{eqnarray}
can, in principle, be focused down to area an area of order $\lambda^2$. Therefore, the maximum photon flux density in phase space is proportional to the ratio of the flux in (\ref{maxfoc}) and an effective  phase space volume of order $\lambda^2$. We thus deduced the parametric dependence of the bending magnet brightness in the limiting case (\ref{siglim}) by means of intuitive arguments, which are in agreement with the rigorous mathematical derivation of Eq. (\ref{Btocomp}) found within the Wigner function formalism.

One can see that on the one hand the intuitively reasonable idea that there is "no need to consider any horizontal angle effects as the light is emitted smoothly over full horizontal angle of $2\pi$ radians" \cite{CLAR} can lead to incorrect results.  On the other hand, however, other intuitive arguments are in agreement with the Wigner function formalism. Only the existence of a rigorous, exact mathematical method guarantees  a well-defined physical meaning for the brightness of a bending magnet.

Another argument that disqualifies Eq. (\ref{Btext}) as an approximation for bending magnet brightness follows from another comparison with exact results. In the beam size and divergence-dominated regime when $\sigma_x^2 \gg \sigma_r^2$, $\sigma_y^2 \gg \sigma_r^2$, $\sigma_{y'}^2 \gg \sigma_{r'}^2$ and at arbitrary beam divergence in horizontal direction Eq. (\ref{Btext}) yields:

\begin{eqnarray}
B = \frac{dF}{d \theta_x} \frac{1}{(2\pi)^{3/2} \sigma_x\sigma_y\sigma_{y'}}~.
\label{otherbug}
\end{eqnarray}
This last expression can be written as

\begin{eqnarray}
B = B_0 \frac{\sigma_r}{\sigma_x}\frac{\sigma_r}{\sigma_y}\frac{\sigma_{r'}}{\sigma_{y'}} ~.
\label{otherbug2}
\end{eqnarray}
Let us now consider rigorous calculations with the help of the Wigner function formalism. As we discussed above, the beam size- and divergence-dominated regime is the simplest geometrical optics asymptote for the undulator case. On the contrary, for a bending magnet, it is the most complicated case to be treated analytically. Intuitively we certainly expect that in this asymptotic limit there should be a competition between effects related with partial coherence (since $\sigma_x^2 \gg \sigma_r^2$) and what we called `aberration' effects (since $\sigma_{x'}^2 \gg \sigma_{r'}^2$). Detailed mathematical analysis confirms such expectation. It can be seen that in this case, depending on the specific ratio between horizontal beam size and divergence, the brightness  is described using functions with completely different  parametric dependence. In particular, Eq. (\ref{otherbug2}) turns out to be parametrically inconsistent when the condition $(\sigma_{x'}/\sigma_{r'})^2 \gg \sigma_x/\sigma_r$ is satisfied.

%
%

It is interesting to see that intuitive and rigorous approaches coincide, at least in the limit when

\begin{eqnarray}
&& \sigma_x^2 \gg \sigma_r^2~, \cr &&
\sigma_y^2 \gg \sigma_r^2~, \cr &&
\sigma_{y'}^2 \ll \sigma_{r'}^2~, \cr &&
\sigma_{y'} \ll \sigma_{r'}^2~.
\label{sondlim}
\end{eqnarray}
In fact, in this quasi-homogeneous case of beam size-dominated regime, which is typical for SR facilities in the X-ray wavelength range, from our definition of brightness we find

\begin{eqnarray}
B = \left(\frac{dF}{d \Omega}\right)_{|_{\theta_y=0}} \frac{1}{2\pi\sigma_x\sigma_y}~  .
\label{okour}
\end{eqnarray}
With the help of simple algebra one can show that the brightness approximated by Eq. (\ref{Btext}) actually coincides with Eq. (\ref{okour}).

\subsection{Discussion}

Although we can always find the brightness by a rigorous mathematical method following the Wigner function approach, it is sometimes possible to get exact results such as Eq. (\ref{okour}) without any calculation. The reasoning leading to such result is based on the fact that in the beam size-dominated regime the brightness is the product of the maximum angular flux density from a single electron and the maximum electron density at the source position. In other words, as already discussed, in the particular case of a quasi-homogeneous source, the brightness can always be interpreted in terms of geometrical optics, and the coordinates in phase space are separable (see Eq. (\ref{Wig2})). The brightness is therefore a product of two positive quantities, Eq. (\ref{okour}). The expression for the photon angular flux density radiated by a single electron in a bending magnet is well known and can be found in any textbook devoted to SR theory.

The arguments we have just given for the bending magnet case can be applied to the undulator case as well. Assuming a beam size-dominated regime and remembering the definition of brightness in the quasi-homogeneous limit as the maximum of the phase space photon flux density, one gets immediately that Eq. (\ref{okour}) gives the exact result Eq. (\ref{BQH2}).  With this, we point out an interesting fact about the angular flux density $dF/d \Omega$ for the undulator radiation. If one calculates the total flux $F$ integrating over the solid angle, one obtains the relation $\max(dF/d \Omega) = (L/\lambda) (F/\pi)$, where $\max(dF/d \Omega)$ is the on axis angular flux density in the diffraction limited case.

We can generalize the method for finding an exact result without any calculations to the asymptotic case where only the beam divergence is dominating. From the above analysis it is not hard to see that the brightness in this limit is given by the product of the maximum angular flux density of the electron beam and the maximum photon flux density radiated from a single electron at the source position.  In other words,  when conditions (\ref{siglim})  hold, for the divergence-dominated case we obtain in analogy with Eq. (\ref{okour}):

\begin{eqnarray}
B = \max\left(\frac{dF}{dS}\right) \frac{1}{2\pi\sigma_{x'}\sigma_{y'}} ~,
\label{ok2}
\end{eqnarray}
where $\max(dF/dS)$ is the maximum photon flux density on the source in the diffraction limited case.

With Eq. (\ref{ok2}) we can calculate the bending magnet brightness in the beam divergence-dominated regime at any wavelength. All we need is an explicit expression for $dF/dS$. In contrast with the function $dF/d \Omega$ in Eq. (\ref{okour}), which is well-known, $dF/dS$ function is much less known. In the undulator case, to the authors' knowledge, the only paper dealing with this issue is \cite{FOUR}. For the bending magnet case, we failed to find, in literature, an expression for $dF/dS$. An explicit formula for $dF/dS$ is derived in section \ref{sec:tre} of this article.

%
%
%
%
%

The previous analysis of the electron beam size- or divergence-dominated regime suggests an interesting question, whether it is always possible, in the geometrical optics limit, to deduce exact results for brightness without any calculations. The answer is negative. We could predict Eq. (\ref{okour}) and Eq. (\ref{ok2}) based on intuitive grounds, and careful mathematical analysis confirms the expectation both for undulator and bending magnet cases. Let us now consider the case when electron beam size and divergence dominate. Following intuitive arguments, the brightness should be proportional to the maximum of the electron beam density in phase space. From the analysis done above it is clear that this expectation is confirmed in the undulator case (see Eq. (\ref{BU1})). However, as described in section \ref{sec:tre}, for a bending magnet in the beam size- and divergence-dominated regime the brightness is never inversely proportional to the product $\sigma_x \sigma_{x'}$. The theory of brightness for a bending magnet is much more difficult than that for an undulator, and only our exact mathematical method automatically includes all features that describe the possibility of focusing the SR beam.

\section{\label{sec:due} Undulator brightness}

We now focus on particular realizations of SR sources of practical interest. In particular, in this section we consider the theory of brightness for undulator sources. Traditionally, all textbooks devoted to SR theory start discussing the characteristics of radiation from dipole magnets. Only after a detailed study of this case they deal with other advanced topics like undulators. In this way, the discussion follows the historical development of the subject, where undulators are presented as a logical extension of dipole magnets. However, the theory of bending magnet radiation is much more difficult than that of undulators. Therefore, here we choose to discuss undulators first. We take advantage of the particular but important situation of perfect resonance, when the polarization direction does not depend on the observation angle and simply reproduces the polarization direction of the undulator field. In the far zone, undulator radiation from a single electron exhibits a finite divergence and a spherical wavefront centered in the middle of the setup. In contrast, bending magnet radiation is emitted over the entire horizontal angle of $2\pi$ radians and, moreover, the state of polarization of the radiation depends on the observation angle and does not exhibit an ideal spherical wavefront in the far zone. These difficulties are reflected in the fact that bending magnet brightness was never described, up to now, in a satisfactory way within the Wigner distribution formalism. In particular, in the usually accepted approximations, the description of the bending magnet brightness turns out to be inconsistent even qualitatively. At variance, the analysis of undulator brightness given by Kim was based from the very beginning on the Wigner distribution formalism. This is reflected in the fact that the approximations he uses for describing the brightness are parametrically consistent with all exact results. However, here we report numerical disagreement between  exact results and approximated results in three geometrical optics limits where the brightness is nothing more than the maximum of the  radiance, which is the photon flux density in phase space.

\subsection{Radiation field in the space-frequency domain}


We consider a planar undulator, so that the transverse velocity of an electron can be written as

\begin{equation}
\vec{v}_\bot(z) = - {c K\over{\gamma}} \sin{\left(k_w z\right)}
\vec{e}_x~, \label{vuzo}
\end{equation}
where $k_w = 2\pi/\lambda_w$ with $\lambda_w$ the undulator period and $K$ the undulator parameter

\begin{equation}
K=\frac{\lambda_w e H_w}{2 \pi m_\mathrm{e} c^2}~, \label{Kpara}
\end{equation}
$m_\mathrm{e}$ being the electron mass and $H_w$ being the maximum of the magnetic field produced by the undulator on the $z$ axis.

We will assume, for simplicity, that the resonance condition with the fundamental harmonic is satisfied. In this way, our treatment leads to an analytical description of undulator radiation at the source position, i.e. in the middle of the undulator, at $z=0$. The resonance condition with the fundamental harmonic is given by

\begin{eqnarray}
\frac{\omega}{2 \gamma^2 c} \left(1+\frac{K^2}{2}\right) =
\frac{2\pi}{\lambda_w}~. \label{rsfirsth}
\end{eqnarray}
A well-known expression for the angular distribution of the first harmonic field in the far-zone (see Appendix A for a detailed derivation) can be obtained from Eq. (\ref{generalfin}). Such expression is axis-symmetric, and can therefore be presented as a function of a single observation angle $\theta$, where

\begin{equation}
\theta^2 = \theta_x^2+\theta_y^2~, \label{thsq}
\end{equation}
$\theta_x$ and $\theta_y$ being angles measured from the undulator $z$-axis in the horizontal and in the vertical direction. One obtains the following distribution for the slowly varying envelope of the electric field:

\begin{eqnarray}
\widetilde{{E}}(z_0, \theta)&=& -\frac{K \omega e L} {2 c^2
z_0 \gamma} A_{JJ}\exp\left[i\frac{\omega z_0}{2 c}
\theta^2\right] \mathrm{sinc}\left[\frac{\omega L\theta^{2}}{4
c}\right] ~,\label{undurad4bis}
\end{eqnarray}
where the field is polarized in the horizontal direction.  Here $L = \lambda_w N_w$ is the undulator length and  $N_w$ the number of undulator periods. Finally, $A_{JJ}$ is defined as

\begin{equation}
A_{JJ} = J_o\left(\frac{K^2}{4+2K^2}\right)
-J_1\left(\frac{K^2}{4+2K^2}\right)~, \label{AJJdef}
\end{equation}
$J_n$ being the n-th order Bessel function of the first kind. Eq.(\ref{undurad4bis}) describes a field with spherical wavefront centered in the middle of the undulator. Eq. (\ref{farzone}) can now be used to calculate the field distribution at the virtual source yielding

\begin{eqnarray}
\widetilde{{E}}(0, r) &=& i \frac{K \omega e} {2 c^2
\gamma} A_{JJ}\left[\pi - 2\mathrm{Si} \left(\frac{\omega {r}^2}{L c}\right)\right]~, \label{undurad5}
\end{eqnarray}

\begin{figure}
\begin{center}
\includegraphics*[trim = 0 280 0 0, clip, width=100mm]{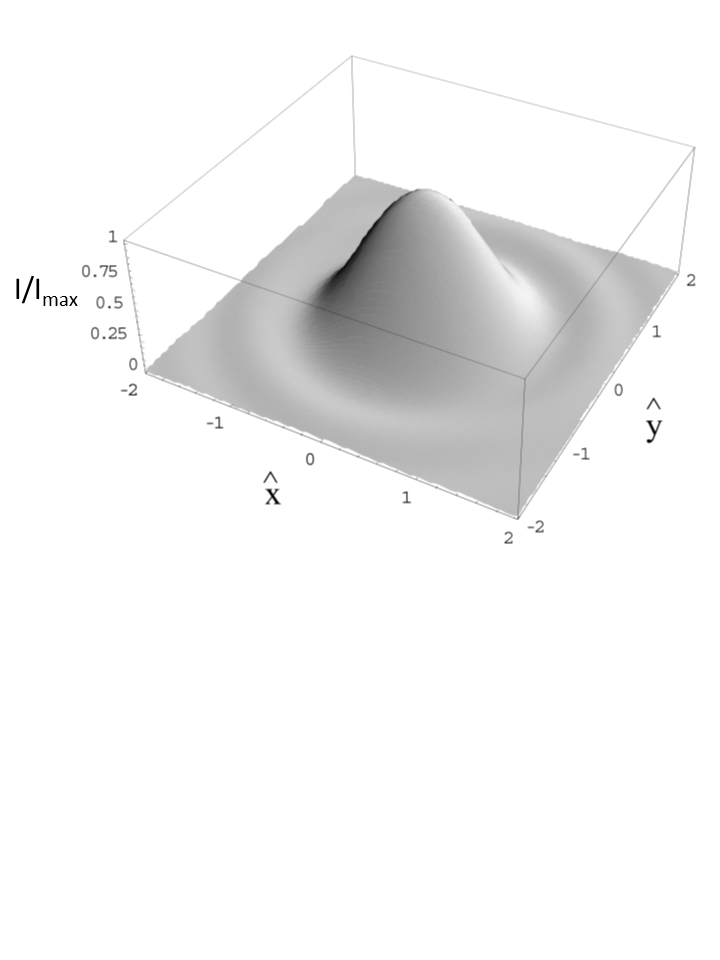}
\caption{\label{virundu} 3D Intensity distribution at the beam waist location $I/I_\mathrm{max}$, as a function of  $x/\sqrt{\lambdabar L}$ and $y/\sqrt{\lambdabar L}$.}
\end{center}
\end{figure}
where $\mathrm{Si}(z)=\int_0^z dt \sin(t)/t$ indicates the sin integral function and $r = |\vec{r}|$ is the distance from the $z$ axis on the virtual-source plane. Note that $\widetilde{E}(0, r)$ is axis-symmetric. Eq. (\ref{undurad5}), that has been already presented in \cite{FOUR}, describes a virtual field with a plane wavefront. Let us compare this virtual field with a laser-beam waist. In laser physics, the waist  is located in the center of the optical cavity. In analogy with this, in our case the virtual source is located in the center of the undulator. Both in laser physics and in our situation the waist has a plane wavefront and the transverse dimension of the waist is much longer than the wavelength. Note that the phase of the wavefront in Eq. (\ref{undurad5}) is shifted of $-\pi/2$ with respect to the spherical wavefront in the far zone. Such phase shift is analogous to the Guoy phase shift in laser physics. Finally, in laser physics, the Rayleigh range for a laser beam is presented in the form $z_R = (\omega/c) w_o^2$, $w_o$ being the radius of the beam at the location of the waist (i.e. at that position along $z$ where the wavefront is flat). This is defined, for example, by requiring that the intensity on the edge of an aperture of radius $w_o$ be one fourth of the intensity at the center of the radiation spot. In the undulator source case, the definition given above amounts to $w_o = 0.9 (c L/\omega)^{1/2}$ and $z_R=0.8 L \simeq L$.  In the case of a laser beam the Rayleigh range is related to the resonator geometrical factor. In analogy with this, in the case of an undulator source the Rayleigh range is related to the undulator geometrical factor. The relative intensity at the virtual source is plotted in Fig. \ref{virundu}.

Eq. (\ref{undurad4bis}) and Eq. (\ref{undurad5}) can be generalized to the case of a particle with a given offset $\vec{l}$ and deflection angle $\vec{\eta}$ with respect to the longitudinal axis, assuming that the magnetic field in the undulator is independent of the transverse coordinate of the particle. Although this can be done using Eq. (\ref{generalfin}) directly, it is sometimes possible to save time by getting the answer with some trick. For example, in the undulator case one take advantage of the following geometrical consideration \cite{FOUR}, which are in agreement with rigorous mathematical derivation. First, we consider the effect of an offset $\vec{l}$ on the transverse plane, with respect to the longitudinal axis $z$. Since the magnetic field experienced by the particle does not change, the far-zone field is simply shifted by a quantity $\vec{l}$.  Eq. (\ref{undurad4bis}), can be immediately generalized by systematic substitution of the transverse coordinate of observation, $\vec{r}_0$ with $\vec{r}_0 -\vec{l}$. This means that $\vec{\theta}=\vec{r}_0/z_0$ must be substituted by $\vec{\theta} - \vec{l}/z_0$, thus yielding

\begin{eqnarray}
\widetilde{{E}}\left(z_0,  \vec{l}, \vec{\theta}\right)&=&
-\frac{K \omega e L} {2 c^2 z_0 \gamma}
A_{JJ}\exp\left[i\frac{\omega z_0}{2 c}
\left|\vec{\theta}-\frac{\vec{l}}{z_0}\right|^2\right]
\mathrm{sinc}\left[\frac{\omega L
\left|\vec{\theta}-\left({\vec{l}}/{z_0}\right)\right|^2}{4
c}\right] ~.\label{undurad4bisgg0}
\end{eqnarray}
Let us now discuss the effect of a deflection angle $\vec{\eta}$. Since the magnetic field experienced by the electron is assumed to be independent of its transverse coordinate, the trajectory followed is still sinusoidal, but the effective undulator period is now given by $\lambda_w/\cos(\eta) \simeq (1+\eta^2/2) \lambda_w$. This induces a relative red shift in the resonant wavelength $\Delta \lambda/\lambda \sim \eta^2/2$. In practical cases of interest we may estimate $\eta \sim 1/\gamma$. Then, $\Delta \lambda/\lambda \sim 1/\gamma^2$ should be compared with the relative bandwidth of the resonance, that is $\Delta \lambda/\lambda \sim 1/N_w$, $N_w$ being the number of undulator periods. For example, if $\gamma=10^3$, the red shift due to the deflection angle can be neglected in all situations of practical relevance. As a result, the introduction of a deflection angle only amounts to a rigid rotation of the entire system. Performing such rotation we should account for the fact that the phase factor in Eq. (\ref{undurad4bisgg0}) is indicative of a spherical wavefront propagating outwards from position $z=0$ and remains thus invariant under rotations. The argument in the $\mathrm{sinc}(\cdot)$ function in Eq. (\ref{undurad4bisgg0}), instead, is modified because the rotation maps the point $(z_0,0,0)$ into the point $(z_0, -\eta_x z_0, -\eta_y z_0)$. As a result, after rotation, Eq. (\ref{undurad4bisgg0}) transforms to

\begin{eqnarray}
&&\widetilde{{E}}\left(z_0, \vec{\eta}, \vec{l},
\vec{\theta}\right)= -\frac{K \omega e L A_{JJ}} {2 c^2 z_0 \gamma}
\exp\left[i\frac{\omega z_0}{2 c}
\left|\vec{\theta}-\frac{\vec{l}}{z_0}\right|^2\right]
\mathrm{sinc}\left[\frac{\omega L
\left|\vec{\theta}-\left({\vec{l}}/{z_0}\right)-\vec{\eta}\right|^2}{4
c}\right] \cr &&\label{undurad4bisgg00}
\end{eqnarray}
Finally, in the far-zone case, we can always work in the limit for $l/z_0 \ll 1$, that allows one to neglect the term ${\vec{l}}/{z_0}$ in the argument of the $\mathrm{sinc}(\cdot)$ function, as well as the quadratic term in $\omega l^2/(2 c z_0)$ in the phase.  Thus Eq. (\ref{undurad4bisgg00}) can be further simplified, giving the generalization of Eq. (\ref{undurad4bis}) in its final form:

\begin{eqnarray}
\widetilde{{E}}\left(z_0, \vec{\eta}, \vec{l},
\vec{\theta}\right)&=& -\frac{K \omega e L A_{JJ}} {2 c^2 z_0 \gamma}
\exp\left[i\frac{\omega}{c}\left( \frac{z_0 \theta^2}{2}-
\vec{\theta}\cdot\vec{l} \right)\right]
\mathrm{sinc}\left[\frac{\omega L
\left|\vec{\theta}-\vec{\eta}\right|^2}{4 c}\right] ~.
\label{undurad4bisgg}
\end{eqnarray}
The expression for the field at virtual source, Eq. (\ref{undurad5}), should be modified accordingly. Namely, one has to plug Eq. (\ref{undurad4bisgg}) into Eq. (\ref{farzone}), which gives

\begin{eqnarray}
\widetilde{{E}}\left(0, \vec{\eta}, \vec{l}, \vec{r}
\right)&=& -\frac{i K \omega^2 e L A_{JJ}} {4 \pi c^3 \gamma} \int
d\vec{\theta} \exp\left[i\frac{\omega}{c}
\vec{\theta}\cdot\left(\vec{r}-\vec{l}\right)\right]
\mathrm{sinc}\left[\frac{\omega L
\left|\vec{\theta}-\vec{\eta}\right|^2}{4 c}\right] \cr &&
\label{undurad4bisggtr}
\end{eqnarray}
yielding

\begin{eqnarray}
\widetilde{{E}}\left(0,\vec{\eta}, \vec{l},
\vec{r}\right) &=& i \frac{K \omega e} {2 c^2 \gamma}
A_{JJ}\exp\left[i \frac{\omega}{c} \vec{\eta} \cdot
\left(\vec{r}-\vec{l}\right) \right]\left[\pi - 2\mathrm{Si}
\left(\frac{\omega \left|\vec{r}-\vec{l}\right|^2}{L
c}\right)\right]~\cr && \label{undurad5gg}
\end{eqnarray}
as final result. The meaning of Eq. (\ref{undurad5gg}) is that offset and deflection of the single electron motion with respect to the longitudinal axis of the system result in a transverse shift and a tilting of the waist plane. The combination $(\vec{r}-\vec{l}~)$ in Eq. (\ref{undurad5gg}) describes the shift, while the phase factor represents the tilting of the waist plane.


To sum up, the diffraction size of the undulator radiation beam is about $\sqrt{\lambdabar L} \gg \lambdabar$. This means that the radiation from an ultra-relativistic electron can be interpreted as generated from a virtual source, which produces a laser-like beam. In principle, such virtual source can be positioned everywhere down the beam, but there is a particular position where it is similar, in many aspects, to the waist of a laser beam. In the case of an undulator this location is the center of the insertion device, where virtual source exhibits a plane wavefront. For a particle moving on-axis, the field amplitude distribution at the virtual source is axially symmetric, Eq. (\ref{undurad5}). When the particle offset is different from zero, the laser-like beam is shifted. When the particle also has a deflection, the laser-like beam is tilted, but the wavefront remains plane. Then, since radiation from a given electron is correlated just with itself, it follows that radiation from an electron beam is an incoherent collection of laser-like beams with different offsets and deflections.

For a filament electron beam of current $I$, the angular spectral flux density in the direction $(\theta_x, \theta_y)$  can be written as

\begin{eqnarray}
\frac{d F}{d \Omega} = \frac{d \dot{N}_{ph}}{d \Omega (d \omega/\omega)}  = \frac{I}{e \hbar} \frac{c z_0^2}{4\pi^2} |\widetilde{E}|^2 ~,
\label{dfdom}
\end{eqnarray}
where $d\dot{N}_{ph}/(d \omega/\omega)d \Omega$ is the number of photons per unit time per unit solid angle per relative frequency bandwidth, and $\widetilde{E}$ is the slowly varying envelope of the electric field produced by a single electron in a planar undulator at the resonance wavelength in the space-frequency domain, Eq. (\ref{undurad4bis}).  Eq. (\ref{dfdom}) can be directly deduced from Eq. (\ref{intW}) considering that $dS/z_0^2 = d \Omega$. It can be shown from that the maximum value of $|\widetilde{E}|^2$ as a function of $\theta_x$ and $\theta_y$ is reached on-axis, for $\theta_x = \theta_y = 0$. The angular spectral flux on-axis is given by

\begin{eqnarray}
\max\left(\frac{dF}{d \Omega}\right) = \frac{I}{e}\alpha K^2 A_{JJ}^2 \frac{L^2}{4\lambda^2\gamma^2}~,
\label{maxdfdom}
\end{eqnarray}
where $\alpha = e^2/(\hbar c)$ is the fine structure constant. The angle-integrated spectral flux $F = d \dot{N}_{ph}/(d \omega/\omega)$ is defined as

\begin{eqnarray}
F = \int \frac{dF}{d \Omega} d \theta_x d \theta_y ~.
\label{angint}
\end{eqnarray}
If we substitute Eq. (\ref{dfdom}) in Eq. (\ref{angint}) we obtain

\begin{eqnarray}
F = \frac{I}{e} \pi \alpha K^2 A_{JJ}^2 \frac{N_w}{2(1+K^2/2)} ~.
\label{Fsub}
\end{eqnarray}
From Eq. (\ref{Fsub}) and Eq. (\ref{maxdfdom}) one derives the following useful relation between the on-axis angular spectral flux and the angle integrated spectral flux

\begin{eqnarray}
\max\left(\frac{dF}{d \Omega}\right) = \frac{F}{\pi} \frac{L}{\lambda}~ .
\label{maxsub}
\end{eqnarray}

\subsection{Geometrical properties of undulator source}

The results presented above for our undulator source analysis is far from trivial. As a matter of fact, our results are in contrast with literature. Let us consider as cardinal example the idea that the calculation of the undulator brightness needs to take into account depth-of-field effects (i.e. the contribution to the apparent source size from different poles). This idea can be found, for example, in \cite{WIED} and is a misconception originating from the analysis of the undulator source based on intuitive geometrical arguments regarded as self-evident.  For example, in the textbook \cite{WIED} one can find: "The actual photon brightness is reduced from the diffraction limit due to betatron motion of the particles, transverse beam oscillation in the undulator, apparent source size on axis and under an oblique angle. All of these effects tend to increase the source size and reduce brightness. The particle beam cross section varies in general along the undulator. We assume here for simplicity that the beam size varies symmetrically along the undulator with a waist in its center. From beam dynamics it is then known that, for example, the horizontal beam size varies like $\sigma_x^2 = \sigma_x^2(0) + \sigma_{x'}^2(0) z^2$, where $\sigma_x(0)$ is the beam size at the waist, $\sigma_{x'}(0)$ the divergence of the beam at the waist and $-L/2 \leq z \leq L/2$ the distance from the waist. The average beam size along the undulator length $L$ is then" given by Eq. (\ref{text}). "Similarly, due to an oblique observation angle $\theta$ with respect to the $(y, z)$-plane or $\psi$ with respect to the $(x, z)$-plane we get a further additive contribution $\theta L/6$ to the apparent beam size. Finally, the apparent source size is widened by the transverse beam wiggle in the periodic undulator field. This oscillation amplitude is $r_w=\lambda_w K/(2 \pi \gamma)$. Collecting all contributions and adding them in quadrature, the total effective beam-size parameters are given by

\begin{eqnarray}
&&\sigma^2_{t,(x,y)} =  \sigma_r^2 + \sigma^2_{x,y}(0) + \left(\frac{\lambda_w K}{2\pi \gamma}\right)^2 + \frac{1}{12} \sigma^2_{x',y'}(0) L^2 + \frac{1}{36} (\theta^2,\psi^2) L^2~, \cr &&
\sigma^2_{t,(x',y')} = \sigma_r'^2 + \sigma^2_{x',y'}(0)~,
\label{wied12}
\end{eqnarray}
where the particle beam sizes can be expressed by the beam emittance and betatron function as $\sigma_{x,y}^2 = \epsilon_{x,y} \beta_{x,y}$, $\sigma_{x',y'}^2 = \epsilon_{x,y}/\beta_{x,y}$, and the diffraction limited beam parameters are $\sigma_r = \sqrt{2 \lambda L}/(4\pi)$ and $\sigma_{r'} = \sqrt{\lambda/(2L)}$"\footnote{Here we have not followed Wiedemann's original notations and definitions.}.

We stress that our criticism  is not only focused on the quantitative aspects the source size analysis in \cite{WIED}, but rather on a contradiction with fundamental facts from classical electrodynamics. In order to better see this, consider an optical setup performing a 1:1 imaging, where we set the object plane at the undulator center. The intensity distribution in the image plane is given by a convolution of the intensity from a single electron, $I_0$, with the electron density distribution in the object plane (which we consider Gaussian, as before):

\begin{eqnarray}
I(x,y) \sim \int_{-\infty}^\infty d l_x \int_{-\infty}^{\infty} d l_y I_0(x - l_x, y -l_y) \exp\left[- \frac{l_x^2}{2\sigma_x^2} - \frac{l_y^2}{2\sigma_y^2}\right]~.
\label{Intensity}
\end{eqnarray}
Here  $I_0(x, y) \sim [\pi -2 \mathrm{Si}(\omega(x^2+y^2)/(c L))]^2$ is the radiation intensity from a single electron on the source (see Eq. (\ref{undurad5gg})). Eq. (\ref{Intensity}) follows from statistical averaging over electron offsets and deflection angles. In fact, tilt angles are small compared to unity and, therefore, the source size is insensitive to the angular beam distribution. It is easy to see that the first two terms in Eq. (\ref{wied12}) for the source size are related with the approximation of the convolution Eq. (\ref{Intensity}) introduced by Kim \cite{KIM1}. However, the last two terms in Eq. (\ref{wied12}), which describe a source size widening due to depth-of-field effects, depend on the angular beam distribution and observation angle, and should not be there. They are rather the result of the misconception discussed above.

There is another objection that could be made to the analysis in \cite{WIED}, related with the third term in Eq. (\ref{wied12}). Under the resonance approximation it is logically inconsistent to present a term of order of $1/N_w$ in the expression for undulator source size. In fact,  on the one hand the radiation diffraction size for a single electron is order of $\sqrt{\lambdabar L}$. On the other hand the electron wiggling amplitude is given by $r_w = (K/\gamma)\lambda_w/(2\pi)$. In particular, at the fundamental harmonic follows that

\begin{eqnarray}
\frac{r_w^2}{\lambdabar L} = \frac{K^2}{\pi N_w (1+K^2/2)} \ll 1~,
\label{sizescomp}
\end{eqnarray}
where we used the resonance condition Eq. (\ref{rsfirsth}). This inequality holds independently of the value of $K$, because under the resonance approximation $N_w \gg 1$. Thus, the electron wiggling amplitude is always much smaller than the radiation diffraction size. It follows that under the resonance approximation the third term in Eq. (\ref{wied12}) should be neglected.

Note that the first-principle computer codes (see e.g. \cite{CHUB0}) have been used quite successful to model advanced SR sources and beamlines without specific analytical simplifications. Results may be obtained using numerical techniques alone, starting from the Lienard-Wiechert expression for the electromagnetic field (in the case of \cite{CHUB0} in the space-frequency domain) using only the ultrarelativistic approximation.  Codes can also be used to treat the case of 1:1 imaging of an undulator source. It is instructive to reconsider the problem of undulator source size prediction by means of numerical simulations alone, which play the same role of an experiment. We might consider two cases at fixed electron beam size:

\begin{itemize}
\item{The usual case with matched beta-function $\beta_0 \sim L$, and additionally $\sigma_x^2 = \epsilon_0 \beta_0$, $\sigma_{x'}^2 = \epsilon_0/\beta_0$, $\sigma_x^2 \gg \sigma_r^2$.}

\item{A case with a tenfold increase in emittance and a tenfold decrease in beta function i.e. $\epsilon = 10 \epsilon_0$ and $\beta = \beta_0/10$.}
\end{itemize}

Of course one is free to choose other numerical cases as well. One can check  that results of simulations confirm our prediction that source image is insensitive to the electron beam divergence. In both cases, with graphical accuracy, we find that the distribution at the image plane is the same, in contrast with the prediction \cite{WIED} of source size widening.

\subsection{\label{43} Wigner distribution and undulator sources}

We now turn to the main topic of this study, namely the analysis of the brightness of SR sources. In this section we apply the considerations developed in section \ref{sec:wiggen} to the case of an undulator source at resonance. First we calculate the Wigner distribution for a filament electron beam, that is a beam with zero emittance.  As second step we take into account the more general case when the electron beam has a finite phase space distribution.

According to the definition in Eq. (\ref{Wig13}), the Wigner distribution in the case of an electron beam with zero emittance is given by

\begin{eqnarray}
W_0(\vec{r}, \vec{\theta}) =  \frac{c}{(2\pi)^4} \frac{I}{e \hbar} \left(\frac{\omega}{c}\right)^2  \int d\Delta\vec{r}~ G_0(\vec{r},\Delta \vec{r}) \exp\left(-i \omega \vec{\theta} \cdot \Delta \vec{r}/c\right)
\label{W0start}
\end{eqnarray}
where $G_0 = \widetilde{E}(\vec{r} + \Delta \vec{r}/2) \widetilde{E}^*(\vec{r} - \Delta \vec{r}/2)$ is the diffraction-limited cross-spectral density and $\widetilde{E}(\vec{r})$ is defined by Eq. (\ref{undurad5}). The peak value of $W_0$ as a function of $\vec{r}$ and $\vec{\theta}$ is reached on-axis for $\vec{r} = 0$ and $\vec{\theta} = 0$. Accordingly, we can compute the undulator brightness. For our diffraction-limited regime at resonance this can be done analytically \cite{KIM1}

\begin{eqnarray}
B = \max(W_0) = \frac{\lambda}{2}^2 F~ ,
\label{Brightund1}
\end{eqnarray}
where $F$ is the angle-integrated spectral flux defined by Eq. (\ref{Fsub}). Here we have only used the axial symmetry of the electric field radiated by a single electron in an undulator at resonance, Eq. (\ref{undurad5}). In particular this symmetry yields the relation $\widetilde{E}(\Delta \vec{r}/2) \widetilde{E}^{*}(-\Delta \vec{r}/2) = |\widetilde{E}(\Delta \vec{r}/2)|^2$. It is reasonable to expect that Eq. (\ref{Brightund1}) is valid for any radiation beam with axial symmetric field distribution. One can easily show that this is indeed the case. One typical example when this fact is verified, is for a Gaussian beam.

The Wigner distribution $W(\vec{r},\vec{\theta})$ for an electron beam with finite emittance  can be presented as a convolution product between the electron phase space distribution $f_\bot(\vec{l}, \vec{\eta})$ and the Wigner distribution for a filament beam $W_0(\vec{r}, \vec{\theta})$ according to Eq. (\ref{Wcorr}) \cite{KIM1}. Note that, as remarked before, Eq. (\ref{Wcorr}) has no full generality and can be used only in the case when focusing elements  are excluded from consideration\footnote{This result can be applied in the case of an undulator without focusing quadrupoles, but not in the case one is interested in calculating the brightness, e.g. the brightness of SR from an XFEL setup, where the undulator is embedded in a FODO lattice.}.

In the following we will make consistent use of dimensional analysis, which allows one to classify the grouping of dimensional variables in a way that is most suitable for subsequent study. Normalized units in the undulator case will be defined as

\begin{eqnarray}
&& \vec{\hat{\eta}} = \frac{\vec{\eta}}{\sqrt{\lambdabar/L}}     \cr
&& \vec{\hat{\theta}} = \frac{\vec{\theta}}{\sqrt{\lambdabar/L}} \cr
&& \vec{\hat{r}} = \frac{\vec{r}}{\sqrt{\lambdabar L}} \cr
&& \vec{\hat{l}} = \frac{\vec{l}}{\sqrt{\lambdabar L}} ~.
\label{normq}
\end{eqnarray}
We assume that the motion of electrons in the horizontal and vertical directions are completely uncoupled. Additionally we assume a Gaussian distribution of the electron beam in phase space. These two assumptions are practically realized, with good accuracy, in storage rings. For simplicity, we also assume that the minimal values of the beta-functions in horizontal and vertical directions are located at the middle of the undulator, at $z = 0$. Then, at that position, the transverse phase space distribution can be expressed as

\begin{eqnarray}
\hat{f}_\bot = f_{\vec{\hat{l}}}\left(\vec{\hat{l}}\right)f_{\vec{\hat{\eta}}}\left(\vec{\hat{\eta}}\right)=f_{\eta_x}(\hat{\eta}_x) f_{\eta_x}(\hat{\eta}_y)
f_{l_x}(\hat{l}_x) f_{l_x}(\hat{l}_y)
\label{fphase}
\end{eqnarray}
with

\begin{eqnarray}
&& f_{\eta_x}(\hat{\eta}_x) = \frac{1}{\sqrt{2\pi D_x}}
\exp{\left(-\frac{\hat{\eta}_x^2}{2 D_x}\right)}~,~~~~
f_{\eta_y}(\hat{\eta}_y)  = \frac{1}{\sqrt{2\pi D_y}}
\exp{\left(-\frac{\hat{\eta}_y^2}{2 D_y}\right)}~,\cr &&
f_{l_x}(~\hat{l}_x) =\frac{1}{\sqrt{2\pi N_x} }
\exp{\left(-\frac{\hat{l}_x^2}{2 N_x}\right)}~,~~~~~
f_{l_y}(~\hat{l}_y)=\frac{1}{\sqrt{2\pi N_y} }
\exp{\left(-\frac{\hat{l}_y^2}{2 N_y}\right)}~.\cr && \label{distr}
\end{eqnarray}
Here

\begin{equation}
D_{x,y} = \frac{\sigma_{x',y'}^2} {{\lambdabar}/L}~,~
N_{x,y} = \frac{\sigma^2_{x,y}} {\lambdabar L}~,\label{enne}
\end{equation}
Parameters $N_{x,y}$ will be called beam diffraction parameters, are analogous to Fresnel numbers and correspond to the normalized square of the electron beam sizes, whereas $D_{x,y}$ represent the normalized square of the electron beam divergences.

We begin by writing the expression for the cross-spectral density at the virtual source:

\begin{eqnarray}
&& G\left(\vec{\hat{r}}, \Delta \vec{\hat{r}}\right) \cr && = \int d \vec{\hat{\eta}} \exp\left(i  \vec{\hat{\eta}}\cdot \Delta \vec{\hat{r}}\right) f_{\vec{\hat{\eta}}}\left(\vec{\hat{\eta}}\right) \int d \vec{\hat{l}} f_{\vec{\hat{l}}}\left(\vec{\hat{l}}\right)
\widetilde{E}\left(\vec{\hat{r}} + \frac{\Delta \vec{\hat{r}}}{2} -\vec{\hat{l}}\right) \widetilde{E}^*\left(\vec{\hat{r}} -\frac{\Delta \vec{\hat{r}}}{2} - \vec{\hat{l}}\right) ~,\cr &&
\label{G0norm}
\end{eqnarray}
where the field is defined by Eq. (\ref{undurad5}). It should be noted that the independent variable in Eq. (\ref{undurad5}) is $\omega r^2/(L c)$, and corresponds to $\hat{r}^2$ in dimensionless units. Thus, the characteristic transverse range of the field at the source in dimensionless units is of order of unity. One sees that the cross-spectral density is the product of two separate factors. The first is the Fourier transform of the distribution of the electrons angular divergence. The second is the convolution of the transverse electron beam distribution with the four-dimensional function $\widetilde{E}(\vec{\hat{r}}+\Delta \vec{\hat{r}}/2)\widetilde{E}^*(\vec{\hat{r}}-\Delta \vec{\hat{r}}/2)$. In fact, after the change of variables $\vec{\phi} = \vec{\hat{r}} - \vec{\hat{l}}$ we have

\begin{eqnarray}
G\left(\vec{\hat{r}},\Delta \vec{r}\right) &=&
\frac{1}{2 \pi \sqrt{N_x N_y}} \exp \left[-\frac{(\Delta \hat{x})^2
D_x}{2}\right] \exp \left[-\frac{(\Delta \hat{y})^2 D_y}{2}\right]\cr &&
\times \int_{-\infty}^{\infty} d \phi_x \int_{-\infty}^{\infty} d
\phi_y \exp\left[-\frac{\left(\phi_x-\hat{x}\right)^2}{2
N_x}\right] \exp\left[-\frac{\left(\phi_y-\hat{y}\right)^2}{2
N_y}\right] \cr && \times
\widetilde{E}\left({\phi_{x}}+\frac{\Delta
\hat{x}}{2},{\phi_{y}}+\frac{\Delta \hat{y}}{2}
\right) \widetilde{E}^*\left({\phi_{x}}-\frac{\Delta \hat{x}}{2}
,{\phi_{y}}-\frac{\Delta \hat{y}}{2}
\right)~.\label{Gnor3}
\end{eqnarray}
It is instructive to examine this expression in the geometrical optics asymptotes. Let us start with beam size- and divergence-dominated regime. In Eq. (\ref{Gnor3}) the range of the variable $\phi_{x,y}$ is effectively limited up to values $|\phi_{x,y}| \sim 1$. In fact $\phi_{x,y}$ enters the expression for $\widetilde{E}$. It follows that at values larger than unity the integrand in Eq. (\ref{Gnor3}) is suppressed. Moreover, in the beam size- and divergence-dominated regime one has $N_{x,y} \gg 1$ and $D_{x,y} \gg 1$, so that we can neglect $\phi_{x,y}$ in the exponential functions in $N_{x,y}$, while from the exponential in $D_{x,y}$ follows that $\Delta \hat{x} \ll 1$ and $\Delta \hat{y} \ll 1$ can be neglected in the field $\widetilde{E}$. As a result we obtain

\begin{eqnarray}
&&G = \frac{1}{2\pi \sqrt{N_x N_y}} \exp\left(- \frac{D_x \Delta \hat{x}^2}{2}- \frac{D_y \Delta \hat{y}^2}{2}\right) \exp\left(- \frac{\hat{x}^2}{2N_x}- \frac{\hat{y}^2}{2N_y}\right) \cr && \times\int_{-\infty}^{\infty} d \phi_x \int_{-\infty}^{\infty} d \phi_y |\widetilde{E}(\phi_x, \phi_y)|^2
\label{GNDlarge}
\end{eqnarray}

or, in dimensional units,

\begin{eqnarray}
&&G = \frac{1}{2 \pi \sigma_x \sigma_y} \exp\left(- \frac{\omega^2 \sigma_{x'}^2 \Delta x^2}{2 c^2} - \frac{\omega^2 \sigma_{y'}^2 \Delta y^2}{2 c^2}\right) \exp\left(-\frac{x^2}{2 \sigma_x^2}-\frac{y^2}{2 \sigma_y^2}\right) \int d\vec{r} ~|\widetilde{E}(\vec{r})|^2~. \cr &&
\label{GNDlargedim}
\end{eqnarray}
One can obtain the Wigner distribution $W$ from the above expression for $G$ by means of Eq. (\ref{Wig13})

\begin{eqnarray}
&&W(\vec{r}, \vec{\theta}) =  \frac{1}{(2 \pi)^2  \sigma_x \sigma_y \sigma_{x'} \sigma_{y'}} \exp\left(- \frac{\theta_x^2}{2 \sigma_{x'}^2 } - \frac{\theta_y^2}{2 \sigma_{y'}^2 }\right) \exp\left(-\frac{x^2}{2 \sigma_x^2}-\frac{y^2}{2 \sigma_y^2}\right) \cr && \times\frac{c}{(2\pi)^2} \frac{I}{e \hbar} \int d\vec{r} ~|\widetilde{E}(\vec{r})|^2 ~.
\label{WNDlargedim}
\end{eqnarray}
The peak value of the Wigner distribution is reached on-axis and is given by

\begin{eqnarray}
B = \max(W)  = \frac{F}{(2\pi)^2 \sigma_x \sigma_y \sigma_{x'} \sigma_{y'}}~,
\label{BNDlarge}
\end{eqnarray}
where $F$ is the spectral flux

\begin{eqnarray}
F= \frac{d \dot{N}_\mathrm{ph}}{d \omega/\omega} = \frac{c}{(2\pi)^2} \frac{I}{e \hbar} \int d \vec{r}~ |\widetilde{E}(\vec{r})|^2
\label{FforBNDlarge}
\end{eqnarray}
radiated by an electron beam with current $I$ in the undulator.

Let us now consider the beam divergence-dominated regime, that is when $D_{x,y} \gg 1 \gg N_{x,y}$. From an analysis of the exponential functions in $D_{x,y}$ in Eq. (\ref{Gnor3}) follows that $\Delta \hat{x} \ll 1$, $\Delta \hat{y} \ll 1$ can be neglected in the expression for the field $\widetilde{E}$. Then, since $N_{x,y} \ll 1$,
it follows that the distributions $\exp[ - (\phi_x - \hat{x})^2/(2N_x)]$ and $\exp[ - (\phi_y  - \hat{y})^2/(2N_y)]$ are respectively sharply peaked about $\phi_x = \hat{x}$ and $\phi_y = \hat{y}$. Hence, we can approximate

\begin{eqnarray}
G = \exp\left(- \frac{D_x \Delta \hat{x}^2}{2}\right) \exp\left(- \frac{D_y \Delta \hat{y}^2}{2}\right) |\widetilde{E}(\hat{x},\hat{y})|^2~.
\label{Gdivdom}
\end{eqnarray}
Finally, from Eq. (\ref{Wig13}) follows an expression for $W$ in dimensional units:

\begin{eqnarray}
W(\vec{r}, \vec{\theta}) = \frac{1}{2\pi \sigma_{x'} \sigma_{y'}} \exp\left(- \frac{\theta_x^2}{2\sigma_{x'}^2}\right) \exp \left(-\frac{\theta_y^2}{2\sigma_{y'}^2}\right)
\frac{c}{(2\pi)^2} \frac{I}{e \hbar} |\widetilde{E}(\vec{r})|^2 ~.
\label{Wdivdom}
\end{eqnarray}
where $\widetilde{E}(\vec{r})$ is the radiation field in Eq. (\ref{undurad5}). The peak value of Wigner distribution is reached on-axis and given by

\begin{eqnarray}
B = \max(W) =  \frac{1}{2 \pi \sigma_{x'} \sigma_{y'}} \max \left(\frac{dF}{dS} \right)~,
\label{Bdivdom}
\end{eqnarray}
where

\begin{eqnarray}
\max\left(\frac{dF}{dS}\right) =  \frac{c}{(2\pi)^2}  \frac{I}{e \hbar} |\widetilde{E}(0)|^2
\label{dfdsforBdivdom}
\end{eqnarray}
is the maximum photon flux density on the source in the diffraction limited case. A straightforward integration of $dF/dS$ over the source area gives the total flux in Eq. (\ref{Fsub}). From Eq. (\ref{undurad5}) and Eq. (\ref{dfdsforBdivdom}) one derives the following relation between on-axis flux density and total flux

\begin{eqnarray}
\max\left(\frac{dF}{dS}\right) = F \frac{\pi}{\lambda L}~.
\label{tflux}
\end{eqnarray}
With this we also point out that in the beam divergence-dominated regime the brightness can be written as

\begin{eqnarray}
B = \frac{F}{2 \sigma_{x'} \sigma_{y'} \lambda L}~ .
\label{BBB}
\end{eqnarray}
A third interesting limiting case can be considered. In the beam size-dominated regime $D_{x,y} \ll 1 \ll N_{x,y}$, Eq. (\ref{Gnor3}) can be simplified as follows.  We can neglect $\phi_{x,y}$ in the exponential functions in $N_{x,y}$. Moreover,  the range of variables $\Delta \hat{x}$ and $\Delta \hat{y}$ is effectively limited up to values of order of unity because they enter the expression for the electric field $\widetilde{E}$. It follows that, when $D_{x,y} \ll 1$, the exponential functions in $D_{x,y}$ can be replaced with unity and we obtain the following expression for the cross-spectral density in dimensional units

\begin{eqnarray}
&&G = \frac{1}{2\pi \sigma_x \sigma_y} \exp\left( - \frac{x^2}{2\sigma_x^2}\right) \exp\left( - \frac{y^2}{2\sigma_y^2}\right) \cr && \times \int_{-\infty}^{\infty} d x' \int_{-\infty}^{\infty} dy' \widetilde{E}\left( x'+\frac{\Delta x}{2}, y'+\frac{\Delta y}{2} \right) \widetilde{E}^*\left( x'-\frac{\Delta x}{2}, y'-\frac{\Delta y}{2}\right)~. \cr &&
\label{Gthird}
\end{eqnarray}
The Wigner distribution in Eq. (\ref{Wig13}) can therefore be written as

\begin{eqnarray}
&&W(\vec{r}, \vec{\theta}) = \frac{1}{2\pi \sigma_x \sigma_y} \exp\left( - \frac{x^2}{2\sigma_x^2}\right)
\exp\left( - \frac{y^2}{2\sigma_y^2}\right) \frac{c}{(2\pi)^{4}} \frac{I}{e \hbar} \left(\frac{\omega}{c}\right)^2 \cr && \times \int d \Delta \vec{r} \exp\left(-i \frac{\omega}{c} \vec{\theta} \cdot \Delta
\vec{r}\right)\int d \vec{r}' \widetilde{E}(\vec{r}' +\Delta \vec{r}/2) \widetilde{E}^*(\vec{r}' - \Delta \vec{r}/2)~.
\label{Wthird}
\end{eqnarray}
Some simplification may be obtained by rewriting the electric field on the source, $\widetilde{E}(\vec{r})$, in the terms of the far field $\widetilde{E}(\vec{\theta})$. In fact, inserting Eq. (\ref{farzone}) into Eq (\ref{Wthird}), performing the integration  and rearranging yields\footnote{If we write $\widetilde{E}(r)$ as the integral in Eq. (\ref{farzone}), after substitution in  Eq. (\ref{Wthird}) we can present results of integration over $\Delta \vec{r}$ and $\vec{r}'$  in terms of a Dirac $\delta$-function and evaluate all integrals analytically.}:

\begin{eqnarray}
W(\vec{r}, \vec{\theta}) = \frac{1}{2\pi \sigma_x \sigma_y} \exp\left(-\frac{x^2}{2\sigma_x^2} -
\frac{y^2}{2\sigma_y^2}\right) \frac{c z_0^2}{(2\pi)^2}  \frac{I}{e \hbar} |\widetilde{E}(\vec{\theta})|^2~,
\label{Wthirdfin}
\end{eqnarray}
where $\widetilde{E}(\vec{\theta})$ is the radiation field from Eq. (\ref{undurad4bis}). The peak value of the Wigner function is given by

\begin{eqnarray}
B = \max(W) = \frac{1}{2\pi\sigma_x\sigma_y} \max \left(\frac{dF}{d \Omega}\right)~,
\label{Bthirdfin}
\end{eqnarray}
where $\max(dF/d \Omega)$ is the maximum of the angular photon flux given by Eq. (\ref{maxsub}).

\subsection{\label{TGS} Third-generation SR source approximation}

Generally, calculation of undulator source brightness in the case of partial coherent radiation involves very complicated and time-expensive evaluations. In fact, one needs to find the maximum of a Wigner distribution
which is function of four variables ($x, y, \theta_x$, $\theta_y$) and depends on four dimensionless problem parameters ($N_{x, y}$, $D_{x, y}$). In some particular cases, however, the brightness can be determined analytically. In section \ref{43} we  have seen how exact results can be obtained for the diffraction-limited case and in geometrical optics limits.

We now focus our discussion on third-generation SR sources. In this case  we can consider $N_x \gg 1$ and $D_x \gg 1$ still retaining full generality concerning the values of $N_y$ and $D_y$, due to the small coupling coefficient between horizontal and vertical emittance. Exploitation of the extra parameters $N_x \gg 1$ and $D_x \gg 1$ specializes our theory to the case of third-generation sources.

With this in mind we  simplify the Wigner distribution calculations beginning with the expression for the cross-spectral density at the virtual source, Eq. (\ref{Gnor3}). In Eq. (\ref{Gnor3}) the range of the integration variable $\phi_x$ is effectively limited up to values $|\phi_x| \sim 1$. In fact, $\phi_x$ enters the expression for the field $\widetilde{E}$. It follows that at values larger than unity the integrand in Eq. (\ref{Gnor3}) is suppressed. Then, since $N_x \gg 1$, we can neglect $\phi_x$ in the exponential function. Moreover $D_x \gg 1$, and from the exponential function in $D_x$ follows that $\Delta \hat{x} \ll 1$ can be neglected in the expression for the field $\widetilde{E}$. As a result, Eq. (\ref{Gnor3}) is factorized in the product of horizontal cross-spectral density $G_x(\hat{x}, \Delta \hat{x})$ and a vertical cross-spectral density $G_y(\hat{y}, \Delta \hat{y})$:

\begin{eqnarray}
G(\hat{x}, \hat{y}, \Delta \hat{x}, \Delta \hat{y}) = G_x(\hat{x}, \Delta \hat{x}) G_y(\hat{y}, \Delta \hat{y})~,
\label{Gfact}
\end{eqnarray}

where

\begin{eqnarray}
G_x = \frac{1}{\sqrt{2\pi N_x}} \exp\left(- \frac{D_x \Delta \hat{x}^2}{2}\right) \exp\left(-\frac{\hat{x}^2}{2N_x}\right) ~,
\label{Gxfact}
\end{eqnarray}

\begin{eqnarray}
&& G_y = \frac{1}{\sqrt{2\pi N_y}} \exp\left(- \frac{D_y \Delta \hat{y}^2}{2}\right)
\int_{-\infty}^{\infty} d \phi_x \int_{-\infty}^{\infty} d \phi_y \exp\left[-\frac{(\phi_y -\hat{y})^2}{2 N_y}\right] \cr &&
\times \widetilde{E}\left(\phi_x , \phi_y +\frac{\Delta \hat{y}}{2}\right) \widetilde{E}^*\left(\phi_x,  \phi_y -\frac{\Delta \hat{y}}{2}\right)  ~.
\label{Gyfact}
\end{eqnarray}
Therefore, the Wigner distribution in dimensional units for an arbitrary state of coherence in the vertical direction can be written as

\begin{eqnarray}
&&W(x, y, \theta_x, \theta_y) = \frac{1}{\sqrt{2\pi}\sigma_x\sigma_{x'}\sigma_y}
\exp\left(-\frac{x^2}{2\sigma_x^2}\right)\exp\left(- \frac{\theta_x^2}{2\sigma_{x'}^2}\right)
\frac{c}{(2\pi)^4} \frac{I}{e \hbar} \frac{\omega}{c} \cr && \times \int_{-\infty}^{\infty} d \Delta y ~\exp\left(-i \frac{\omega}{c}\theta_y\Delta y\right) \exp\left(-\frac{\omega^2 \sigma_{y'}^2 \Delta y^2}{2 c^2} \right) \cr && \times \int_{-\infty}^{\infty} d x' \int_{-\infty}^{\infty} dy' \exp\left[- \frac{(y'-y)^2}{2 \sigma_y^2}\right] \widetilde{E}\left( x',  y' + \frac{\Delta y}{2}\right)\widetilde{E}^*\left( x', y' - \frac{\Delta y}{2}\right)
\label{WdimV}
\end{eqnarray}
%
%
%

It is instructive to check this expression in the case of (vertical) beam size- and (vertical) divergence-dominated regime. Since $D_y \gg 1$, one can neglect $\Delta y$ in the expression for the amplitude of the electric field. Moreover, $N_y \gg 1$ and the exponential function $\exp[-(y'-y)^2/(2 \sigma_y^2)]$ is smoothly varying in the region of integration of importance. Therefore, we can take it out of the integral sign, and Eq. (\ref{WdimV}) yields back Eq. (\ref{WNDlargedim}) and Eq. (\ref{BNDlarge}) as it must be.

Some other limiting forms of Eq. (\ref{WdimV}) are of interest, and will be discussed here. In the diffraction-limited regime for the vertical direction one has $N_y \ll 1$ and $D_y \ll 1$. The distribution $1/(\sqrt{2 \pi} \sigma_y)\exp[-(y'-y)^2/(2 \sigma_y^2)]$ is sharply peaked about $y' = y$, and can be approximated by a Dirac $\delta$-function. Moreover, the variable $\Delta y'$ enters the expression for the electric field $\widetilde{E}$, and in the region of integration of importance the exponential function in $\Delta y^2$ can be replaced by unity. Finally, the field has the symmetry property $\widetilde{E}(x,y) = \widetilde{E}(x,-y)$. Therefore, on-axis,  Eq. (\ref{WdimV}) can be simplified to


%
%

\begin{eqnarray}
B = W(0, 0, 0, 0) = \frac{F}{2\pi \sigma_x \sigma_{x'}} \left( \frac{2}{\lambda} \right)~.
\label{W0000dif}
\end{eqnarray}
This exact result coincides with the result expected with the help of the approximate formula Eq. (\ref{Bundug}).

Let us now consider the beam divergence-dominated regime. When $N_y \ll 1 \ll D_y$ the variable $\Delta y$ can be neglected in the expression for the field and the exponential function in $N_y$, multiplied by $1/\sqrt{2\pi N_y}$, can be approximated by a Dirac $\delta$-function under the convolution integral in Eq. (\ref{Gyfact}). Therefore, the expression for the Wigner distribution on-axis can be simplified as

\begin{eqnarray}
W(0, 0, 0, 0) =  \frac{\sqrt{2\pi}}{\sigma_x\sigma_{x'}\sigma_{y'}} \frac{I}{e \hbar} \frac{c}{(2\pi)^4} \int_{-\infty}^{\infty} dx' \left|\widetilde{E}(x',0)\right|^2~
\label{B1item}
\end{eqnarray}
where $\widetilde{E}(x,y)$ is the radiation field in Eq. (\ref{undurad5}). Accounting for Eq. (\ref{tflux}) the brightness may be written as

\begin{eqnarray}
B = \max(W) = W(0, 0, 0, 0) =  \left[\frac{F}{\pi \sigma_x\sigma_{x'}\sigma_{y'} \sqrt{\lambda L} }\right] \frac{1}{4}  \int_{-\infty}^{\infty} d\phi S_0^2(\phi)~
\label{B2item}
\end{eqnarray}
where the dimensionless field distribution $S_0$ is defined by

\begin{eqnarray}
S_0(\phi) = \frac{1}{\pi} [\pi - 2 \mathrm{Si}(\phi^2)]
\label{S0}
\end{eqnarray}
The expression in the first parenthesis in Eq. (\ref{B2item}) represents the brightness expected by the estimate in Eq. (\ref{Bundug}) with the parameter choice in (\ref{sigrrp}). The number

\begin{eqnarray}
\frac{1}{4} \int_{-\infty}^{\infty} d\phi S_0^2(\phi) \simeq 0.35
\label{numberboh}
\end{eqnarray}
represents the disagreement between the exact expression and the estimated maximum of the photon flux density in the phase space.
%
%
%
%
%
%

%
%

A second geometrical optics limit for third generation SR sources is the beam size-dominated regime, when $D_y \ll 1 \ll N_y$. The  Wigner distribution Eq. (\ref{WdimV}) can be simplified following the same line of reasoning as for Eq. (\ref{Wthird}). From the analysis of the exponential function in $N_y$ in Eq. (\ref{Gyfact}) follows that we can take it out from under the integral over $\phi_y$. Moreover, the exponential function in $D_y$ can be replaced by unity and Eq. (\ref{WdimV}) can be written on-axis as

\begin{eqnarray}
&&W(0,0,0,0) = \frac{1}{\sqrt{2\pi}\sigma_x\sigma_{x'}\sigma_y}\frac{c}{(2\pi)^4} \frac{I}{e\hbar}
\frac{\omega}{c} \cr && \times \int_{-\infty}^{\infty} d \Delta y \int_{-\infty}^{\infty} dx' \int_{-\infty}^{\infty} dy'~ \widetilde{E}\left(x', y'+\frac{\Delta y}{2}\right)\widetilde{E}^*\left(x',y'-\frac{\Delta
y}{2}\right) ~. \label{Wsizdomax}
\end{eqnarray}
If we write the electric field amplitude $\widetilde{E}(x,y)$ as the integral in Eq. (\ref{farzone}), we can present result of the integration over  $x'$,$y'$ and  $\Delta y$  in Eq. (\ref{Wsizdomax}) as Dirac $\delta$-functions. Performing the integration and rearranging yields:

\begin{eqnarray}
W(0,0,0,0) = \frac{1}{\sqrt{2\pi} \sigma_x \sigma_{x'}\sigma_y} \frac{I}{e\hbar} \frac{c z_0^2}{(2\pi)^3}  \int_{-\infty}^{\infty} d \theta_x |\widetilde{E}(\theta_x,0)|^2  ~,
\label{Wsizdomax2}
\end{eqnarray}
where the  expression for the far field $\widetilde{E}(\theta_x,\theta_y)$ is given in Eq. (\ref{undurad4bis}). Accounting for Eq. (\ref{maxsub}) the brightness can be written as

\begin{eqnarray}
&&B = \max(W) = W(0,0,0,0) \cr && = \left[ \frac{F}{2\sqrt{2}\pi^2\sigma_x\sigma_{x'}\sigma_y} \sqrt{\frac{L}{\lambda}}\right]\frac{1}{\pi \sqrt{2}}\int_{-\infty}^{\infty} d \phi~ \mathrm{sinc}^2(\phi^2/4)]~,
\label{Wsizdomax3}
\end{eqnarray}
where $\mathrm{sinc}(\phi^2/4)$ is the dimensionless far field distribution. The factor in square brackets represents the brightness expected by the estimate in Eq. (\ref{Bundug}) with the parameter choice in (\ref{sigrrp}). The number

\begin{eqnarray}
\frac{1}{\pi \sqrt{2}}\int_{-\infty}^{\infty} d \phi~ \mathrm{sinc}^2(\phi^2/4)] = \frac{4}{3}\sqrt{\frac{2}{\pi}} \simeq  ~ 1.06
\label{phiphi}
\end{eqnarray}
represents, instead, the disagreement between the exact expression and the maximum photon flux density in the phase space.

Thus we have found that the exact result in Eq. (\ref{Wsizdomax3}) is naturally different from the estimate in Eq. (\ref{Bundug}), although the difference is not large. The difference between the value found in Eq. (\ref{phiphi}), which is close to unity and that in Eq. (\ref{numberboh}), which is quite different from unity, is striking. One can observe that this difference is in close relation with other comparisons between the estimate in Eq. (\ref{Bundug}) and exact results. In fact, Eq. (\ref{Bundug}) coincides with the exact result Eq. (\ref{Bthirdfin}) in the limit for $D_{x,y} \ll 1 \ll N_{x,y}$ and at the same time was proven to overestimate the exact result by four times in the limit for $N_{x,y} \ll 1 \ll D_{x,y}$.

\section{\label{sec:tre} Bending magnet brightness}

\begin{figure}
\begin{center}
\includegraphics*[width=90mm]{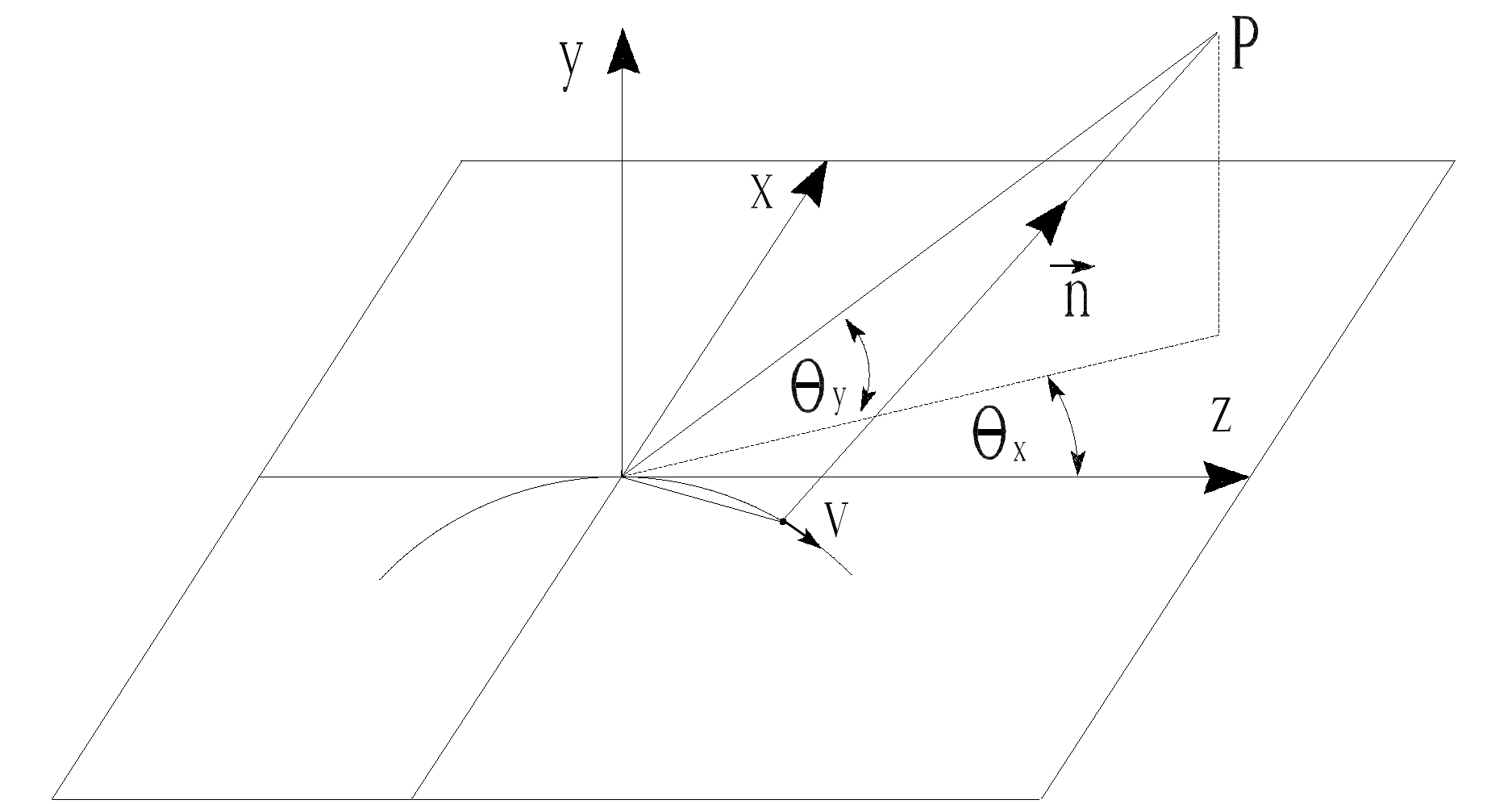}
\caption{\label{srgeo} Geometry for the analysis of bending magnet brightness.}
\end{center}
\end{figure}

\begin{figure}
\begin{center}
\includegraphics*[width=90mm]{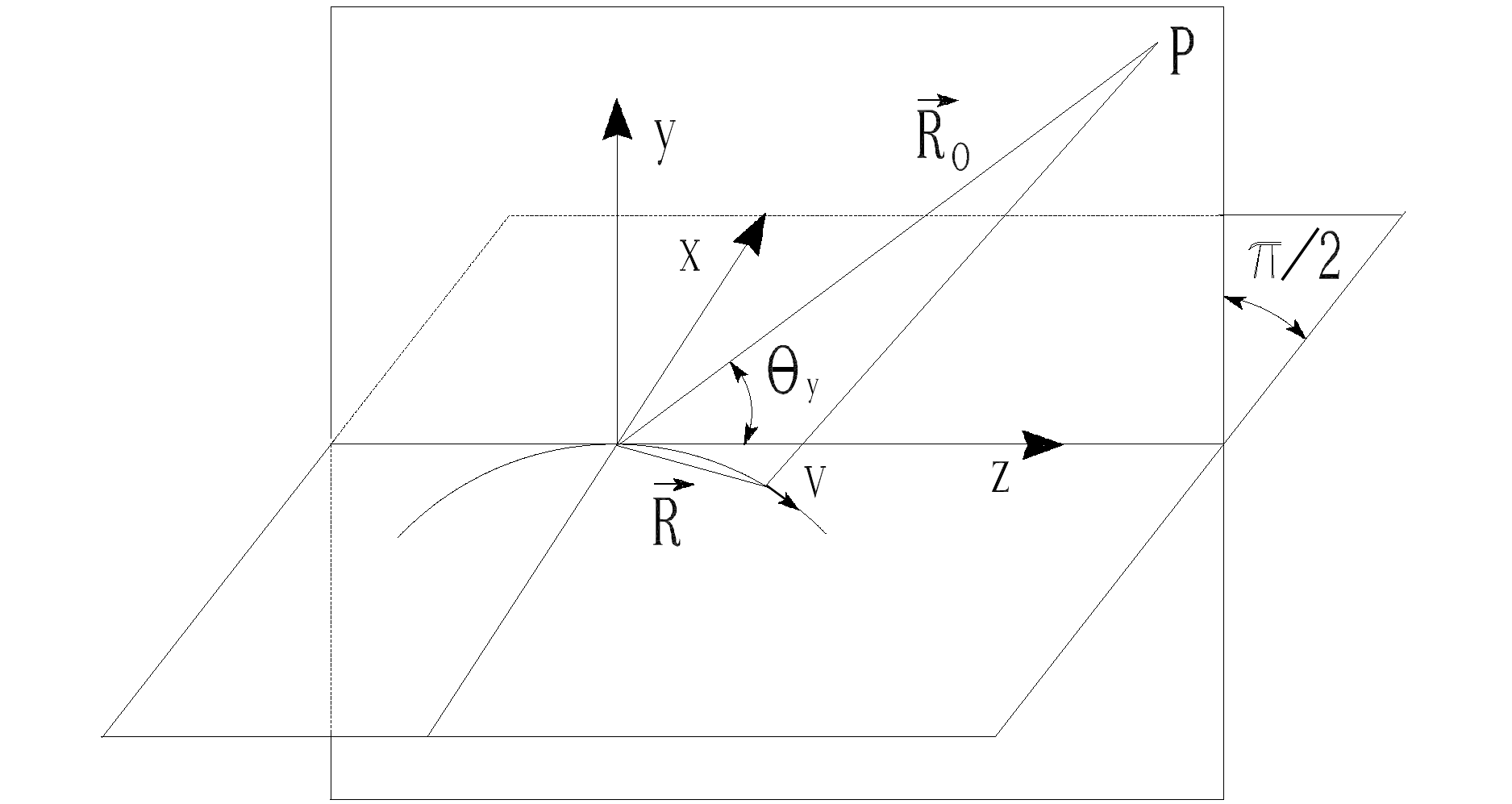}
\caption{\label{srstand} Standard geometry for the analysis of synchrotron radiation from a bending magnet.}
\end{center}
\end{figure}
Consider a single relativistic electron moving on a circular orbit and an observer as sketched in Fig. \ref{srgeo}. It is worth to underline the difference between the geometry depicted in Fig. \ref{srgeo} and the geometry used in most SR textbooks  for the treatment of bending magnet radiation depicted in Fig. \ref{srstand}. The observer in Fig. \ref{srstand} is assumed to be located in a vertical plane tangent to the circular trajectory at the origin, at an angle $\theta$ above the level of the orbit. In other words, in this geometry the $z$ axis is not fixed, but depends on the observer's position. Note that the geometry of the electron motion has a cylindrical symmetry, with the vertical axis going through the center of the circular orbit.  Because of this symmetry, in order to calculate spectral and angular photon distributions, it is not necessary to consider an observer at a more general location. However, since the wavefront is not spherical, this way of proceeding can hardly help to obtain the phase of the field distribution on a plane perpendicular to a fixed $z$ axis. This is required, for instance, if one needs to calculate the Wigner distribution, as in our case.

\subsection{Radiation field in space-frequency domain}

We can use Eq. (\ref{generalfin}) to calculate the far zone field of radiation from a relativistic electron moving along an arc of a circle. Assuming a geometry with a fixed $z$ axis as in Fig. \ref{srgeo}, we can write the transverse position of the electron as a function of the curvilinear abscissa $s$ as

\begin{equation}
\vec{r}(s) = -R\left(1-\cos(s/R)\right) \vec{e_x}
\label{trmot}
\end{equation}
and

\begin{equation}
z(s) = R \sin(s/R) \label{zmot}
\end{equation}
where $R$ is the bending radius.

Since the integral in Eq. (\ref{generalfin}) is performed along $z$ we should invert $z(s)$ in Eq. (\ref{zmot}) and find the explicit dependence $s(z)$:

\begin{equation}
s(z) = R \arcsin(z/R) \simeq z + {z^3\over{6R^2}} \label{sz}
\end{equation}
so that

\begin{equation}
\vec{r}(z) = - {z^2\over{2 R}} \vec{e_x}~,\label{rpdis}
\end{equation}
where the expansion in Eq. (\ref{sz}) and Eq. (\ref{rpdis}) is justified, once again, in the framework of the paraxial approximation.

%
%

With Eq. (\ref{generalfin}) we obtain the radiation field amplitude in the far zone:

\begin{eqnarray}
\vec{\widetilde{E}}= {i \omega e\over{c^2 z_0}}
\int_{-\infty}^{\infty} dz' {e^{i  \Phi_T}} \left({
z'+R\theta_x\over{R}}\vec{e_x}
+\theta_y\vec{e_y}\right)~\label{srtwo}
\end{eqnarray}
where

\begin{eqnarray}
&&\Phi_T = \omega \left[
\left({\theta_x^2+\theta_y^2\over{2c}}z_0\right)
+\left({1\over{2\gamma^2c}} +
{\theta_x^2+\theta_y^2\over{2c}}\right)z' \right.  \cr && \left. +
\left({\theta_x\over{2Rc}}\right)z'^2 +
\left(1\over{6R^2c}\right)z'^3\right]~.\label{phh2}
\end{eqnarray}

One can easily reorganize the terms in Eq. (\ref{phh2}) to obtain

\begin{eqnarray}
&& \Phi_T = \omega\left[
\left({\theta_x^2+\theta_y^2\over{2c}}z_0\right)-{R\theta_x\over{2c}}\left({1\over{\gamma^2}}
+{\theta_x^2\over{3}} +\theta_y^2\right) \right. \cr && \left.
+\left({{1\over{\gamma^2}}+\theta_y^2}\right){\left(z'+R\theta_x\right)\over{2c}}
+ {\left(z'+R\theta_x\right)^3\over{6 R^2 c
}}\right]~.\label{phh2b}
\end{eqnarray}
With redefinition of $z'$ as $z' + R \theta_x$ under integral we obtain the final result \cite{BOS2}, \cite{TAKA}, \cite{CHUB}:

\begin{eqnarray}
&& \vec{\widetilde{E}}= {i \omega e\over{c^2 z_0}} e^{i\Phi_s}
e^{i\Phi_0} \int_{-\infty}^{\infty} dz'
\left({z'\over{R}}\vec{e_x}+\theta_y\vec{e_y}\right) \cr &&
\times
\exp\left\{{i\omega\left[{z'\over{2\gamma^2c}}\left(1+\gamma^2\theta_y^2\right)
+{z'^3\over{6R^2c}}\right]}\right\}~,\label{srtwob}
\end{eqnarray}
where
\begin{equation}
\Phi_s ={\omega z_0\over{2c}}\left(\theta_x^2+\theta_y^2
\right)\label{phis}
\end{equation}
and

\begin{equation}
\Phi_0 = -{\omega R \theta_x\over{2c}}\left( {1\over{\gamma^2}}
+{\theta_x^2\over{3}} +\theta_y^2 \right)~.\label{phio}
\end{equation}
In standard treatments of bending magnet radiation, the phase term $\exp(i\Phi_0)$ is absent. In fact, the horizontal observation angle $\theta_x$ is always equal to zero in the reference system sketched in Fig. \ref{srstand}. The reason for this is that most textbooks focus on the calculation of the intensity radiated by a single electron in the far zone,  which involves the square modulus of the field amplitude but do not analyze, for instance, situations like source imaging.

Our next goal is to evaluate the integral in Eq. (\ref{srtwob}). It is convenient to introduce dimensionless geometrical quantities

\begin{eqnarray}
&&\vec{\hat{\theta}} = \frac{\vec{\theta}}{(\lambdabar/R)^{1/3}} \cr &&
\vec{\hat{r}} = \frac{\vec{r}}{(R \lambdabar^2)^{1/3}}~,
\label{normqr}
\end{eqnarray}
and the dimensionless parameter

\begin{eqnarray}
\xi =\left (\frac{\lambda_c}{\lambda}\right)^{2/3}~.
\label{xiii}
\end{eqnarray}
If we then go through the algebra we can simplify Eq. (\ref{srtwob}) to

\begin{eqnarray}
&& \vec{\widetilde{E}}(z_0,\vec{\hat{\theta}}) = - \frac{2 e \gamma \xi^{1/2}}{\sqrt{3} c
z_0} \exp[i \Phi_s]\exp\left[-\frac{i\hat{\theta}_x}{2}\left(\xi +
\frac{\hat{\theta}_x^2}{3}+\hat{\theta}_y^2\right)\right]\cr && \times\Bigg\{ \vec{e}_x
\left[(\xi + \hat{\theta}_y^2) K_{2/3} \left(\frac{1}{3}
(\xi+\hat{\theta}_y^2)^{3/2}\right)\right]\cr && - i \vec{e}_y \left[(\xi +
\hat{\theta}_y^2)^{1/2} \theta_y K_{1/3}\left(\frac{1}{3}
(\xi+\hat{\theta}_y^2)^{3/2}\right)\right] \Bigg\}~,\label{EF}
\end{eqnarray}
where $K_{1/3}$ and $K_{2/3}$ are the modified Bessel functions. Eq (\ref{EF}) is  equivalent to Eq. (\ref{srtwob}), but is expressed in  a more suitable form for calculating the field distribution  at the virtual source, which is assumed to be located in the $(x,y)$-plane, perpendicular to the circular trajectory at the origin (see Fig. \ref{srgeo}). After substitution of Eq. (\ref{EF}) in Eq. (\ref{farzone}) we can write result in terms of the Airy functions

\begin{eqnarray}
&& \widetilde{E}_{x}\left(\vec{\hat{r}}\right) = -\frac{i e \gamma^2 \xi}{c R}
\frac{2^{4/3}}{\sqrt{3}}\cr && \times \int_{-\infty}^{\infty} d
\hat{\theta}_y \exp[2 i \hat{y} \hat{\theta}_y] (\xi+\hat{\theta}_y^2) K_{2/3}
\left[\frac{1}{3} (\xi +\hat{\theta}_y^2)^{3/2}\right]
\mathrm{Ai}\left[\frac{1}{2^{2/3}} (\xi - 2 \hat{x} +
\hat{\theta}_y^2)\right] \cr && \label{E0X}
\end{eqnarray}
and

\begin{eqnarray}
&& \widetilde{E}_{y}\left(\vec{\hat{r}}\right) = -\frac{e \gamma^2 \xi}{c R}
\frac{2^{4/3}}{\sqrt{3}}\cr && \times \int_{-\infty}^{\infty} d
\hat{\theta}_y ~ \hat{\theta}_y \exp[2 i \hat{y} \hat{\theta}_y]  K_{1/3} \left[\frac{1}{3}
(\xi +\hat{\theta}_y^2)^{3/2}\right] \mathrm{Ai}\left[\frac{1}{2^{2/3}}
(\xi - 2 \hat{x} + \hat{\theta}_y^2)\right]~. \label{E0Y}
\end{eqnarray}

\begin{figure}[tb]
\includegraphics[width=0.5\textwidth]{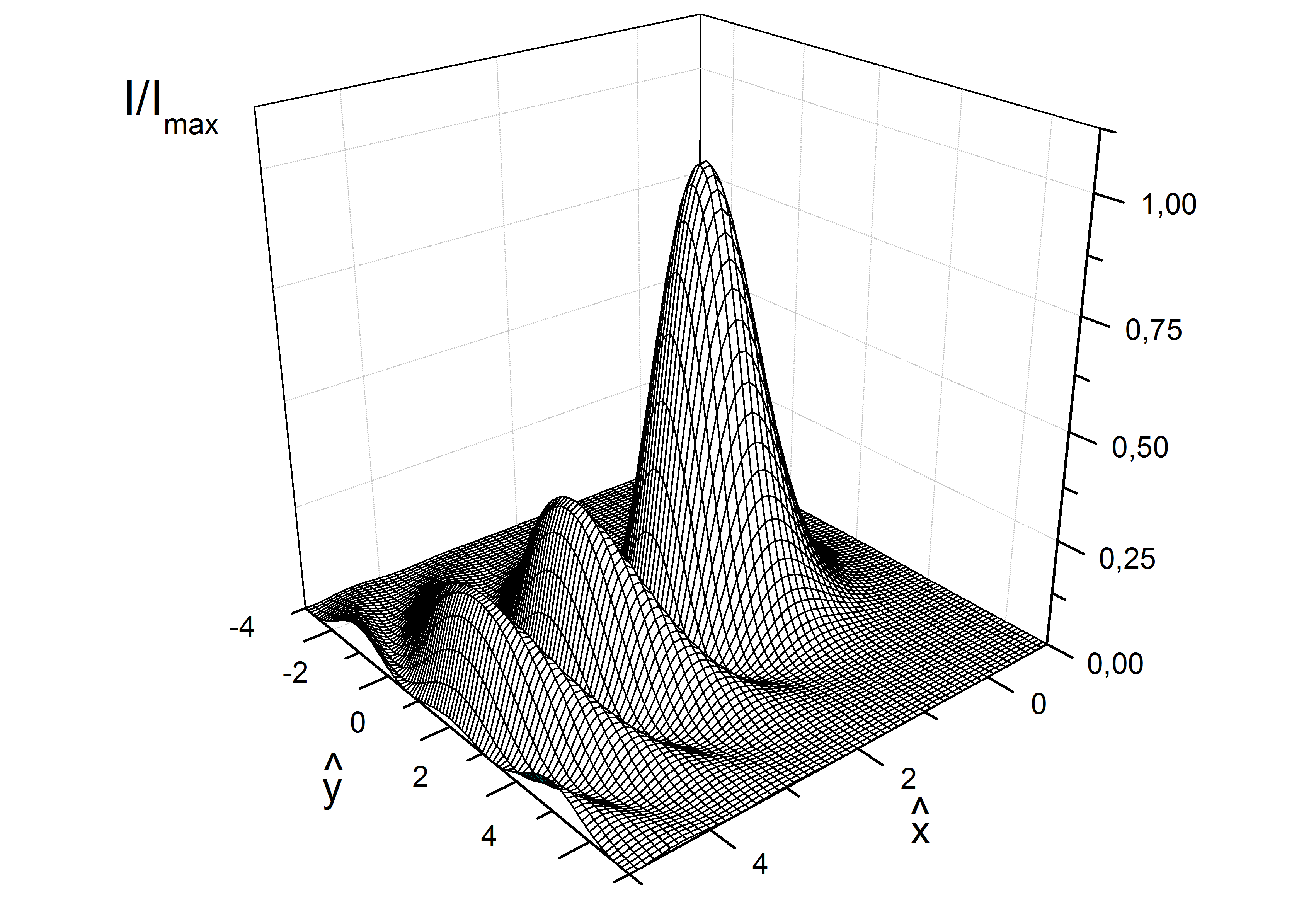}
\includegraphics[width=0.5\textwidth]{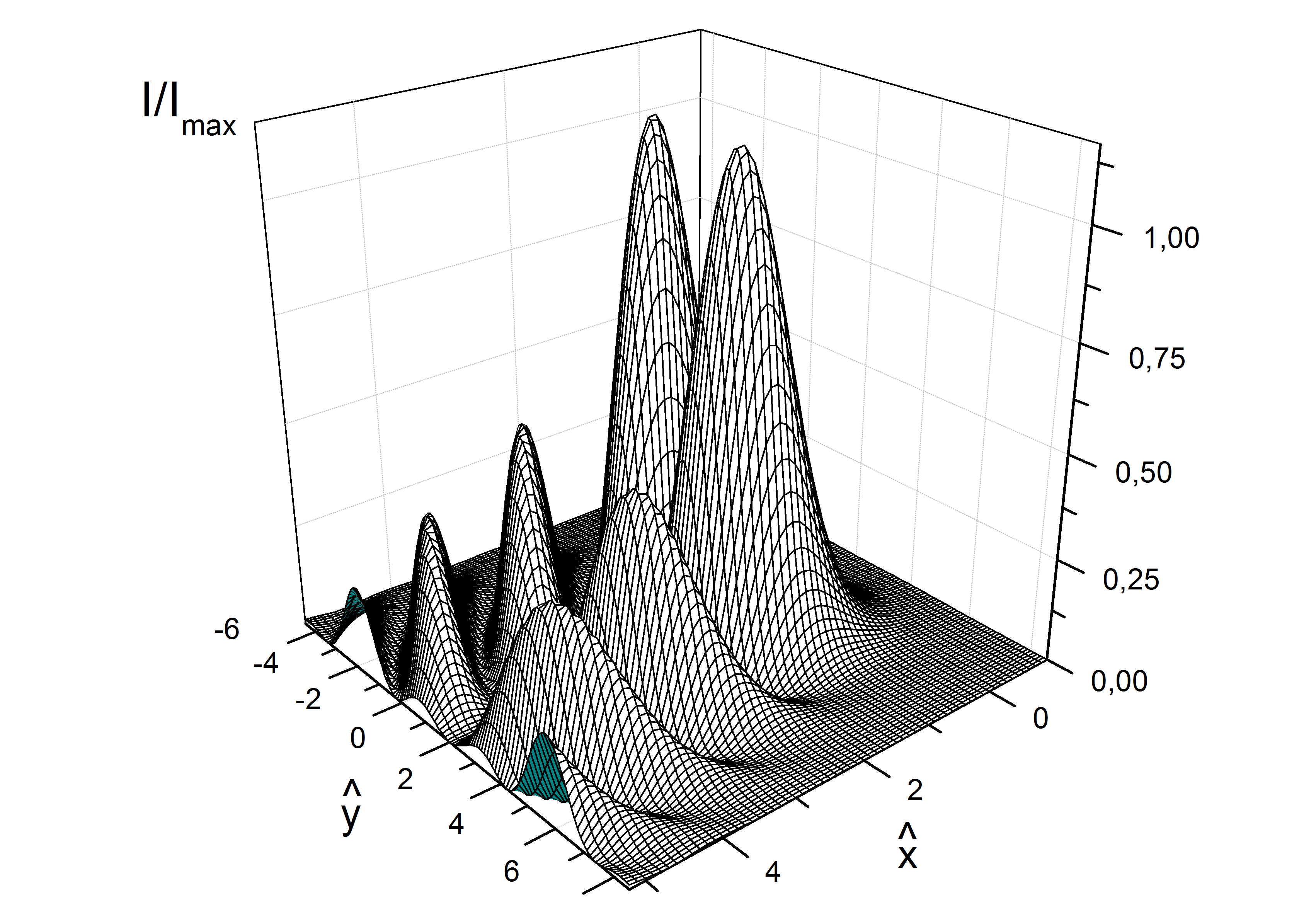}
\caption{Intensity profiles for the virtual source $I/I_\mathrm{max}$, as a function of  $x/(R^{1/3} \lambdabar^{2/3})$ and $y/(R^{1/3} \lambdabar^{2/3})$ for the
horizontal (left plot) and vertical (right plot) polarization components.} \label{VSXY}
\end{figure}

The integral in Eq. (\ref{E0X}) and Eq. (\ref{E0Y}) must be calculated numerically.

Computational results of the intensity distribution of horizontal and vertical polarization components in the 1:1 image plane by perfect thin lens \cite{SMOL} are presented in Fig.\ref{VSXY}. Due to phase differences of the bending magnet radiation in the far zone from a spherical wavefront, the wavefront is aberrated. For a single electron this aberration appears when the horizontal aperture is compatible with the natural opening angle $\sigma_{r'}$, and it becomes severe as the horizontal aperture increases further.

Up to this point we considered an electron moving along a circular trajectory that lies in the $(x,z)$-plane and following a path through origin  of the coordinate system sketched in Fig. \ref{srgeo}, tangent to the $z$ axis. The phase difference in the fields will be determined by the position of the observer position  and by the electron trajectory. Let us now discuss the bending magnet radiation from a single electron with arbitrary angular deflection and offset with respect to the nominal orbit. Such an expression was first calculated, starting from the Lienard-Wiechert fields, in \cite{TAKA}.

The meaning of horizontal and vertical deflection angles $\eta_x$ and $\eta_y$ is clear once we specify the electron velocity

\begin{eqnarray}
&&\vec{v}(s) = v\left[-\sin\left({s\over{R}}+\eta_x\right)\cos(\eta_y)
\vec{e_x} + \sin(\eta_y) \vec{e_y} +
\cos\left({s\over{R}}+\eta_x\right)\cos(\eta_y) \vec{e_z}
\right]~,\cr && \label{veloangle}
\end{eqnarray}
so that the trajectory can be expressed as a function of the curvilinear abscissa $s$ as

\begin{eqnarray}
&&x(s) \vec{e}_x + y(s) \vec{e}_y + z(s) \vec{e}_z = \cr &&  \left[l_x +
R\cos\left({s\over{R}}+\eta_x\right)\cos(\eta_y) - R\cos(\eta_x)\cos(\eta_y) \right]
\vec{e_x} \cr && + \left[l_y+s \sin(\eta_y)\right]  \vec{e_y}
\cr &&+ \left[ R\sin\left({s\over{R}}+\eta_x\right)\cos(\eta_y) -
R\sin(\eta_x)\sin(\eta_y) \right] \vec{e_z} \cr &&
\label{trajangle}
\end{eqnarray}
Here we have introduced, also, an arbitrary offset $(l_x,l_y,0)$
in the trajectory. Using Eq. (\ref{trajangle}) an approximated
expression for $s(z)$ can be found:

\begin{equation}
s(z) = z+ {z^3\over{6 R^2}}+{z^2 \eta_x\over{2 R}} +{z
\eta_x^2\over{2}}+{z \eta_y^2\over{2}} \label{szangle}
\end{equation}
so that

\begin{equation}
\vec{v}(z) =   \left(- {v z\over{R}}+ v \eta_x  \right)
\vec{e_x} + \left(v \eta_y  \right)
\vec{e_y}~\label{vapprangle}
\end{equation}
and

\begin{equation}
\vec{r}(z) = \left(- {z^2\over{2R}}+ \eta_x z + l_x\right)
\vec{e_x} + \left(\eta_y z +l_y \right)
\vec{e_y}~.\label{trmot2}
\end{equation}
%
%
%

It is evident that the offsets $l_x$ and $l_y$ are always subtracted from $x_0$ and $y_0$  respectively: a shift in the particle trajectory on the vertical plane is equivalent to a shift of the observer in the opposite direction. With this in mind we introduce angles $\bar{\theta}_x = \theta_x -l_x/z_0$ and $\bar{\theta}_y = \theta_y - l_y/z_0$ to obtain

\begin{eqnarray}
\vec{\widetilde{E}}= {i \omega e\over{c^2 z_0}}
\int_{-\infty}^{\infty} dz' {e^{i \Phi_T}} \left({
z'+R(\bar{\theta}_x-\eta_x)\over{R}}\vec{e_x}
+{(\bar{\theta}_y-\eta_y)}\vec{e_y}\right)~\label{srtwoang}
\end{eqnarray}
and

\begin{eqnarray}
&& \Phi_T =
\omega \left({\bar{\theta}_x^2+\bar{\theta}_y^2\over{2c}}z_0 \right)
+{\omega\over{2c}}\left({1\over{\gamma^2}} +
\left(\bar{\theta}_x-\eta_x\right)^2  +
\left(\bar{\theta}_y-\eta_y\right)^2\right)z' \cr && +
\left({\omega\bar{\theta}_x\over{2Rc}}\right)z'^2 +
\left(\omega\over{6R^2c}\right)z'^3~.\label{phh2ang}
\end{eqnarray}
One can easily reorganize the terms in Eq. (\ref{phh2ang}) to
obtain

\begin{eqnarray}
&&\Phi_T =
\omega\left({\bar{\theta}_x^2+\bar{\theta}_y^2\over{2c}}z_0\right)-
{\omega R(\bar{\theta}_x-\eta_x)\over{2c}} \cr &&\times
\left({1\over{\gamma^2}} +(\bar{\theta}_y-\eta_y)^2
+{(\bar{\theta}_x-\eta_x)^2\over{3}}\right) \cr &&
+\left({{1\over{\gamma^2}}+(\bar{\theta}_y-\eta_y)^2}\right)
{\omega\left(z'+R(\bar{\theta}_x-\eta_x)\right)\over{2c}} \cr  &&+
{\omega \left(z'+R (\bar{\theta}_x-\eta_x)\right)^3\over{6 R^2 c
}}~.\label{phh2angfin}
\end{eqnarray}
Redefinition of $z'$ as $z'+R(\bar{\theta}_x-\eta_x)$ gives the
result

\begin{eqnarray}
&&\vec{\widetilde{E}}= {i \omega e\over{c^2 z_0}} e^{i \Phi_s} e^{i
\Phi_0} \int_{-\infty}^{\infty} dz'
\left({z'\over{R}}\vec{e_x}+(\bar{\theta}_y-\eta_y)\vec{e_y}\right)
\cr && \times
\exp\left\{{i\omega\left[{z'\over{2\gamma^2c}}\left(1+\gamma^2
(\bar{\theta}_y-\eta_y)^2\right)
+{z'^3\over{6R^2c}}\right]}\right\}~,\label{srtwoang2}
\end{eqnarray}
where
\begin{equation}
\Phi_s = {\omega z_0
\over{2c}}\left(\bar{\theta}_x^2+\bar{\theta}_y^2
\right)\label{phisang}
\end{equation}
and

\begin{equation}
\Phi_0 = - {\omega R(\bar{\theta}_x-\eta_x)\over{2c}}
\left({1\over{\gamma^2}} +(\bar{\theta}_y-\eta_y)^2
+{(\bar{\theta}_x-\eta_x)^2\over{3}}\right)~.\label{phioang}
\end{equation}
In the far zone we can neglect terms in $l_x/z_0$ and $l_y/z_0$, which leads to

\begin{eqnarray}
&&\vec{\widetilde{E}}= {i \omega e\over{c^2 z_0}} e^{i \Phi_s} e^{i
\Phi_0} \int_{-\infty}^{\infty} dz'
\left({z'\over{R}}\vec{e_x}+\left(\theta_y-\eta_y
\right)\vec{e_y}\right) \cr && \times
\exp\left\{{i\omega\left[{z'\over{2\gamma^2c}}\left(1+\gamma^2
\left(\theta_y-\eta_y\right)^2\right)
+{z'^3\over{6R^2c}}\right]}\right\}~,\label{srtwoang2bis}
\end{eqnarray}
where
\begin{equation}
\Phi_s = {\omega z_0 \over{2c}}\left(\theta_x^2+\theta_y^2
\right)\label{phisangbis}
\end{equation}
and

\begin{eqnarray}
\Phi_o \simeq - {\omega R({\theta_x}-\eta_x)\over{2c}}
\left({1\over{\gamma^2}} +(\theta_y-\eta_y)^2 
+{(\theta_x-\eta_x)^2\over{3}}\right)-{\omega\over{c}}(l_x
\theta_x+l_y\theta_y) ~.\label{phioangbis}
\end{eqnarray}
Note that Eq. (\ref{srtwoang2bis}) and Eq. (\ref{srtwob}) satisfy the equality in Eq. (\ref{tiltshiftfar}). Therefore, also in the bending magnet case the statistical average can be simplified to the convolution integral Eq. (\ref{Wcorr}).

\subsection{Wigner distribution and bending magnet sources}

We first calculate the brightness for a filament beam and, as a second step, we account for a finite phase space distribution for the electron beam. The Wigner distribution in the case of an electron beam with zero emittance is given by Eq. (\ref{W0start}), where $\widetilde{E}(\vec{r})$ is related to the electric field radiated by a single electron in a bending magnet, Eq. (\ref{E0X}), where we now restrict our attention to the horizontal polarization component. According to Eq. (\ref{W0start}) and Eq. (\ref{E0X}), the Wigner distribution in the diffraction-limited case is a function of four geometrical variables $\hat{x}$, $\hat{y}$, $\hat{\theta}_x$, $\hat{\theta}_y$ and one parameter $\xi = (\lambda_c/\lambda)^{2/3}$. We can compute the bending magnet brightness numerically, as a function of these variables. The peak value of the Wigner distribution at $\xi =1$ as a function of $\vec{\hat{r}}$ and $\vec{\hat{\theta}}$ is reached for $\hat{x} = 1.15$, $\hat{y} = 0$, $\hat{\theta}_x = 0$ and $\hat{\theta}_y = 0$. Therefore, the expression for the brightness is given by

\begin{eqnarray}
B = \max(W_0) = W_0(1.15, 0, 0, 0) = 0.59  \left[\frac{4}{\lambda^2} \frac{I}{e} \alpha \right]
\label{BBM1}
\end{eqnarray}

and, according to the usual normalization, it yields a number of photons per relative bandwidth per unit time per unit area per unit solid angle. Apart from a numerical factor, it equals the theoretical maximum concentration of the photon flux on the sample on the basis of qualitative arguments. The number of  photons with $\lambda \sim \lambda_c$ emitted from a formation length $\sim R/\gamma$ per unit time can be estimated as $(I/e) \alpha$. This radiation concentrates around the axis within a solid angle of about $\sigma_{r'}^2$ and, as a result, it has an $M^2$ factor close to unity. In literature it is often noted that the coherent flux from radiation with an ideal wavefront can be ultimately focused down to a spot-size of dimension $\lambda^2/4$. It is clear that the numerical factor $0.59$ is related with the particular choice of wavelength, and will be different for different choices of $\xi$.

The analysis of the effects of a finite electron beam emittance can be made using the same approach as in section \ref{sec:due}. The calculations are easier if we make use of dimensionless variables and parameters for the electron beam distribution. Normalized units in the bending magnet case are given by Eq. (\ref{normqr}) and its analogues

\begin{eqnarray}
&&\vec{\hat{\eta}} = \frac{\vec{\eta}}{(\lambdabar/R)^{1/3}} \cr &&
\vec{\hat{l}} = \frac{\vec{l}}{(R \lambdabar^2)^{1/3}} ~.
\label{normq1}
\end{eqnarray}
As for the undulator case, in order to make our analysis treatable we assume a Gaussian distribution of the electron beam in the transverse phase space, which can be factorized at the position of the virtual source (at $z = 0$). In this case, the electron beam distribution at the source position can be expressed as in Eq. (\ref{fphase}) and Eq. (\ref{distr}). In the bending magnet case,  the dimensionless parameters $N_{x,y}$ and $D_{x,y}$  are given by

\begin{eqnarray}
&& N_{x,y} = \frac{\sigma_{x,y}^2}{(R \lambdabar^2)^{2/3}}  ~,\cr &&
D_{x,y} = \frac{\sigma_{x',y'}^2}{(\lambdabar/R)^{2/3}}~.
\label{ND2}
\end{eqnarray}
We are now in the position to calculate the cross-spectral density at the virtual source by taking advantage of our dimensionless analysis. The expression for the cross-spectral density $G$ is in Eq. (\ref{Gnor3}), where now dimensionless variables and parameters are given by Eq. (\ref{normqr}), Eq. (\ref{normq1}) and Eq. (\ref{ND2}). The expression for the two polarization components of the electric field, Eq. (\ref{E0X}) and Eq. (\ref{E0Y}), allow for an explicit calculation of $G(\vec{\hat{r}},\Delta\vec{\hat{ r}})$. The final step consists in the calculation of the Wigner distribution $W$. The relation between $W$ and $G$ is expressed by the transformation in Eq. (\ref{Wig13}), so that

\begin{eqnarray}
W = \frac{c}{(2\pi)^4} \frac{I}{e \hbar} \left(\frac{R}{\lambdabar}\right)^{2/3} \int d \Delta \vec{\hat{r}} \exp(-i \vec{\hat {\theta}} \cdot \Delta \vec{\hat{r}})
G(\vec{\hat{r}},\Delta \vec{\hat{r}})~.
\label{WGdless}
\end{eqnarray}
The integral in Eq. (\ref{WGdless}) is  a function of four dimensionless variables $\hat{x},~\hat{y},~\hat{\theta}_x,~\hat{\theta}_y$, and  five dimensionless parameters $N_{x,y}$, $D_{x,y}$, and $\xi$. In order to calculate the brightness we need to find the maximum of the Wigner distribution. Suppose that the maximum of $W$ is reached at $\vec{\hat{r}} = \vec{\hat{r}}_M$ and $\vec{\hat{\theta}} = \vec{\hat{\theta}}_M$. Then one has

\begin{eqnarray}
B = \frac{c}{(2\pi)^4} \frac{I}{e \hbar} \left(\frac{R}{\lambdabar}\right)^{2/3} f\left(\vec{\hat{\theta}}_M, \vec{\hat{r}}_M, N_x, N_y, D_x, D_y , \xi\right),
\label{Balgo}
\end{eqnarray}
where $f$ is the integral in Eq. (\ref{WGdless}). Here we have, at last, a well-defined procedure for computing the brightness from a bending magnet source. Thus, in principle, we have solved the problem of determining the brightness of a given SR setup.  In particular, in the geometrical optics limits, when parameters $N_{x,y}$ or $D_{x,y}$  are large, calculations become simple and it is possible to calculate the brightness analytically. When the parameters $N_{x,y}$ and $D_{x,y}$ are order of unity the situation becomes more complicated, and must be solved numerically. We will not further pursue the matter of determining the brightness with the help of numerical techniques. For the discussion of the next section \ref{GeoSRlim} attention will be restricted to the geometrical optics limit only.

\subsection{\label{GeoSRlim} Geometrical optics limit}

It is our purpose here to demonstrate how a straightforward application of Eq. (\ref{Balgo}) yields analytical expressions for the brightness in several geometrical optics limits. We begin our analysis of geometrical optics asymptotes with the beam divergence-dominated regime, for $N_{x,y} \ll 1 \ll D_{x,y}$, and with the beam size-dominated regime for $D_{x,y} \ll 1 \ll N_{x,y}$.

Before proceeding, however, we should first make a few remarks to discuss the choice of these examples. In fact, in section \ref{43} we considered the beam size- and divergence-dominated regime as the simplest geometrical optics asymptote for the undulator case.  In contrast with this, in the bending magnet case the beam size- and divergence-dominated regime, corresponding to $N_{x,y} \gg 1$ and $D_{x,y} \gg 1$, is the most complicated situation for analytical treatment. On intuitive grounds we certainly expect that in this asymptotic limit there is a competition between  effects related with partial coherence (since $N_{x,y} \gg 1$) and what we called `aberration' effects (since $D_{x,y} \gg 1$), and careful mathematical analysis confirms such expectation. Given its complexity, we will discuss this asymptote only as a final example.

From a mathematical viewpoint there are two differences between the bending magnet case and the undulator case. The first difference stems from the fact that the range of the variables $\phi_{x,y}$ in Eq. (\ref{Gnor3})   is effectively limited up to values $|\phi_{x,y}| \sim 1$ in the undulator case. In fact, $\phi_{x,y}$ enters the expression for electric field Eq. (\ref{undurad5}), and at values larger than unity the integrand in Eq. (\ref{Gnor3})  is suppressed. Therefore, when $N_{x,y} \gg 1$ we can neglect $\phi_{x,y}$ in the exponential functions under the integral. At variance, in the bending magnet case the expression for the field is given by Eq. (\ref{E0X}) and Eq. (\ref{E0Y}), and the range of the variable $\phi_x$ in Eq. (\ref{Gnor3}) is not limited. Therefore, Eq. (\ref{Gnor3})  cannot be simplified by neglecting $\phi_x$ in the exponential function in $N_x$ under the integral sign only based on the assumption $N_x \gg 1$. The second difference can be found in the fact that, in the undulator case for $D_{x,y} \gg 1$, from the exponents in $D_{x,y}$ in Eq. (\ref{Gnor3}) follows that $\Delta x \ll 1$ and $\Delta y \ll 1$, and can be neglected in the expression for the field. In contrast with this, in the bending magnet case, from the expression for the electric field on the source position (Eq. (\ref{E0X}) and Eq. (\ref{E0Y})) it is clear that there are strong oscillations with a decreasing `local period' along the $x$ direction. Thus, it can be seen that $\Delta x$ cannot be ignored at $D_x \gg 1$, and the integral in Eq. (\ref{Gnor3})  cannot be simplified to an integral of the flux density over the source surface, as was done for the undulator case, Eq. (\ref{WNDlargedim}).

Let us now consider the beam divergence-dominated regime for $N_{x,y} \ll 1 \ll D_{x,y}$. From the analysis of the exponential functions in $D_{x,y}$ in Eq. (\ref{Gnor3}) follows that these functions have a significant magnitude only when $|\Delta x|  \lesssim 1/\sqrt{D_x}$ and  $|\Delta y| \lesssim 1/\sqrt{D_y}$. Then, since $N_{x,y} \ll 1$, it follows that the distributions $\exp[- (\phi_x - \hat{x})^2/(2N_x)]$ and $\exp[-(\phi_y -\hat{y})^2/(2N_y)]$ are sharply peaked about $\phi_x =  \hat{x}$ and $\phi_y = \hat{y}$, respectively. The integrals along the $x$ axis and the $y$ axis in Eq. (\ref{Gnor3}),  including the two polarization components for the field $\widetilde{E}_{x,y}(\hat{x}, \hat{y})$ given by Eq. (\ref{E0X}) and Eq. (\ref{E0Y}), can therefore be simplified by the same line of thought as for the undulator case. However, the important difference with the undulator case is that now the inequality $N_x \ll 1 \ll D_x$ cannot be called upon for ignoring $\Delta \hat{x}$, thus approximating the exponential function in $N_x$, multiplied by $1/\sqrt{2\pi N_x}$, with a Dirac $\delta$-function under the integral sign at any point $\hat{x}$ and $\hat{y}$ of the source surface. In fact, based on the expressions for the electric field amplitude at the source position, Eq. (\ref{E0X}) and Eq. (\ref{E0Y}), we might argue that the exponential function in $N_x$, multiplied by $1/\sqrt{2\pi N_x}$, can be approximated by a Dirac $\delta$-function and $\Delta \hat{x}$ can be neglected only if $\sqrt{N_x}$ and $1/\sqrt{D_x}$ are small compared to the `local period' of oscillation of the field along the $x$ axis. This requirement (at fixed $N_x$, $D_x$) would set an upper limit on the $x$-coordinate when the integrand can be simplified as described above.

The situation can be summarized by saying that the asymptotic expression for the Wigner  distribution in the beam divergence-dominated regime in Eq. (\ref{Wdivdom}) can be used in the bending magnet case at least in the region where $\hat{x} \lesssim 1$ and $\hat{y} \lesssim 1$. This is in contrast with the undulator case, when Eq. (\ref{Wdivdom}) can be used at any point on the source plane. According to Eq. (\ref{E0X}), for the horizontal polarization component  in the diffraction-limited case, the peak value of the square modulus of the electric field amplitude on the source at $\xi = 1$ is reached at $\hat{y} = 0$ and $\hat{x} = 1.54$ (see Fig. \ref{VSXY}). This point is well within the region of applicability of Eq. (\ref{Wdivdom}) for $N_{x,y} \ll 1 \ll D_{x,y}$. Thus, the expression for the brightness of the horizontal polarization component is given by\footnote{The similarity with the corresponding Eq. (\ref{Bdivdom}) for the undulator case is striking here.}:

\begin{eqnarray}
B = \max(W) = \frac{1}{2\pi \sigma_{x'} \sigma_{y'}} \max\left(\frac{dF}{dS}\right)~,
\label{BXbend}
\end{eqnarray}
where

\begin{eqnarray}
\max \left(\frac{dF}{dS}\right) = \frac{I}{e \hbar} \frac{c}{4\pi^2}|\widetilde{E}(1.54, 0)|^2
\label{maxdfds}
\end{eqnarray}
is the maximum photon flux density for the horizontal polarization component of the field of the bending magnet source in the diffraction-limited case.  For $\lambda = \lambda_c$ Eq. (\ref{BXbend}) can also be written as

\begin{eqnarray}
B = 0.76 \cdot \left[\frac{4}{\lambda^2}\frac{I}{e} \alpha \frac{\sigma_{r'}}{\sigma_{x'}}\frac{\sigma_{r'}}{\sigma_{y'}} \right] ~,
\label{Bmaxfluxd}
\end{eqnarray}
where $\sigma_{r'}$ is the so-called  `vertical opening angle' \cite{CLAR} introduced already in section \ref{oldres}. At $\lambda = \lambda_c = 2\pi R/\gamma^3$  one finds, for the horizontal polarization component, that $\sigma_{r'} = 0.67/\gamma$.  The numerical factor  is related with the particular choice of wavelength, and will be different for different choices of $\xi$, while the parametric dependence in square brackets remains unchanged.

A second, simple limiting case can now be considered. In the beam size-dominated regime, when $D_{x,y} \ll 1 \ll N_{x,y}$, Eq. (\ref{Gnor3}) can be simplified following the same line of reasoning as in the case of an undulator. The only difference is in the region of applicability of this simplification.  We start asserting the simplifications that can be made and examining their implications. Variables $\Delta \hat{x}$ and $\Delta \hat{y}$ in Eq. (\ref{Gnor3}) are effectively limited up to values $|\Delta \hat{x}| \sim 1$ and $|\Delta \hat{y}| \sim 1$, if we limit the observation angles up to $|\hat{\theta}_x| \sim 1$, $|\hat{\theta}_y| \sim 1$. Therefore, the exponential functions in $D_{x,y}$ in Eq. (\ref{Gnor3})  can be replaced with unity. Moreover $N_{x,y} \gg 1$, so that we can neglect $\phi_{x,y}$ in the exponential function in $N_{x,y}$. The Wigner distribution can therefore be written as Eq. (\ref{Wthird}). The integral in this equation can be simplified by rewriting the electric field at the source, Eq. (\ref{E0X}) and Eq. (\ref{E0Y}), in the terms of the far field given by Eq. (\ref{EF}). In fact, inserting Eq. (\ref{tiltshiftfar}) into Eq. (\ref{Wthird}), performing integration and rearranging yields Eq. (\ref{Wthirdfin}), where now the electric field amplitude in the far zone is given by Eq. (\ref{EF}).

A formal way to understand this simplification procedure is the following. The Wigner distribution $W$ is proportional to the two-dimensional Fourier transform of the cross-spectral density $G$, and the coordinates $\hat{\theta}_{x,y}$ can be interpreted as reduced spatial frequencies. Suppose now that we have $|\hat{\theta}_x| \sim 1$. The significant values of $\Delta \hat{x}$ in the integral correspond, then, to an argument of the electric field amplitude $\widetilde{E}$ of order of unity. This is because the reduced `local period' of the electric field amplitude oscillations  is of order of unity near the origin and decreases along the $x$ axis. In the case when spatial frequencies are of order unity, the main contribution to the integral comes from the region near $\Delta \hat{x} = 0$ and $\phi_x = 0$, and its width is of order $|\Delta \hat{x}| \sim 1$, $|\phi_x| \sim 1$. Since the function $\exp[-(\phi_x - \hat{x})^2/(2N_x)]$ is smoothly varying in this region we can replace it by $\exp[-\hat{x}^2/(2N_x)]$, and take it out the integral sign. We can also replace the exponential function in $D_{x,y}$ with unity, and the integral is then approximately given by  Eq. (\ref{Wthird}).

Up to now, for our calculations we used the Wigner distribution method at the virtual source position. From the discussion concerning the beam size-dominated regime for bending magnet sources, it is obvious that such choice is not convenient in this asymptotic. In contrast to the beam divergence-dominated regime, the approximation procedure described above is not straightforward, and involves a number of subtleties.  Calculations of $W$ can be performed in two different ways, both consistently leading to the same result, and a comparison between the different calculation procedures is instructive. The approximation procedure in the beam size-dominated regime is easily understood if, instead of the virtual source, one uses the far zone for calculating $W$. This approach also offers an opportunity to easily find the condition for applicability of the approximation made.

As explained in section \ref{sec:wiggen}, the standard way of deriving the expression for the Wigner distribution makes use of the expression for the cross-spectral density $G$. In this way, our problem is reduced to the calculation of the cross-spectral density in the far zone. It can be interesting to explore the properties of the cross-spectral density and see that an explicit expression in the far zone can be derived from Eq. (\ref{Gnor3}), describing $G$ at the source position, by symmetry under the duality transformations:

\begin{eqnarray}
&& (\hat{\theta}_x, \hat{\theta}_y) \longrightarrow (\hat{x}, \hat{y})~ ,
\cr
&& (\Delta \hat{x}, \Delta \hat{y}) \longrightarrow (\Delta \hat{\theta}_x, \Delta \hat{\theta}_y) ~,
\cr
&& N_{x,y} \longrightarrow D_{x,y}
\cr
&& D_{x,y} \longrightarrow N_{x,y}
\label{transform}
\end{eqnarray}
The cross-spectral density $G$ in the far zone is therefore given by

\begin{eqnarray}
\hat{G}\left(\vec{\hat{\theta}},\Delta \vec{\hat{\theta}}\right) &=&
\frac{1}{2 \pi \sqrt{D_x D_y}} \exp \left[-\frac{(\Delta \hat{\theta}_x)^2
N_x}{2}\right] \exp \left[-\frac{(\Delta \hat{\theta}_y)^2 N_y}{2}\right]\cr &&
\times \int_{-\infty}^{\infty} d \phi_x \int_{-\infty}^{\infty} d
\phi_y \exp\left[-\frac{\left(\phi_x-\hat{\theta}_x\right)^2}{2
D_x}\right] \exp\left[-\frac{\left(\phi_y-\hat{\theta}_y\right)^2}{2
D_y}\right] \cr && \times
\widetilde{E}\left({\phi_{x}}+\frac{\Delta
\hat{\theta}_x}{2},{\phi_{y}}+\frac{\Delta \hat{\theta}_y}{2}
\right)  \widetilde{E}^*\left({\phi_{x}}-\frac{\Delta \hat{\theta}_x}{2},{\phi_{y}}-\frac{\Delta \hat{\theta}_y}{2}
\right)~,\cr &&\label{Gnor3far}
\end{eqnarray}
where the amplitude of the electric field $\widetilde{E}(\hat{\theta}_x,\hat{\theta}_y)$ in the far zone is given by Eq. (\ref{EF}). Note that the mathematical rule that we just gave to calculate the cross-spectral density in the far zone differs from the usual approach, where the angular and the position representations of the field, $\widetilde{E}(\vec{r})$ and $\widetilde{E}(\vec{\theta})$ are related via a spatial Fourier transform and the fields are given in the form of angular spectrum of plane waves. In that way, the statistical properties of an ensemble of fields at the source position is fully reflected in the ensemble of the angular spectrum amplitudes, and the cross-spectral density is a measure of the correlation between the fields of the plane-wave modes propagating at $\vec{\theta}-\Delta \vec{\theta}$ and  $\vec{\theta} + \Delta \vec{\theta}$. However,  the correlation between plane waves is a purely mathematical quantity that depends, for example, on the normalization convention used.  In contrast with this, in our approach we do not make any use of spatial Fourier transforms, because the angular and the position representations of the field are related via the Fresnel propagation Eq. (\ref{farzone}) and Eq. (\ref{farzonef}). The field propagation in free space is a physical process, and does not depend on the definition of the spatial Fourier transform.

We now return to our quantitative discussion concerning the beam size-dominated example. If $D_{x,y} \ll 1$,  we can replace the exponential functions in $D_{x,y}$ multiplied by $1/\sqrt{2\pi D_{x,y}}$  with Dirac $\delta$-functions, so that the expression for $G$ in the far zone, Eq. (\ref{Gnor3far}), can be approximated with

\begin{eqnarray}
&&G = \exp \left[-\frac{(\Delta \hat{\theta}_x)^2
N_x}{2}\right] \exp \left[-\frac{(\Delta \hat{\theta}_y)^2 N_y}{2}\right] \cr && \times \widetilde{E} \left(\hat{\theta}_x + \frac{\Delta \hat{\theta}_x}{2}, \hat{\theta}_y + \frac{\Delta \hat{\theta}_y}{2}\right) \widetilde{E}^*\left(\hat{\theta}_x - \frac{\Delta \hat{\theta}_x}{2}, \hat{\theta}_y - \frac{\Delta \hat{\theta}_y}{2}\right) ~.
\label{Gfarappr}
\end{eqnarray}
In order to proceed it is necessary to determine the behavior  of the $\widetilde{E}(\hat{\theta}_x, \hat{\theta}_y)$ along the horizontal axis. Notice that in the far zone, according to Eq. (\ref{EF}), the module of the electric field does not depend on $\hat{\theta}_x$, while its phase varies as $\hat{\theta}_x^3$ at $\hat{\theta}_x \gg 1$. Consequently, Eq. (\ref{Gfarappr}) depends on $\Delta \hat{\theta}_x$ only through the phase, and one can further approximate the phase as $\sim \hat{\theta}_x^2 \Delta \hat{\theta}_x$. If $N_x \gg 1$, because of the exponential function $\exp(-N_x \Delta \hat{\theta}_x^2/2)$ in Eq. (\ref{Gfarappr}), the region of $\Delta \hat{\theta}_x$ for which $G$ is large is  near the point $\Delta \hat{\theta}_x = 0$, and its width is of order $1/\sqrt{N_x}$. Suppose we have $\hat{\theta}_x^2 \ll \sqrt{N_x}$. We can then approximate $\hat{\theta_x}^2 \Delta \hat{\theta}_x \ll 1$, and consequently the phase factor in Eq. (\ref{EF}) is about equal to unity.  We thus obtain


\begin{eqnarray}
&& G = \exp\left(- \frac{\omega^2 \sigma_x^2  \Delta {\theta}_x^2}{2 c^2} \right)\exp\left(- \frac{\omega^2 \sigma_y^2  \Delta {\theta}_y^2}{2 c^2} \right) \exp\left[i \frac{\omega}{c} \left(\theta_x \Delta \theta_x z_0 + \theta_y \Delta \theta_y z_0 \right)\right] \cr && \times
 |\widetilde{E}({\theta}_x, {\theta}_y)|^2~.
\label{Gfarappr2}
\end{eqnarray}
In the geometrical optics limit, the cross-spectral density at the source position can be written in the form of Eq. (\ref{qh}), meaning that variables $\vec{r}$ and $\Delta \vec{r}$ are separable. However, according to Eq. (\ref{Gfarappr2}), in the far zone there is no separation of variables $\vec{\theta}$ and $\Delta \vec{\theta}$ in $G$, even in the geometrical optics limit. This is contrast with the traditional definition of cross-spectral density in the far zone as a correlation function between plane-wave modes. The non-separability in Eq. (\ref{Gfarappr2}) is a consequence of the fact that in our definition of cross-spectral density, Eq. (\ref{Gnor3far}), we use the complete field amplitudes in the far zone, which include a spherical phase factor.

The condition for the applicability of Eq. (\ref{Gfarappr2}) is $\hat{\theta}_x^2 \ll \sqrt{N_x}$ for $D_{x,y} \ll 1 \ll N_{x,y}$. For our purposes it is preferable to express the Wigner distribution at the source position in terms of the cross-spectral density in the far-zone. The field at $z = 0$ is related to the field in the far zone by  Eq. (\ref{farzonef}). Inserting this expression into Eq. (\ref{Wig13}), one finds

\begin{eqnarray}
&& W =  \frac{c  z_0^2}{(2\pi)^4}\frac{I}{e\hbar} \left(\frac{\omega}{c}\right)^2 \int d \Delta \vec{{\theta}} \exp\left(i \frac{\omega}{c} {\vec{r}} \cdot \Delta \vec{{\theta}}\right) \exp\left(-i \frac{\omega z_0}{c } \vec{{\theta}} \cdot{\Delta \vec{{\theta}}}\right)  \cr && \times G(\vec{{\theta}}, \Delta \vec{{\theta}})~.
\label{Wigfarfin}
\end{eqnarray}
If we now substitute Eq. (\ref{Gfarappr2}) in Eq. (\ref{Wigfarfin}) we obtain Eq. (\ref{Wthirdfin}) as must be,  but with the alternate derivation above we specified  more precisely  the conditions of applicability of Eq. (\ref{Wthirdfin}).

Let us finally investigate the more complicated beam size- and divergence-dominated regime. Formally we shall consider the limiting case when the dimensionless parameters of the electron beam distribution $N_{x,y}$ and $D_{x,y}$ are much larger than unity. However, even at $D_{x,y} \gg 1$ and $N_{x,y} \gg 1$ we have to distinguish between two limiting expressions for the brightness of a bending magnet, at variance with the single result obtained for undulators. In fact, for bending magnets, the dimensionless parameter $D_x/\sqrt{N_x}$ plays an important role, and one should additionally consider two limiting cases for $D_x/\sqrt{N_x}$ much smaller or much larger than unity. To explore the nature of this extra parameter, we begin by noting that another situation where a similar parameter exists was already discussed above, when treating the condition of applicability for the approximation of the Wigner distribution in the beam size-dominated regime.  From that case one can show that the condition $\hat{\theta_x}^2 \Delta \hat{\theta}_x \ll 1$  is a consequence of the small `aberration' influence within a window centered in the far zone at an angle $\hat{\theta}_x$ with a horizontal opening angle $\Delta \theta_x$. In fact, the electron motion in a bend has cylindrical symmetry with the vertical axis going through the center of the circular orbit. Therefore, an observer on-axis (Fig. \ref{srstand}) receives as much radiation from an electron with horizontal deflection angle $\hat{\theta}_x$ as an observer looking at an electron with zero deflection angle from an angle $\hat{\theta}_x$. It follows that the existence of the parameter $D_x/\sqrt{N_x}$  is a consequence of the small `aberration' influence within a window placed in the far zone about the $z$-axis, with a horizontal opening angle $\Delta \hat{\theta}_x \sim 1/\sqrt{N_x}$. Finally, it is worth emphasizing that the horizontal opening angle $1/\sqrt{N_x}$ is the angular dimension of a coherent area in the far zone. Therefore, one can summarize the previous observations by saying that the SR emitted within a solid angle of about $1/(\sqrt{N_x}\sqrt{N_y})$ is fully transversely coherent and additionally, at

\begin{eqnarray}
\frac{D_x}{\sqrt{N_x}} \ll 1 ~,
\label{condM2}
\end{eqnarray}
has an $M^2$ factor close to unity. Such fraction of radiation can be ultimately focused down to a spot size of dimension of about $\lambda^2$. In light of this, one sees that $D_x/\sqrt{N_x}$ is an important problem parameter that is required in order to calculate the bending magnet brightness in the beam size- and divergence-dominated regime. In particular, based on this intuitive reasoning we expect that in the asymptotic limit Eq. (\ref{condM2}) the brightness should be proportional to the coherent solid angle $1/(\sqrt{N_x}\sqrt{N_y}) \sim 1/(\sigma_x \sigma_y)$, but should not depend on the electron beam divergence in the horizontal direction, even when $D_x \gg 1$. In the following we will demonstrate that rigorous mathematical analysis confirms such expectation.

We begin by approximating the expression for the cross-spectral density in the far zone, Eq. (\ref{Gnor3far}), in the limiting case for $D_{x,y} \gg 1$ and $N_{x,y} \gg 1$. Suppose that condition Eq. (\ref{condM2}) is satisfied. From the analysis of the exponential functions in $N_{x,y}$ in Eq. (\ref{Gnor3far}) follows that $\Delta \theta_x$ and $\Delta \theta_y$ can be neglected in the expression for the field. Thus, the cross-spectral density is a product of two separate factors. The first is the Fourier transform of the transverse electron beam distribution. The second is the convolution of the electron beam angular distribution with the two dimensional function $|\widetilde{E}(\phi_x, \phi_y)|^2$, which is proportional to the angular intensity distribution of the radiation. Moreover, the range of the variable $\phi_y$ in this convolution integral is effectively limited up to values $|\phi_y| \sim 1$. In fact, $\phi_y$ enters the expression for the modulus of the electric field (see Eq. (\ref{EF})), and at values larger than unity the integrand in the convolution is suppressed. It follows that when $D_y \gg 1$ we can neglect $\phi_y$ in the exponential function $\exp[-(\phi_y - \hat{\theta}_y)^2/(2 D_y)]$ under the integral sign.  Eq. (\ref{Gnor3far}) is therefore approximated by

\begin{eqnarray}
&&G = \exp\left(-\frac{N_x \Delta \hat{\theta}_x^2}{2}\right)\exp\left(-\frac{N_y \Delta \hat{\theta}_y^2}{2}\right) \cr &&
\times \frac{1}{2 \pi\sqrt{D_x D_y}} \exp\left(-\frac{\hat{\theta}_y^2}{2D_y}\right)
\int_{-\infty}^{\infty} d \phi_x \int_{-\infty}^{\infty} d \phi_y \exp\left[-\frac{(\phi_x-\hat{\theta}_x)^2}{2D_x}\right] |\widetilde{E}(\phi_x, \phi_y)|^2~. \cr &&
\label{Gnor3farappr}
\end{eqnarray}
Since $\int d \phi_y |\widetilde{E}(\phi_x, \phi_y)|^2$ does not depend on $\phi_x$, we can take it out from under the integral over $\phi_x$. If we  now substitute this expression in the definition of the Wigner distribution, Eq. (\ref{Wigfarfin}) and perform the integral over $\Delta \hat{\theta}_{x,y}$ we obtain in dimensional units

\begin{eqnarray}
&&W(x, {y}, {\theta}_x, {\theta}_y) = \frac{1}{(2\pi)^{3/2} \sigma_x \sigma_y \sigma_{y'}}
\exp\left(-\frac{{x}^2}{2 \sigma_x^2}\right) \exp\left(-\frac{{y}^2}{2 \sigma_y^2}\right) \exp\left(-\frac{{\theta}_y^2}{2 \sigma_{y'}^2}\right)
\frac{dF}{d {\theta}_x}~ \cr &&
\label{Wfinfin}
\end{eqnarray}
where

\begin{eqnarray}
\frac{dF}{d {\theta}_x} =  \frac{d N_{ph}}{d {\theta}_x (d \omega/\omega)} = \frac{I}{e\hbar} \frac{c z_0^2}{(2\pi)^2} \int_{-\infty}^{\infty} d \theta_y |\widetilde{E}({\theta}_x, \theta_y)|^2 = \mathrm{constant}
\label{fluxdtheta}
\end{eqnarray}
is the photon flux per unit time per unit horizontal angle per relative spectral bandwidth. The brightness approximated by Eq (\ref{Btext}) coincides with the maximum of $W$ in Eq. (\ref{Wfinfin})  in the case when $D_{x,y} \gg 1$ and $N_{x,y} \gg 1$ and at arbitrary electron beam divergence in horizontal direction (i.e. at arbitrary $D_x$).  Now we demonstrated that the condition for validity of Eq. (\ref{Wfinfin}) is Eq. (\ref{condM2}).

Let us now consider the asymptotic case opposite to (\ref{condM2}):

\begin{eqnarray}
\frac{D_x}{\sqrt{N_x}} \gg 1 ~.
\label{condM2poor}
\end{eqnarray}
Up to now, for our calculations in the beam size- and divergence-dominated regime ($N_{x,y} \gg 1$ and $D_{x,y} \gg 1$) we used the Wigner function method in the far zone. In the asymptotic case (\ref{condM2poor}) it is more convenient to use the virtual source for calculating $W$. The cross-spectral density at the virtual source position is given by Eq. (\ref{Gnor3}). From the analysis of the exponential function in $D_{x,y}$ in Eq. (\ref{Gnor3}) follows that $\Delta \hat{x}$ and $\Delta \hat{y}$ can be neglected in the expression for the field. Thus, the cross-spectral density is a product of two separate factors. The first is the Fourier transform of the angular electron beam
distribution. The second is the convolution of the electron beam spatial distribution with the two dimensional function $|\widetilde{E}(\phi_x, \phi_y)|^2$, which is proportional to the intensity distribution of the radiation from a single electron at the source position. The range of the variable $\phi_y$ in the convolution integral is effectively limited up to values $|\phi_y| \sim 1$, because $\phi_y$ enters the expression for module of electric field
Eq. (\ref{E0X}). Therefore, when $N_y \gg 1$ we can neglect $\phi_y$ in the exponential function $\exp[-(\phi_y - \hat{y})^2/(2 N_y)]$ under the convolution integral. The approximation of Eq. (\ref{Gnor3}) is then given by

\begin{eqnarray}
&& G = \frac{1}{2 \pi \sqrt{N_x N_y}} \exp\left[-\frac{(\Delta \hat{x})^2 D_x }{2}\right]\exp\left[-\frac{(\Delta \hat{y})^2 D_y }{2}\right]
\exp\left(-\frac{\hat{y}^2}{2 N_y}\right) \cr && \times \int_{-\infty}^{\infty} d \phi_x \int_{-\infty}^{\infty} d \phi_y \exp\left[-\frac{(\phi_x-\hat{x})^2}{2 N_x}\right] |\widetilde{E}(\phi_x, \phi_y)|^2 ~,
\label{Gapprlast}
\end{eqnarray}
where $\widetilde{E}$ is the radiation field in Eq. (\ref{E0X}). Since $\int d \phi_y |\widetilde{E}(\phi_x,\phi_y)|^2$ depends on $\phi_x$, we cannot take it out from under the integral. If we substitute this approximation of $G$ in the definition of the Wigner function Eq. (\ref{WGdless}) and perform the prescribed integration we obtain the following result in dimensional units

\begin{eqnarray}
&& W(x,y, \theta_x, \theta_y) = \frac{1}{(2\pi)^{2}\sigma_{x}\sigma_{x'}\sigma_{y}\sigma_{y'}} \frac{I}{e \hbar}\frac{c}{(2\pi)^2} \cr && \times
\exp\left(- \frac{\theta_x^2}{2\sigma_{x'}^2}\right)\exp\left(-\frac{\theta_y^2}{2\sigma_{y'}^2}\right)
\exp\left(-\frac{y^2}{2\sigma_y^2}\right) \cr && \times \int_{-\infty}^{\infty} dx' \int_{-\infty}^{\infty} dy' ~|\widetilde{E}(x', y')|^2 \exp\left[-\frac{(x-x')^2}{2\sigma_x^2}\right]
\label{Wfin5p3}
\end{eqnarray}
Comparing Eq. (\ref{Gapprlast}) and Eq. (\ref{Wfin5p3}) one can see that in the beam size- and divergence-dominated regime, depending on the specific ratio between horizontal beam size and divergence, the Wigner distribution is described using functions  with completely different parametric dependence.  The brightness approximated by Eq. (\ref{Btext}) does not coincide with the maximum of $W$ in the case when condition (\ref{condM2poor}) is satisfied.  Finally, we should note that the usually accepted approximation for bending magnet brightness turns out to be parametrically inconsistent  not only in the intermediate geometrical optics asymptote when the beam divergence dominates over the diffraction angle, but also in the `simplest' geometrical optics asymptotic case when both beam size and beam divergence dominate over diffraction size and diffraction angle.

\section{\label{sec:quattro} Conclusions}

This document discusses the relation between statistical optics and the electromagnetic theory of SR radiation. A basic problem pertaining this relation is the definition of the SR brightness in terms of electromagnetic fields and their statistical properties. We consider the brightness defined with the help of a Wigner distribution. We propose a mathematical scheme and formulate a novel theory of SR brightness defined as the maximum of the Wigner distribution.

Formulating the theory of brightness in the language of Wigner distributions  has only one guideline, a particular correspondence principle. The conceptual foundation of this correspondence principle is based on the assumption that the formalism involved in the calculation of brightness must include radiometry as a limiting case. We use the classical definition of radiance to obtain a correct proportionality factor in the definition of brightness. In this way, in the geometrical optics limit, the brightness can be represented as the maximum value of the radiance. In classical radiometry the maximum of the radiance is the maximum of the photon flux density in phase space. Since the correspondence principle is integrated once and for all in the very foundation of our theory, one expects to eliminate the necessity of its recurrent application to every individual problem.

We compute various geometrical optics limits according to our definition of brightness, and we compare results with expectations from theories currently used in literature. In many cases we find a significant disagreement between exact calculations of the maximum photon flux density in phase space and the usually accepted approximations for undulator and bending magnet brightness.

\section{Acknowledgements}

We thank Oleg Gorobtsov for carefully reading our manuscript and for useful discussions.


\newpage

\section*{Appendix A. Undulator radiation in resonance approximation. Far zone.}

Calculations pertaining undulator radiation  are well established. see e.g. \cite{ELLE}. In this appendix we present a simple derivation of the frequency representation of the radiated field produced by an electron in an undulator. For the electron transverse velocity we assume

\begin{eqnarray}
v_x(z) = -{c \theta_s} \sin(k_w z) = -\frac{c \theta_s}{2
i}\left[\exp(ik_w z)-\exp(-i k_w z) \right]~. \label{vxpl_ap1}
\end{eqnarray}
Here  $k_w=2\pi/\lambda_w$, and $\lambda_w$ is the undulator
period. Moreover, $\theta_s=K/\gamma$, where $K$ is the deflection
parameter defined as

\begin{eqnarray}
K = \frac{e\lambda_w H_w}{2 \pi m_e c^2}~,\label{Kpar_ap1}
\end{eqnarray}
$m_e$ being the electron mass at rest and $H_w$ being the maximal
magnetic field of the undulator on axis.

We write the undulator length as $L = N_w \lambda_w$, where
$N_w$ is the number of undulator periods. With the help of Eq.
(\ref{generalfin}) we obtain an expression, valid in the far zone:

\begin{eqnarray}
&& {\vec{\widetilde{E}}}= {i \omega e\over{c^2 z_0}}
\int_{-L/2}^{L/2} dz' {\exp\left[i
\Phi_T\right]\exp\left[i\frac{\omega \theta^2 z_0}{2c}\right]}
\left[{K\over{\gamma}} \sin\left(k_w z'\right)\vec{e}_x
+\vec{\theta}\right]~. \cr && \label{undurad_ap1}
\end{eqnarray}
Here

\begin{eqnarray}
&& \Phi_T = \left({\omega \over{2 c \bar{\gamma}_z^2}}+ {\omega
\theta^2 \over{2  c }}\right) z' -
{K\theta_x\over{\gamma}}{\omega\over{k_w c}}\cos(k_w z') -
{K^2\over{8\gamma^2}} {\omega\over{k_w c}} \sin(2 k_w z')
~,\cr &&\label{phitundu_ap1}
\end{eqnarray}
where the average longitudinal Lorentz factor $\bar{\gamma}_z$ is
defined as

\begin{equation}
\bar{\gamma}_z = \frac{\gamma}{\sqrt{1+K^2/2}}~. \label{bargz_ap1}
\end{equation}
The choice of the integration limits in Eq. (\ref{undurad_ap1})
implies that the reference system has its origin in the center of
the undulator.

Usually, it does not make sense to calculate the intensity
distribution from Eq. (\ref{undurad_ap1}) alone, without extra-terms
(both interfering and not) from the other parts of the electron
trajectory. This means that one should have complete information
about the electron trajectory and calculate extra-terms to be
added to Eq. (\ref{undurad_ap1}) in order to have the total field from
a given setup. Yet, we can find \textit{particular situations} for
which the contribution from Eq. (\ref{undurad_ap1}) is dominant with
respect to others. In this case Eq. (\ref{undurad_ap1}), alone, has
independent physical meaning.

One of these situations is  when the resonance approximation is
valid. This approximation does not replace the paraxial one, based
on $\gamma^2 \gg 1$, but it is used together with it. It takes
advantage of another parameter that is usually large, i.e. the
number of undulator periods $N_w \gg 1$. In this case, the
integral in $dz'$ in Eq. (\ref{undurad_ap1}) exhibits simplifications,
independently of the frequency of interest due to the long
integration range with respect to the scale of the undulator
period.

In all generality, the field in Eq. (\ref{undurad_ap1}) can
be written as

\begin{eqnarray}
&&{\vec{\widetilde{E}}}= \exp\left[i\frac{\omega \theta^2
z_0}{2c}\right] \frac{i \omega e}{c^2 z_0} \cr && \times \int_{-L/2}^{L/2}
dz'\left\{\frac{K}{2 i \gamma}\left[\exp\left(2 i k_w
z'\right)-1\right]\vec{e}_x +\vec{\theta}\exp\left(i k_w
z'\right)\right\} \cr &&\times \exp\left[i \left(C + {\omega
\theta^2 \over{2 c }}\right) z' -
{K\theta_x\over{\gamma}}{\omega\over{k_w c}}\cos(k_w z')  -
{K^2\over{8\gamma^2}} {\omega\over{k_w c}} \sin(2 k_w z') \right]
~. \cr &&\label{undurad2_ap1}
\end{eqnarray}
Here $\omega = \omega_r + \Delta \omega$, $C =  k_w \Delta\omega/\omega_r$ and

\begin{eqnarray}
\omega_r = 2 k_w c \bar{\gamma}_z^2~, \label{res_ap1}
\end{eqnarray}
is the fundamental resonance frequency.

Using the Anger-Jacobi expansion:

\begin{equation}
\exp\left[i a \sin(\psi)\right] = \sum_{p=-\infty}^{\infty} J_p(a)
\exp\left[ip\psi\right]~, \label{alfeq}
\end{equation}
where $J_p(\cdot)$ indicates the Bessel function of the first kind
of order $p$, to write the integral in Eq. (\ref{undurad2_ap1}) in a
different way:

\begin{eqnarray}
&&{\vec{\widetilde{E}}}= \exp\left[i\frac{\omega \theta^2
z_0}{2c}\right] \frac{i \omega e}{c^2 z_0} \sum_{m,n=-\infty}^\infty
J_m(u) J_n(v) \exp\left[\frac{i \pi n}{2}\right] \cr && \times
\int_{-L/2}^{L/2} dz'\exp\left[i \left(C + {\omega \theta^2
\over{2 c }}\right) z'\right] \cr &&\times \left\{\frac{K}{2 i \gamma}
\left[\exp\left(2 i k_w z'\right)-1\right]\vec{e}_x
+\vec{\theta}\exp\left(i k_w z'\right)\right\}
\exp\left[i (n+2m) k_w z'\right] ~,\cr &&\label{undurad3_ap1}
\end{eqnarray}
where\footnote{Here the parameter $v$ should not be confused with the velocity.}

\begin{equation}
u = - \frac{K^2 \omega}{8 \gamma^2 k_w c}~~~~\mathrm{and}~~~v = -
\frac{K \theta_x \omega}{\gamma k_w c}~. \label{uv}
\end{equation}
Up to now we just re-wrote Eq. (\ref{undurad_ap1}) in a different way.
Eq. (\ref{undurad_ap1}) and Eq. (\ref{undurad3_ap1}) are equivalent. Of
course, definition of $C$ is suited to
investigate frequencies around the fundamental harmonic but no
approximation is taken besides the paraxial approximation.

Whenever

\begin{equation}
C  + \frac{\omega \theta^2}{{2 c}} \ll k_w \label{eqq_ap1} ~,
\end{equation}
the first phase term in $z'$ under the integral sign in Eq.
(\ref{undurad3_ap1}) is varying slowly on the scale of the undulator
period $\lambda_w$. As a result, simplifications arise when $N_w
\gg 1$, because fast oscillating terms in powers of  $\exp[i k_w
z']$ effectively average to zero. When these simplifications are
taken,  resonance approximation is applied, in the sense that one
exploits the large parameter $N_w \gg 1$. This is possible under
condition (\ref{eqq_ap1}). Note that (\ref{eqq_ap1}) restricts the range
of frequencies for positive values of $C$ independently of the
observation angle ${\theta}$, but for any value $C<0$ (i.e. for
wavelengths longer than $\lambdabar_r = c/\omega_r$) there is
always some range of $\theta$ such that Eq. (\ref{eqq_ap1}) can be
applied. Altogether, application of the resonance approximation is
possible for frequencies around $\omega_r$ and lower than
$\omega_r$. Once any frequency is fixed, (\ref{eqq_ap1}) poses
constraints on the observation region where the resonance
approximation applies. Similar reasonings can be done for
frequencies around higher harmonics with a more convenient
definition of the detuning parameter $C$.

Within the resonance approximation we further select frequencies
such that

\begin{eqnarray}
\frac{|\Delta \omega|}{\omega_r} \ll 1~,~~~~ \mathrm{i.e.}~~|C|
\ll k_w ~.\label{resext_ap1}
\end{eqnarray}
Note that this condition on frequencies automatically selects
observation angles of interest $\theta^2 \ll 1/\gamma_z^2$. In
fact, if one considers observation angles outside the range
$\theta^2 \ll 1/\gamma_z^2$, condition (\ref{eqq_ap1}) is not
fulfilled, and the integrand in Eq. (\ref{undurad3_ap1}) exhibits fast
oscillations on the integration scale $L$. As a result, one
obtains zero transverse field, $\vec{\widetilde{E}} = 0$,
with accuracy $1/N_w$. Under the constraint imposed by
(\ref{resext_ap1}), independently of the value of $K$ and for
observation angles of interest $\theta^2 \ll 1/\gamma_z^2$, we
have

\begin{equation}
|v|={K|\theta_x|\over{\gamma}}{\omega\over{k_w c}} =
\left(1+\frac{\Delta \omega}{\omega_r}\right) \frac{2 \sqrt{2}
K}{\sqrt{2+K^2}} \bar{\gamma}_z |\theta_x| \lesssim
\bar{\gamma}_z |\theta_x| \ll 1~. \label{drop}
\end{equation}
This means that, independently of $K$, $|v| \ll 1$ and we may
expand $J_n(v)$ in Eq. (\ref{undurad3_ap1}) according to $J_n(v)
\simeq [2^{-n}/\Gamma(1+n)]~v^n$, $\Gamma(\cdot)$ being the Euler
gamma function

\begin{eqnarray}
\Gamma(z) = \int_0^\infty dt~t^{z-1} \exp[-t] ~.\label{geule}
\end{eqnarray}
Similar reasonings can be done for frequencies around higher
harmonics with a different definition of the detuning parameter
$C$. However, around odd harmonics, the before-mentioned
expansion, together with the application of the resonance
approximation for $N_w \gg 1$ (fast oscillating terms in powers of
$\exp[i k_w z']$ effectively average to zero), yields
extra-simplifications.

Here we are dealing specifically with the first harmonic.
Therefore, these extra-simplifications apply. We
neglect both the  term in $\cos(k_w z')$ in the phase of Eq. (\ref{undurad2_ap1})
and the term in $\vec{\theta}$ in Eq. (\ref{undurad2_ap1}). First,
non-negligible terms in the expansion of $J_n(v)$ are those for
small values of $n$, since $J_n(v) \sim v^n$, with $|v|\ll 1$. The
value $n=0$ gives a non-negligible contribution $J_0(v) \sim 1$.
Then, since the integration in $d z'$ is performed over a large
number of undulator periods $N_w\gg 1$, all terms of the expansion
in Eq. (\ref{undurad3_ap1}) but those for $m=-1$ and $m=0$ average to
zero due to resonance approximation. Note that surviving
contributions are proportional to $K/\gamma$, and can be traced
back to the term in $\vec{e}_x$ only, while the
term in $\vec{\theta}$ in Eq. (\ref{undurad3_ap1}) averages to zero
for $n=0$. Values $n=\pm 1$ already give negligible contributions.
In fact, $J_{\pm 1}(v) \sim v$. Then, the term in $\vec{e}_x$ in
Eq. (\ref{undurad3_ap1}) is $v$ times the term with $n=0$ and is
immediately negligible, regardless of the values of $m$. The
term in $\vec{\theta}$ would survive averaging when $n=1,
~m=-1$ and when $n=-1, ~m=0$. However, it scales as $\vec{\theta}
v$. Now, using condition (\ref{resext_ap1}) we see that, for
observation angles of interest $\theta^2 \ll 1/\gamma_z^2$,
$|\vec{\theta}|~ |v| \sim (\sqrt{2}~ K~/\sqrt{2+K^2}~)
~\bar{\gamma}_z \theta^2 \ll K/\gamma$. Therefore, the
term in $\vec{\theta}$ is negligible with respect to the term in $\vec{e}_x$
for $n=0$, that scales as $K/\gamma$. All terms corresponding to
larger values of $|n|$ are negligible.

Summing up, all terms of the expansion in Eq. (\ref{alfeq}) but
those for $n=0$ and $m=-1$ or $m=0$ give negligible contribution.
After definition of

\begin{eqnarray}
A_{JJ} = J_0\left(\frac{\omega K^2}{8 k_w c \gamma^2}\right) -
J_1\left(\frac{\omega K^2}{8 k_w c \gamma^2}\right) ~,\label{AJJ}
\end{eqnarray}
that can be calculated at $\omega = \omega_r$ since $|C| \ll k_w$,
we have

\begin{eqnarray}
&&{\vec{\widetilde{E}}}= - \frac{K \omega e }{2
c^2 z_0 \gamma} A_{JJ} \exp\left[i\frac{\omega \theta^2 z_0}{2c}\right]
\int_{-L/2}^{L/2} dz' \exp\left[i \left(C + {\omega \theta^2
\over{2  c }}\right) z' \right] \vec{e}_x~,\cr &&\label{undurad5}
\end{eqnarray}
yielding the well-known free-space field distribution:

\begin{eqnarray}
&&{\vec{\widetilde{E}}}(z_0, \vec{\theta}) = -\frac{K \omega e
L  }{2 c^2 z_0 \gamma} A_{JJ}\exp\left[i\frac{\omega \theta^2
z_0}{2c}\right] \mathrm{sinc}\left[\frac{L}{2}\left(C+\frac{\omega
\theta^2}{2c} \right)\right] \vec{e}_x~,\cr && \label{generalfin4}
\end{eqnarray}
where $\mathrm{sinc}(\cdot) \equiv \sin(\cdot)/(\cdot)$.
Therefore, the field is horizontally polarized and azimuthal
symmetric.

\end{document}